 \documentclass[10pt,preprint]{aastex}  
 \usepackage{natbib}
\def\gapprox{\lower.4ex\hbox{$\;\buildrel >\over{\scriptstyle\sim}\;$}}
\def\lapprox{\lower.4ex\hbox{$\;\buildrel <\over{\scriptstyle\sim}\;$}}
\usepackage{graphics}
\usepackage{graphicx}
\usepackage{epsfig}
\usepackage{times}
\usepackage{makeidx}

\shortauthors{ASCHWANDEN}
\shorttitle{Self-Organization}

\begin{document}

\title{         Order out of Randomness : Self-Organization Processes 
		in Astrophysics }

\author{        Markus J. Aschwanden$^1$}
\affil{		$^1)$ Lockheed Martin, 
		Solar and Astrophysics Laboratory, 
                Org. A021S, Bldg.~252, 3251 Hanover St.,
                Palo Alto, CA 94304, USA;
                e-mail: aschwanden@lmsal.com }

\author{	Felix Scholkmann$^2$}
\affil{		$^2)$ Research Office for Complex Physical 
		and Biological Systems,
                Mutschellenstr. 179, 8038 Z\"urich, Switzerland;
                e-mail: felix.scholkmann@gmail.com }         

\author{	William B\'ethune$^3$}
\affil{		$^3)$ DAMTP, University of Cambridge,
		CMS, Wilberforce Road,
		Cambridge CB3 0WA, United Kingdom;
		e-mail: wb288@damtp.cam.ac.uk }	

\author{	Werner Schmutz$^4$}
\affil{		$^4)$ Physikalisch Meteorologisches Observatorium Davos
                and World Radiation Center,
		Dorfstrasse 33, 7260 Davos Dorf, Switzerland ; 
		e-mail: werner.schmutz@pmodwrc.ch }

\author{	Valentina Abramenko$^{5}$}
\affil{		$^{5})$ Crimean Astrophysical Observatory,
		Russian Academy of Science, Nauchny,
		Bakhchisaray 298409, Crimea ;
		e-mail: vabramenko@gmail.com }

\author{        Mark Cheung$^{6}$}
\affil{		$^{6}$ Lockheed Martin, 
		Solar and Astrophysics Laboratory, 
                Org. A021S, Bldg.~252, 3251 Hanover St.,
                Palo Alto, CA 94304, USA;
                e-mail: cheung@lmsal.com }

\author{	Daniel M\"uller$^{8}$}
\affil{		$^{8})$ European Space Agency ESTEC (SCI-9),
		P.O.Box 299,
		NL-2200 AG, The Netherlands;  
		e-mail: dmueller@cosmos.esa.int }

\author{	Arnold Benz$^{9}$}
\affil{		$^{9})$ Institute for Astronomy, ETH, Zurich;
		Institute of 4D Technologies, FHNW, Windisch,
		Switzerland ;
		e-mail: benz@astro.phys.ethz.ch }

\author{	Juergen Kurths$^{10}$}
\affil{		$^{10})$ Potsdam Institute for Climate Research,
		14473 Potsdam, Germany ;
		e-mail: juergen.kurths@pik-potsdam.de }

\author{	Guennadi Chernov$^{11}$}
\affil{		$^{12})$ Pushkov Institute of Terrestrial Magnetism,
		Ionosphere and Radio Wave Propagation,
		Russian Academy of Sciences, Ismiran, Russia ;
		e-mail: gchernov@izmiran.ru }
		 
\author{	Alexei G. Kritsuk$^{12}$}
\affil{		$^{12})$ Physics Department and
		Center for Astrophysics and Space Sciences,
		University of California, San Diego,
		La Jolla, CA 92093-0424, USA ;
		e-mail: akritsuk@ucsd.edu }

\author{	Jeffrey D. Scargle$^{13}$}
\affil{		$^{13})$ NASA Ames Research Center,
		Astrobiology and Space Science Division,
		Moffett Field, CA 94035, USA ;
		e-mail: jeffrey.d.scargle@nasa.gov }

\author{	Andrew Melatos$^{14}$}
\affil{		$^{14})$ School of Physics,
		University of Mebourne,
		Parkville, VIC 3010, Australia ;  
		e-mail: amelatos@unimelb.edu.au }

\author{	Robert, V. Wagoner$^{15}$}
\affil{		$^{15})$ Dept. of Physics and KIPAC,
		Stanford University,
		Stanford, CA 94305-4060, USA ;  
		e-mail: wagoner@stanford.edu}

\author{	Virginia Trimble$^{16}$}
\affil{		$^{16})$ University of California, Irvine, 
		Department of Physics and Astronomy,
		Irvine, CA 92697-4575, USA ; 
		e-mail: vtrimble@astro.umd.edu }

\author{	William Green$^{17}$}
\affil{		$^{17})$ Department of Physics,
		Florida State University,
		Tallahassee, Florida 32306, USA ;
		e-mail: whgreen@fsu.edu }

\begin{abstract}
Self-organization is a property of dissipative nonlinear processes 
that are governed by a global driving force and a local positive feedback 
mechanism, which creates regular geometric and/or temporal patterns,  
and decreases the entropy locally, in contrast to random processes.
Here we investigate for the first time a comprehensive number of 
(17) self-organization processes that operate in planetary physics, 
solar physics, stellar physics, galactic physics, and cosmology.
Self-organizing systems create spontaneous {\sl ``order out of randomness''}, 
during the evolution from an initially disordered system to an 
ordered quasi-stationary system, mostly by quasi-periodic limit-cycle dynamics,
but also by harmonic (mechanical or gyromagnetic) resonances. 
The global driving force can be due to gravity, electromagnetic forces,
mechanical forces (e.g., rotation or differential rotation), thermal pressure, 
or acceleration of nonthermal particles, while the positive 
feedback mechanism is often an instability, such as the 
magneto-rotational (Balbus-Hawley) instability, the convective 
(Rayleigh-B\'enard) instability, turbulence, vortex attraction, magnetic 
reconnection, plasma condensation, or a loss-cone instability. Physical 
models of astrophysical self-organization processes require hydrodynamic, 
magneto-hydrodynamic (MHD), plasma, or N-body simulations.  Analytical 
formulations of self-organizing systems generally involve coupled differential 
equations with limit-cycle solutions of the Lotka-Volterra or Hopf-bifurcation 
type.
\end{abstract}


.  \clearpage

\section{	INTRODUCTION			}

{\sl Self-organization is the spontaneous often seemingly purposeful
formation of spatial, temporal, spatio-temporal structures or
functions in systems composed of few or many components.
In physics, chemistry, and biology, self-organization occurs
in open systems driven away from thermal equilibrium. The process
of self-organization can be found in many other fields also, such
as economy, sociology, medicine, technology} (Haken 2008).
Self-organization creates {\sl ``order out of randomness''} that is
opposite to random processes with increasing entropy.
Self-organization is a spontaneous process that does not need
any control by an external force. It is often initiated by
random fluctuations where the local reaction is amplified
by a positive feedback mechanism. It can evolve into a 
stationary cyclic dynamics governed by a (strange) attractor, 
and develops, as a result of many microscopic interactions,
a macroscopic regular geometric spatial pattern 
(Nicolis and Prigogine 1977; Kauffman 1993, 1996).
In this review we compile for the first time 
a comprehensive set of self-organizing systems 
observed or inferred in astrophysics. 
For each astrophysical self-organizing system 
we discuss a physical model, generally in terms of a
system-wide driving force and a positive feedback mechanism,
which by mutual interactions evolve into a self-organized
quasi-stationary pattern that is different from a random structure.
Note that the term ``self-organization'' should not be confused 
with the term ``self-organized criticality'' (Bak et al.~1987;
Pruessner 2012; Aschwanden et al.~2016), which is just one
(of many) self-organizing complex systems, producing 
power law-like size distributions of scale-free avalanche events, 
whereas self-organizing systems usually evolve into a specific 
quantized (spatial or temporal) scale that is not scale-free.

Physical models of self-organization involve non-equilibrium 
processes, mechanical resonances, magneto-convection, 
plasma turbulence, superconductivity, phase transitions, 
or chemical reactions. In planetary physics, 
the principle of self-organization has been applied to harmonic orbit 
resonances (Aschwanden 2018; Aschwanden and Scholkmann 2018), 
Jupiter's or Saturn's rings and moons (Peale 1976), 
protoplanetary disks (Kunz and Lesur 2013; B\'ethune et al.~2016),
Jupiter's Red Spot (Marcus 1993), and the planetary entropy
balance (Izakov 1997).
In solar physics, it was applied to photospheric granulation 
(Krishan 1991, 1992),
solar magnetic fields (Vlahos and Georgoulis 2004; Kitiashvili et al.~2010),
the magnetic solar cycle (Hale 1908; Consolini et al.~2009), 
the evaporation-condensation cycle of flares
(Krall and Antiochos 1980; Kuin and Martens 1982), 
and to quasi-periodic solar radio bursts 
(Zaitsev 1971; Aschwanden and Benz 1988). 
In astrophysics, it was applied to
galaxy and star formation (Bodifee 1986; Cen 2014).
An overview of 17 self-organization processes operating
in astrophysical environments is given in Table 1,
which lists also the underlying physical driving forces 
and feedback mechanisms. 
Besides the astrophysical applications, the process of
self-organization can be found in many other fields, 
such as magnetic reconnection in laboratory physics
(Yamada 2007; Yamada et al.~2010; Zweibel and Yamada 2009;
http://cmso.uchicago.edu),
plasma turbulence (Hasegawa 1985), magnetospheric physics 
(Valdivia et al.~2003; Yoshida et al.~2010), 
ionospheric physics (Leyser 2001),
solid state physics and material science (M\"uller and Parisi 2015), 
chemistry (Lehn 2002), sociology (Leydesdorff 1993),
cybernetics and learning algorithms (Kohonen 1989; Geach 2012),
or biology (Camazine et al.~2001). A more specific overview
of self-organization processes in non-astrophysical fields
is provided in Table 2. 

In this review we discuss 17 different astrophysical processes that
exhibit self-organization. For the definition of the term
``self-organization'' we proceed pragmatically. A nonlinear dissipative 
process qualifies to be called a ''self-organization'' process if it
fulfills at least one of the following six criteria:
(S) A spatially ordered pattern that is significantly different
from a random pattern; (T) a temporally ordered (e.g., quasi-periodic)
structure that is significantly different from random time intervals;
(E) a system with negative entropy change ($dS < 0$);
(LC) a nonlinear dissipative system with limit-cycle behavior
(which by definition produces quasi-periodic temporal oscillations);
(R) a nonlinear dissipative system with resonances; and
(I) a nonlinear dissipative system that is driven by an external
force and counter-acted by a positive feedback force, triggered by 
an instability or turbulence. We classify the 17 analyzed self-organization 
processes with these defining criteria (S, T, E, LC, R, I) in Table 1.

The structure of this review is organized by the various subfields
in astrophysics, such as planetary physics (Section 2), solar physics
(Section 3), stellar physics (Section 4), galactic physics (Section 5),
and cosmology (Section 6). A discussion of randomness, self-organization,
and self-organized criticality processes (Section 7) and a summary
of the conclusions (Section 8) is given at the end. 
Each description of the 17 self-organization processes is 
annotated with a critical assessment at the end of each Section. 
The selected 17 cases are all explicitly
addressed as self-organization processes by the authors of the 
cited studies, but we are aware that there are many more phenomena
in astrophysics that implicitly qualify as self-organization processes,
although they are often not labeled as such in the original literature.
As a caveat, our review thus contains some bias towards citations with 
author-identified self-organization processes.

\clearpage

\section{	PLANETARY PHYSICS				}

\subsection{	Planetary Spacing 				}

Our solar system exhibits planet distances $R_i, i=1,...,n$ from the Sun
that are not randomly distributed, but rather follow a regular
pattern that has been quantified with the Titius-Bode law, 
known since 250 years. The original Titius-Bode law approximated the
planet distance ratios by a factor of two, i.e., $R_{i+1}/R_i
\approx 2$, while a generalized Titius-Bode law specified 
the relationship with a logarithmic spacing and a constant 
geometric progression factor $Q$, i.e., $R_{i+1}/R_i=Q$ (Blagg 1913). 
According to Kepler's third law, a distance ratio $Q$ corresponds
to a period ratio $q=T_{i+1}/T_i=Q^{(3/2)}$ of the orbital time periods $T$.
However, both the original and the generalized Titius-Bode law 
represent empirical laws without a physical model. 

Planet spacing with low harmonic ratios $q$ of their orbital time periods $T$, 
such as q=(m:n), with $n=1,2,3$ and $m \ge n+1$, have been interpreted as harmonic 
(mechanical) orbit resonances and are expected to occur in a self-organizing 
system with stable long-lived orbits (Laplace 1829; Peale 1976), 
especially in systems with resonant chains (e.g., Mills et al.~2016).
One of the main heuristic understandings of chaos and instability is that
a planet system is generated by overlapping resonances (Wisdom 1980),
which explains why not all (or even the majority of) planet systems have
exact (2-body) harmonic resonances (Daniel Fabrycky, private communication). 

\begin{figure}
\centerline{\includegraphics[width=0.9\textwidth]{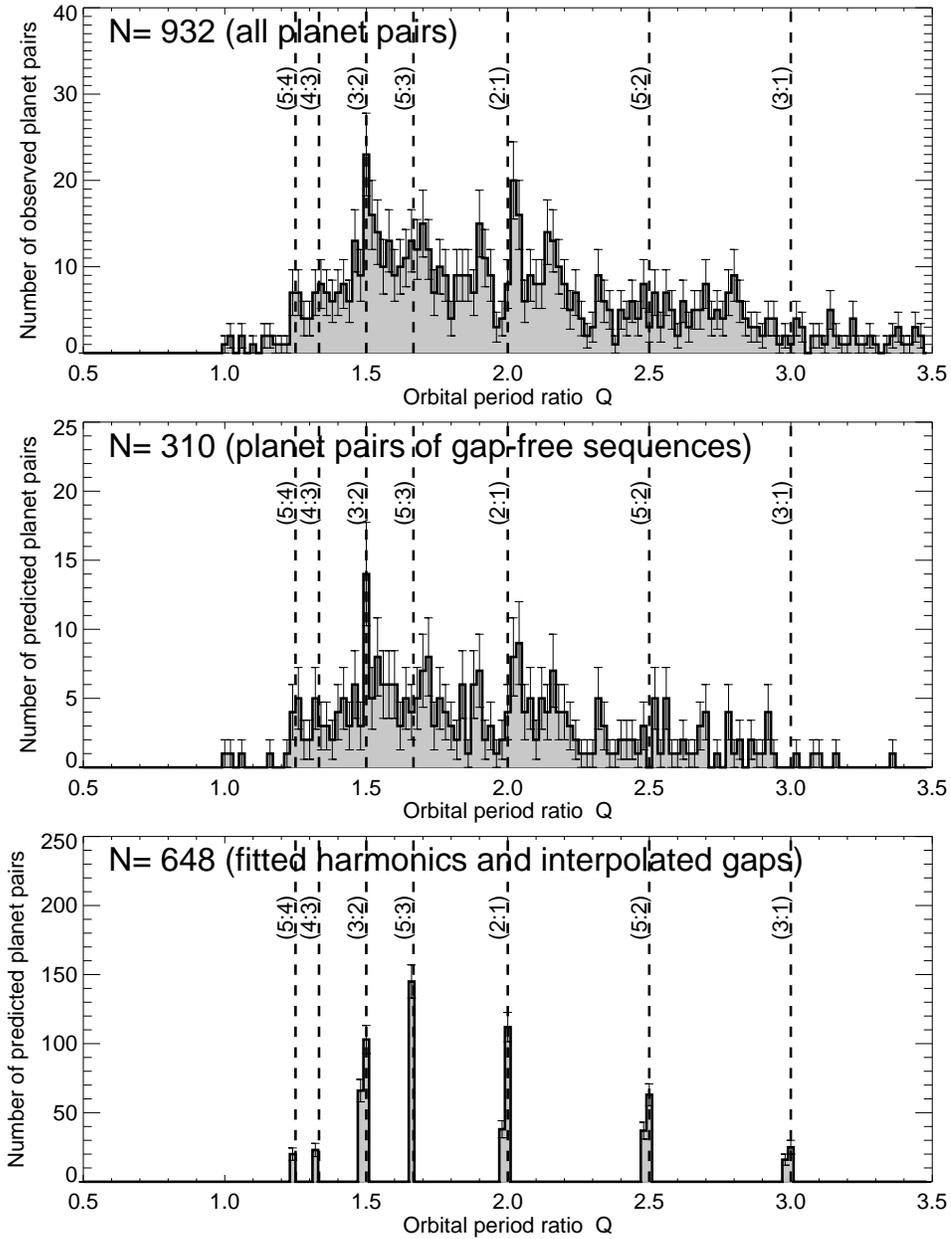}}
\caption{{\sl Top:} The distribution of orbital period ratios in
932 pairs of exo-planets observed with the KEPLER mission. 
{\sl Middle:} The distribution of 310 orbital period ratios in
gap-free sequences of exo-planets.
{\sl Bottom:} Quantized distribution of best-fit harmonic period 
ratios, including interpolation in ``gappy sequences''
[adapted from Aschwanden and Scholkmann 2018].}
\label{f_exo}
\end{figure}

Recently, the planet spacing has been fitted with 5 low-harmonic ratios 
(Aschwanden 2018), or with 7 low-harmonic ratios $q=(2:1), (3:1), (3:2), 
(4:3), (5:2), (5:3), (5:4)$ 
that were found to fit 648 pairs of exo-planet distances (Aschwanden and 
Scholkmann 2018), using observations of the KEPLER mission. A distribution of
the 7 best-fit harmonic ratios of orbital periods is shown for
detected and (interpolated) missing exo-planet pairs in Fig.~(\ref{f_exo}).
In other studies with Kepler data, resonances with low-harmonic ratios were 
found to be uncommon among small planets with periods shorter than a few years 
(Fabrycky et al.~2014; Winn and Fabrycky 2015).
Gaps with ratios $q > 3$ were interpreted as missing planets and interpolated
with low-harmonic ratios in the analysis of Aschwanden and Scholkmann (2018).

Most recently, the Laplacian 3-body resonances have been studied 
in great detail in the TRAPPIST-1 exo-planet system
(e.g., Luber et al.~2017; Scholkmann 2017),
which contains 7 planets and is continuously monitored by the Kepler mission.
The 6 planet spacings of TRAPPIST-1 closely match the low-harmonic ratios
$q=(4:3), (3:2), (5:3)$ within an accuracy of $\lapprox 1\%$ 
(Luger et al.~2017; Scholkmann 2017; Aschwanden and Scholkmann 2017). 

Harmonic planet orbits represent a special solution
of the general N-body problem in celestial mechanics, 
which can be expressed by $n$ second-order differential
equations,
\begin{equation}
	m_i {d^2 {\bf R}_i \over dt^2}
	=G \sum_{j=1}^{n} {m_i m_j \over r_{ij}^3}
	{\bf r}_{ij} \quad i=1,...,n, \quad i \neq j  \ ,
	\label{eq_grav}
\end{equation}
where $G$ is the Newton gravitational constant, $m_i$ and $m_j$
are two different masses, ${\bf R}_i$ and ${\bf R}_j$ are their spatial
vectors in a Cartesian coordinate system, and ${\bf r}_{ij}=
({\rm R}_j - {\rm R}_i)$ are the vectors between two bodies,
with ${\bf r}_{ij} = - {\bf r}_{ji}$. 

The dynamics of two planets orbiting the Sun can be
formulated with a N-body problem (with $N=3$), 
\begin{eqnarray}
  \ddot{x_1} &=& - G m_2 {(x_1-x_2) \over |x_1-x_2|^3} 
                 - G m_3 {(x_1-x_3) \over |x_1-x_3|^3} \label{eq_grav1} \\ 
  \ddot{x_2} &=& - G m_3 {(x_2-x_3) \over |x_2-x_3|^3} 
                 - G m_1 {(x_2-x_1) \over |x_2-x_1|^3} \label{eq_grav2} \\ 
  \ddot{x_3} &=& - G m_1 {(x_3-x_1) \over |x_3-x_1|^3} 
                 - G m_2 {(x_3-x_2) \over |x_3-x_2|^3} \label{eq_grav3}  
	\ ,
\end{eqnarray}
The 3-body problem is treated in the textbook {\sl Solar 
System Dynamics} by Murray and Dermott (1999) and recently 
reviewed in Lissauer and Murray (2007) and
Musielak and Quarles (2015), building on the work
of Isaac Newton, Jean le Rond d'Alembert, Alexis Clairaut,
Joseph-Louis Lagrange, Pierre-Simon Laplace, 
Heinrich Bruns, Henri Poincar\'e, and Leonard Euler. 
Some restricted solutions yield stationary orbits in the 
Lagrangian points L1 to L5.  Numerical searches for
periodic orbits and resonances based on approximations 
to harmonic oscillators (similar to the physical model 
of coupled pendulums) yield the following
nominal resonance location $a_3$ for a third body that
orbits between the primary and secondary body (internal
resonance) (Murray and Dermott 1999),
\begin{equation}
	a_3 = \left({ k \over k+l} \right)^{2/3}
		\left( {m_1 \over m_1 + m_2} \right) a_2 \ ,
	\label{eq_res1}
\end{equation}
where $m_1$ is the mass of the first body (e.g., the Sun),
$m_2$ the mass of the secondary body (e.g., Venus),
$a_2$ is the semi-major axis of the secondary body, 
$a_3$ the distance of the third body (e.g., Mercury)
that orbits between the first and second body, 
$l$ is the order of the resonance, and $(k,l)$ are 
integer numbers. Since the planet masses are
much smaller than the solar mass, the relationship (Eq.~\ref{eq_res1})
simplifies to,
\begin{equation}
	a_3 \approx \left({ k \over k+l} \right)^{2/3} a_2 \ ,
	\label{eq_res2}
\end{equation}
where the exponent $(2/3)$ results from Kepler's third law, e.g., 
$a \propto T^{(2/3)}$, with $T$ the orbital period, while orbital 
periods have harmonic integer ratios $q=T_2/T_3 \propto (k+l)/k$. 
This yields the ratios $q=(2:1)$, (3:2), (4:3), (3:1), (5:3), (4:1), 
(5:2) for the lowest orders $l=1, 2, 3$. In Table 3 we list
the harmonic ratios (H1:H2) from our solar system, or the order 
of the resonances [l,k]  that fit the observed orbital periods best,
which includes the harmonic ratios (2:1), (3:1), (3:2), (5:2), 
and (5:3). The resulting planet distance ratios agree with the 
observed semi-major axis with an accuracy of about 2\%,
($a_{harm}/a_{obs}=0.99 \pm 0.02$, i.e., see mean and standard
deviation of ratios in last column of Table 3), 
which clearly demonstrates
that the spacing of planets obeys a regular pattern that is not
consistent with random locations. In the terminology of
self-organizing systems, the driver of the system is the
gravitational force, while the feedback mechanism
that creates order out of random is the orbit stabilization
that occurs at low harmonic ratios. Planets may have been formed initially
at ``chaotic'' distances from the Sun, but the long-term stable 
orbits survive in the end, which apparently require 
gravitational resonances at low-harmonic orbital ratios. 

The planetary spacing can be described in terms of two hierarchical
self-organization processes: (i) the Keplerian orbital motion, and 
(ii) the secular precession. The Keplerian orbital motion is driven 
by the gravitational force, while the balance with the centrifugal 
force represents the feedback mechanism, resulting into an ellipse 
trajectory with some eccentricity (Appendix A). This can be considered
as a self-organizing system with a limit cycle that corresponds to the
orbital period. If the planet has a large eccentricity, the Sun-planet
distance varies around the equilibrium value, while a circular motion 
corresponds to a fixed limit cycle with a constant distance from the Sun. 
We show a phase diagram of the planet velocity $v$ as a function of the
distance $R$ in Fig.~(\ref{f_kepler}). On top of the Keplerian motion we have
gravitational disturbances from other planets that vary the secular
motion of the planet. Gravitational disturbances are then the driving
forces, while the low-harmonic resonances represent the feedback
mechanisms that self-organize multiple planet distances into a quantized
(non-random) spatial pattern. This is illustrated by the harmonic
ratios of the planet distances shown in Fig.~(\ref{f_exo}). In essence, two
self-organization mechanisms control the orbits of planets. 

Alternative mechanisms besides gravitational N-body resonance 
self-organization have been proposed also, such as: 
(i) Hierarchical self-organization processes based on sequential 
resonance accretion (starting with the accretion of massive objects 
first) and 2-body resonance capture of planetesimals in the 
primordial solar nebula (Patterson 1987); 
(ii) plasma self-organization driven by
the development to minimum energy states of the generic solar
plasma during protostar formation (Wells 1989a, 1989b, 1990);
(iii) susequent mass ejections into planetary rings around a central
rotating body with magnetic field properties predicted by stochastic 
electrodynamics (Surdin et al.~1980), 
(iv) retarded gravitational 2-body resonance, i.e., macroscopic 
quantization of orbital parameters due to finite gravitational 
propagation speed (Gine 2007); or
(v) quantization of orbital periods in terms of the quantum-mechanical
Schr\"odinger equation (Perinov et al.~2007; De Neto et al.~2004;
Scardigli 2007; Chang 2013).

{\sl \underbar{Critical Assessment:} The spacing of planets, moons, or 
exo-planets exhibit quantized values that correspond to low-harmonic ratios
according to some studies, in which large period ratios of planet pairs are
interpreted as gaps with missing (un-detected or non-existing) planets.
A regular pattern of orbital periods (T), produced by low-harmonic ratios of 
orbital resonances (R), causes then also a regular pattern  
in planetary spacings (S), via Kepler's third law. 
Other studies find that harmonic ratios are rare for exo-planets with 
orbital periods of less than a few years. 
The physical model of Lagrangian mean-motion resonances 
predicts exact harmonic ratios (in resonant chains), but secular disturbances,
planet migrations, and overlapping resonances (Wisdom 1980) may cause 
slowly-varying deviations. Nevertheless, the fact that harmonic ratios 
fit the planet orbital periods in the order of a few percents, 
strongly indicates the presence of a self-organizing system, opposed 
to randomness (Table 1: qualifiers R,S,T).} 

\subsection{	Planetary Rings and Moons			}

Planetary systems with moons and rings can be considered as
miniature versions of solar (or stellar) systems, as noted
by Galileo, and thus may have a similar formation process 
and are governed by the same celestial mechanics. For instance,
the mean motions of the inner three Galilean satellites of Jupiter 
(Io, Europa, Ganymede) exhibit harmonic orbits with a very
high precision (by nine significant digits; Peale 1976),
a property that has been interpreted by Laplace (1829) as evidence
for the high stability of resonant orbits. Besides the
Galilean satellites, further orbital resonance commensurabilities
were found for Saturn moons (Franklin et al.~1971; Sinclair 1972; 
Greenberg 1973; Colombo et al.~1974; Peale 1976), and for 
asteroids-Jupiter resonances such as the Trojans
(Brown and Shook 1933; Takenouchi 1962; Schubart 1968; 
Sinclair 1969; Marsden 1970; Lecar and Franklin 1973;  
Franklin et al.~1975; Peale 1976). Planetary rings have been found for
all giant planets (Jupiter, Saturn, Uranus, and Neptune).
A reconstruction of the Saturn ring system from Cassini
observations is shown in Fig.~(\ref{f_saturn}).

If we hypothesize that planets and moons are preferentially
located at low-harmonic orbits, how do we explain the existence
of gaps in a ring system, such as the Cassini division or the
Encke gap in Saturn's ring system? If moons form by
accretion of planetesimals that orbit in close proximity to
the accreting moon, a gap will result after sweeping over 
many nearby orbits, with the growing moon sitting in the middle of 
the gap. Therefore, gaps and moons are essentially cospatial 
in a long-term stable system. The most prominent ``shepherding 
moon'' in Saturn's ring system is the satellite Mimas, which
is responsible for the strongest resonance, i.e., 
the Cassini Division, a 4700-km gap between Saturn's A and B
rings (Porco and Hamilton 2000; McFadden et al.~1999, 2007). 
The two smaller moons Janus
and Epimetheus cause the sharp outer edge of the A ring.
The 320-km Encke gap in the outer A ring is believed to be
controlled by the 20-km diameter satellite Pan. At Uranus,
Cordelia and Ophelia have the role of ``shepherding moons''.
The moon Galatea plays a similar role in Neptune's Adams ring.

The idea of self-organization in planetary rings has already
been raised by Gor'kavyji and Fridman (1991). Gravitational
forces and collisional deflection represent the drivers,
while harmonic orbit resonances produce a feedback mechanism
to organize the flat planetary ring plane into discret rings,
mostly because harmonic orbits tend to be more stable 
statistically. If the phenomenon of harmonic orbit resonances
would not exist, randomized collisions only would determine
the dynamics of ring particles, leading to a smooth and 
homogeneous planetary disk (or an asteroid belt or 
Oort cloud), rather than to quantized rings. 
The spatial pattern of rings with quantized ratios in their
distance from the center of a planet (e.g., Saturn, Jupiter,
Uranus, Neptune) thus is a manifestation of a self-organizing
``ordered'' scheme beyond a random pattern.

{\sl \underbar{Critical Assessment:} The argument to explain
the harmonic structure of planetary rings is idential to the
previously discussed case of planetary distances, because both
are believed to be produced by the same stabilizing effect of
orbital resonances with low-harmonic ratios. The arrangement of
rings in quantized distances reveals a regular pattern in space (S) 
and time (T) that is beyond randomness, governed by mechanical 
resonances (R). These properties (R,S,T) argue in favor of a
self-organization system (Table 1: qualifiers R,S,T).}

\subsection{	Protoplanetary Disks  				}

The formation of planets can obviously be seen as a self-organizing process, 
creating "order out of randomness". The interstellar gas, initially randomly 
distributed in a molecular cloud, collapses under its own gravity to form 
a young stellar object. Unless it loses its angular momentum, the gas cannot 
directly fall onto the newly born star: its angular velocity would increase 
and matter would be centrifugally expelled at larger radii. In the frame 
co-rotating with the gas, the effective gravitational potential is minimal 
in a plane, where dissipative processes allow the protoplanetary 
(or circum-stellar) disk to form. Dozens of such disks have now been observed over 
a range of wavelengths (McCaughrean and O'dell 1996), and their link to planet 
formation casts no doubt, since planets have been observed in older 
``debris disks'' (Kospal et al.~2009).  
The imaging of dust emission, whether thermal or scattered, has revealed a 
number of large-scale structures in protoplanetary disks. Such features 
include spiral arms (Muto et al.~2012; Benisty et al.~2015) or cavities in the 
innermost regions of the disk (e.g., Andrews et al.~2011). The former 
are generally attributed to the excitation of density waves by massive 
planets, while the latter could result from accretion and/or photo-evaporation 
of the inner disk (Alexander et al.~2006; Koepferl et al.~2013). Horseshoe-shaped 
dust concentrations have also been identified in several disks 
(Fukagawa et al.~2013; van der Marel et al.~2013); it is commonly agreed that 
these could correspond to large-scale anticyclonic vortices in the gas flow 
(Birnstiel et al.~2013).  

The most puzzling structures remain the axisymmetric dust gaps and rings 
observed in some disks, see Fig.~(\ref{f_hltauri}). It is tempting to 
attribute them to gaps carved by protoplanets and their gravitational 
resonances (Crida et al.~2006; Baruteau and Papaloizou 2013), 
but it is unclear how several massive 
planetary bodies could already be formed in such young disks. One class 
of mechanisms relies on the coupling between the gas and large-scale 
magnetic fields. Magnetic fields are thought to drive the bipolar jets 
emitted perpendicularly to the disk plane (Cabrit et al.~2011). The 
coupling of magnetic fields with the electrically neutral gas in the 
outer disk is still possible via collisions with the few charged species 
(e.g., Wardle and Ng 1999). Of particular relevance for this review, 
the magneto-hydrodynamic (MHD) mechanism identified by Kunz and Lesur (2013)
and further investigated by B\'ethune et al.~(2016) generates self-organized, 
regularly spaced axisymmetric structures in the gas flow. Such structures 
would affect the migration of dust grains and could produce dust rings 
and gaps.  MHD processes have received an increasing interest after 
realizing that for perfectly ionized Keplerian disks, arbitrarily weak 
magnetic fields could trigger a linear instability, the magneto-rotational 
instability (MRI) (Balbus and Hawley 1991), saturating in a turbulent state. 
In this turbulent flow, angular momentum could be ``viscously'' transported 
outwards (Shakura and Sunyaev 1973), thus allowing the observed accretion of gas 
onto the star. In weakly ionized plasmas, this instability can be damped 
(Jin 1996; Kunz and Balbus 2004)
or modified in nature (Balbus and Terquem 2001; Kunz 2008). 
The transport of magnetic field in weakly ionized disks can be described 
via a modified induction equation:
\begin{equation} 
	{\partial B \over \partial t}
	= \nabla \times \left[ v \times B - \eta_{\mathrm{O}} J - 
	\eta_{\mathrm{H}} J \times e_B + \eta_{\mathrm{A}} 
	\left( J \times e_B \right) \times e_B \right] \ ,
	\label{eq_protoplanet1}
\end{equation}
where $B$ is the magnetic field locally along $e_B$, $J = \nabla \times B$ 
is the electric current density, and $\eta_{\mathrm{O},\mathrm{H},\mathrm{A}}$ 
are the Ohmic, Hall and ambipolar diffusivities. Ohmic and ambipolar diffusions 
are indeed dissipative terms, respectively caused by collisions of electrons 
and ions. The Hall term is not a dissipative one: it describes the collisionless 
drift between electrons and ions and can only transport magnetic energy via 
whistler waves. Retaining only the ideal and Hall terms amounts to neglecting 
the ion dynamics, following the induction of magnetic field by electrons only. 
In this limit, a linear instability remains that could sustain the turbulent 
transport of angular momentum in accretion disks (Wardle 1999). 
Early simulations including the Hall term showed that the Hall-MRI would still 
saturate in a turbulent state (Sano and Stone 2002a,b), though with varying 
effective viscosities. However, the Hall term might largely dominate the ideal 
induction term in the midplane of protoplanetary disks (Kunz and Balbus 2004). 
In this regime, the Hall-shear instability still operates in Keplerian disks,
but with a different outcome (Kunz and Lesur 2013). After a phase of 
linear growth, the instability breaks into a non-linear and disordered regime. 
From this turbulent phase, high magnetic flux regions progressively merge 
together, ultimately separating contiguous regions of strong magnetic field 
from regions devoid of magnetic flux. 

This behavior can be understood as follows. The linear instability requires 
a magnetic field that is sufficiently weak, such that the shear rate of the 
flow is larger than the whistler waves frequency at a given scale. 
Note, if the flow along $v_y$ is sheared in the $x$ direction, then 
the shear rate is defined as $\partial_x v_y$. 
Besides, the instability generates a 
magnetic stress $\mathcal{M} = - B \otimes B$, i.e. a tension of the magnetic 
field lines that can exchange momentum with the plasma. Retaining only the 
Hall term, Eq.~(\ref{eq_protoplanet1}) can be recast
\begin{equation} 
	{\partial B \over \partial t}
	= \ell_{\mathrm{H}}\ \nabla \times \left[\nabla \cdot \mathcal{M} \right] \ ,
	\label{eq_protoplanet2}
\end{equation}
where $\ell_{\mathrm{H}} = \eta_{\mathrm{H}} / v_{\mathrm{A}}$, the effective 
Hall diffusivity divided by the Alfv\'en velocity, happens to be independent 
of the magnetic field intensity (e.g., Lesur et al.~2014); this coefficient, 
analogous to an ion skin depth, was assumed to be constant for simplicity. 
Projected on the direction normal to the disk, this equation implies that 
magnetic flux is transported away from stress maxima, and this opens a route 
to self-organization. In the limit of weak magnetic flux, the linear 
instability has accordingly small growth rates and does not generate a 
significant stress. In the limit of strong magnetic flux, whistler waves can 
propagate despite the strong shear, when the Keplerian flow becomes linearly 
stable. For intermediate intensities of the magnetic flux, the instability 
generates a magnetic stress that effectively pushes magnetic flux away. 
If the magnetic flux locally increases, the flow can be stabilized, the 
magnetic stress becomes locally minimal, and therefore the stabilized region 
becomes a sink for magnetic flux. Eventually, these magnetic concentrations 
grow and spread in the azimuthal direction. If something tries to spread 
the magnetic flux radially, this will decrease its intensity down to the 
point where the linear instability is triggered again; as a feedback, the 
instability generates magnetic stress, thus confining magnetic flux again. 
Given the total magnetic flux through the disk, the turbulent and ordered 
phases are two available outcomes for the flow. The Hall effect, when 
strong enough, allows a spontaneous transition from the turbulent phase 
to an ordered equilibrium featuring large-scale and long-lived structures. 
Its relevance to astrophysical disks is uncertain though. The main caveat 
of these studies is the neglect of vertical stratification, i.e. the 
transition from the dense disk to its dilute and strongly magnetized 
coronna. Results from numerical simulations (Fig.~\ref{f_bethune})
including all three non-ideal MHD terms in Eq.~(\ref{eq_protoplanet1})
suggest that self-organization is inhibited by the density stratification 
(Lesur et al.~2014; Bai 2015; B\'ethune et al.~2017). Still, 
striped structures have been observed in stratified simulations of strongly 
magnetized disks (Moll 2012); axisymmetric magnetic accumulations 
could be a generic feature of MHD turbulent disks (Bai and Stone 2014;
Ruge et al.~2016), most apparent in the presence of ambipolar diffusion 
(B\'ethune et al.~2017; Simon et al.~2017). At the moment, this behavior 
lacks a robust explanation. 

{\sl \underbar{Critical Assessment:} The argument of a self-organization 
process in the evolution of a protopoanetary disk is mostly made in
terms of the spatially emerging order (S), which starts from random-like 
turbulent flows with a complex fine structure and ends up in almost
equidistantly ordered rings. From the 3-D MHD simulations it appears
that the Hall-shear instability (I) acts as a feedback mechanism 
to organize an initially ``chaotic'' axis-symmetric disk into an ordered 
system of bands. These properties (I,S) argue for a
self-organization process (Table 1: qualifiers I,S).}

\subsection{	Jupiter's Red Spot		}

Jupiter exhibits a stable {\sl Great Red Spot} since 187 years 
(or possibly since 350 years), which indicates a high-pressure zone 
of a persistent anticyclonic storm (Fig.~\ref{f_jupiter}). 
The vortex-like velocity field in Jupiter's Red Spot has been derived
and rendered in Fig.~(\ref{f_jupiter_vortex}) by Simon et al.~(2014). 
The temperature of Jupiter's 
atmospheres above the Great Red Spot is measured to be hundreds of degrees 
warmer than simulations based on solar heating alone can explain 
(O'Donoghue et al.~2016).  The Great Red Spot has a width of $\approx 16,000$ km 
and rotates counter-clockwise with a period of $\approx 3$ days. The longitude
of the Great Red Spot oscillated with a 90-day period (Link 1975; Reese and Beebe 1977).
Why can such an ordered, stable, long-lived structure exist in the (randomly) 
turbulent atmosphere of a gas giant ? Why would it not decay into similar
turbulent structures as observed in the surroundings ? There exists 
a similar feature in Neptune's atmosphere, visible during 1989-1994, 
called the {\sl Great Dark Spot}.

Early interpretations associated Jupiter's Great Spot with
a Korteweg-de Vries soliton solution (Maxworthy and Redekopp 1976),
a solitary wave solution to the intermediate-geostrophic equations
(Nezlin et al.~1996), a Taylor column, a Rossby wave, or a hurricane
(Marcus 1993). A geostrophic wind or current results from the balance 
between pressure gradients and Coriolis forces.
One theoretical explanation that was put forward is the
self-organization of vorticity in turbulence: 
The Jovian vortices reflect the behavior of quasi-geostrophic
vortices embedded in an east-west wind with bands of uniform
potential vorticity (Marcus 1993).
Numerical simulations based on the quasi-geostrophic equations
for a Boussinesq fluid in a uniformly rotating and stably
stratified environment indicated the self-organization of the
flow into a large population of coherent vortices 
(McWilliams et al.~1994). This scenario suggests an evolution from 
initial turbulent (random) to coherent (ordered) large-scale structures.

The vortex-solution of the Rossby wave equation gives not only a 
solution resembling of the Great Red Spot but is also similar to 
a drift soliton in plasma where the Coriolis force (in a rotating 
atmosphere) is replaced by the Lorenz force (in a magnetized plasma) 
(Petviashvili, 1980). The similarity of these phenomena not only 
indicates similar self-organizing principle behind them, but also may 
hint that the Great Dark Spot is the result of a MHD process. 

{\sl \underbar{Critical Assessment:} The argument of Jupiter's Great Red
Spot being a self-organizing structure is mostly based on the emergence
of a stable ordered large-scale structure (S), which is opposite to
random-like turbulent small-scale structures. It is also the
longevity of this large-scale structure that sets the Great Red Spot
appart from short-lived small-scale turbulent structures. The physical
process has been modeled with MHD simulations, essentially showing
an inverse MHD turbulent cascade (from small to large scales),
as it is known in 2-D turbulence (I). Thus, self-organization is
established based on the properties I,S (Table 1: qualifiers I,S).}

\subsection{  	Saturn's Hexagon 		 	}

Saturn's north pole exhibits at 77$^\circ$ N a hexagonal cloud pattern
that was first discovered in the 1980s by the Voyager mission (Godfrey 1988),
which was later imaged with high resolution by the Cassini Orbiter 
(Baines et al.~2009). The images obtained by Cassini revealed that the 
structure consists of two elements: a hexagonal circumpolar jet-stream and 
a North Polar vortex (NPV), see Fig.~(\ref{f_saturn_hexagon}).
Recently, Rostami et al.~(2017) showed by 
computational simulations that the cloud pattern can be described as a 
coupled dynamical system consisting of the hexagonal circumpolar jet-stream 
and the NPV, resulting in a self-organized stable hexagonal pattern. 
The hexagonal shape is formed in a specific region of the turbulent flow 
between the jet-stream and the NPV that rotate with different speeds; 
the hexagonal shape is stabilized by the NPV. The concentric ring 
structure surrounding the vortices at the north and south pole and 
their peculiar temperature distribution (Fletcher et al.~2008), 
the occurrence of auroras at the poles (Dyudina et al.~2016), as well
as an electrodynamic coupling of Saturn with his moons, e.g. Enceladus 
(Pontius and Hill 2006; Tokar et al.~2006), indicate that the cloud 
structures seem on the 
poles may be also related to plasma-physical and electrical phenomena. 
Indeed, laboratory studies of plasma discharge showed structures 
occurring at the diocotron instability (analogous to the Kelvin-Helmholtz 
instability in fluid mechanics) that resemble structures (discharge and 
cloud formations) at planetary poles (Parett et al., 2007).

{\sl \underbar{Critical Assessment:} The argument of Saturn's hexagon 
self-organizing structure is, as in the case of Jupiter's Great Spot, 
mostly based on the emergence of a stable ordered large-scale structure 
(S), which is opposite to random-like turbulent small-scale structures. 
The modelling of the cloud structures based on fluid dynamics or plasma 
physics shows that instabilities (I) are involved in the organization 
process. The self-organization is thus established by properties I and S.}

\subsection{  	Planetary Entropy 		 	}

Random processes increase the entropy according to the second
thermodynamic law, while self-organizing processes decrease the
entropy, which is also expressed as an increase of negentropy
(negative entropy). The entropy of a nonequilibrium system
can be defined by the Gibbs formula,
\begin{equation}
	dS = {dE \over T} + {p dV \over T} \ ,
	\label{eq_entropy1} 
\end{equation}
where $dS$ is the entropy flux of an open system, $E$ is the internal
energy flux, $T$ is the temperature, $p$ is the pressure, and 
$V$ is the volume. For planets, volume changes $dV$ can be 
neglected. For energy balance one needs to include the
solar radiation (or energy flux) $E_s$ absorbed by the planet, and the
infrared radiation (or energy flux) $E_p$ emitted by the planet (Izakov 1997),
\begin{equation}
	dE = E_s - E_p = f_s (1-A)\pi r^2 - 4\pi r^2 f_p \ ,
	\label{eq_entropy2}
\end{equation}
where $f_s$ is the incident solar radiation per unit area (or irradiance),
$E_s$ is the incident energy flux from the Sun,
$E_p$ is the outgoing energy flux from the planet,
$f_p \approx \sigma_B T_e^4$ is the infrared radiation
emitted from the unit area of the planet's surface,
$T_e$ is the equilibrium temperature, $A$ is the integral 
spherical albedo of the planet, and $r$ is the radius of
the planet. The average energy flux imbalance $dE$ of the Earth 
at the top of the atmosphere is a crucial number characterizing
the status of climate change. In practice, it is very difficult
to measure the imbalance accurately. For the approach here, 
it is sufficient to note that it is found
to be approximately zero. The energy flux balance of incoming and outgoing 
energy fluxes in the Earth's atmosphere is depicted in Fig.~(\ref{f_climate})
(energy flux imbalance numbers at lower left).

An interesting consequence of the energy flux balance $dE \approx 0$
is the amount of negentropy flux that flows into a planet system.
The difference of the entropy flux input from the Sun and the
entropy flux output from the planet, representing the amount that goes
into self-organization processes, can be estimated to be 
\begin{equation}
	dS = {4 \over 3} \left( {E_s \over T_s} - {E_p \over T_p} 
	\right) \ ,
	\label{eq_entropy3}
\end{equation}
which is found to be negative $(dS < 0)$ for energy flux balance
$E_p \approx E_s$ and blackbody equilibrium temperatures
$T_s \gg T_p$, since the Sun is much hotter than the planet.
In the following estimates we adapt nominal solar and terrestrial
quantities from Pr\v sa (2016). 
Approximating with blackbody temperatures, we have $T_s=5772$ K
for the Sun, $T_p=255$ K for the temperature of the Earth's 
thermal radiation, $f_s=1361$ W m$^{-2}$ for the solar
irradiance, and $A=0.29$ for the albedo, yielding a
negentropy flux of $-dS = 9 \times 10^{14}$ W K$^{-1}$.
The greenhouse effect, which yields a higher temperature of
$T_0=288$ K near the surface than the equilibrium
temperature $T_e=255$ K, ensures the existence of water
and the biosphere on the Earth. About 70\% of the
negentropy flux inflowing to Earth accounts for the maintenance
of the thermal regime on the planet. About 25\% of the
negentropy flux is spent on the evaporation of water, mostly
from the surface of the oceans, supplying clouds and
rainfall for the vegetation. Only a
small fraction of about 5\% goes into flows of mass and
heat, tsunamis, hurricanes, etc. On Venus, where no
water is, a larger fraction of negentropy flux goes into the
dynamics of the atmosphere. Therefore,
{\sl the greenhouse effect, the hydrologic cycle of water,
the global circulation of the atmosphere and oceans,
are essentially dissipative structures supported by the
supply of negentropy and making up the global
self-organizing system whose characteristic is the
climate on the Earth} (Izakov 1997). 

Global energy flux budgets and {\sl Trenberth diagrams}
for the climates of terrestrial and gas giant planets are
given in Read et al.~(2016). 

{\sl \underbar{Critical Assessment:} Since the entropy flux is 
increasing in random processes, we can conclude that
processes with decreasing entropy fluxes are non-random processes,
which is one of the definitions of self-organization here. 
The entropy flux calculation of the Earth's atmosphere is made
by assuming energy flux balance between the incoming solar
radiation and the outgoing infrared emission from Earth.
Based on this estimate of the entropy flux change (E) we can
conclude that the atmosphere including its weather and
climate changes have self-organizing capabilities
(Table 1: qualifier E).} 

\clearpage

\section{	SOLAR PHYSICS				}

\subsection{	Photospheric Granulation		}

The solar photosphere exhibits a pattern of ``bubbling'' cells
(like boiling water in a frying pan), which is called 
``photospheric granulation'' (Fig.~\ref{f_granulation}) and has been interpreted
in terms of hydrodynamic convection cells. The central part of
a granulation cell is occupied with upflowing plasma, which
then cools down and descends in the surrounding edges,
which consequently appear to be darker than the center, because
a cooler temperature corresponds to fainter white-light emission.
The photospheric temperature is $T_s=5780$ K, the typical size
of a granule is $w \approx 1500$ km, and the life time is about
8-20 min. 

The underlying physical mechanism of convection has been
studied in great detail in terms of the Rayleigh-B\'enard
instability, known as Lorenz model (Lorenz 1963), 
described also in the monographs of Chandrasekhar (1961) 
and Schuster (1988). The basic ingredients of the 
(hydrodynamic) Lorenz model are the Navier-Stokes equation, 
the equation for heat conduction, and the continuity equation,
\begin{eqnarray}
        \rho {d {\bf v} \over dt} &=& 
	{\bf F} - \nabla p + \mu \nabla^2 {\bf v} \label{eq_navier1}\\
	{d T \over dt} &=& \kappa \nabla^2 T      \label{eq_navier2}\\
	{d \rho \over dt} &=& - \nabla \cdot (\rho {\bf v}) \label{eq_navier3}
\end{eqnarray}
where $\rho$ is the density of the fluid, $\mu$ is the viscosity,
$p$ is the pressure, $\kappa$ is the thermal conductivity,
${\bf F} = - \rho g {\bf e}_z$ is the external force in the
$e_z$ direction due to gravity, and the boundary conditions
are $T(x,y,z=0, t) = T_0 + \Delta T$ and $T(x,y,z=h)= T_0$
for a temperature gradient in vertical direction.
For the special case of translational invariance in $y$-direction,
using the Boussinesq approximation, and retaining only the lowest
order terms in the Fourier expansion, we obtain the much
simpler form of the Lorenz model,
\begin{equation}
 \begin{array}{ll}
  \dot X &= - \sigma X + \sigma Y \label{eq_lorenz1}\\
  \dot Y &= - X Z + r X - Y       \label{eq_lorenz2}\\
  \dot Z &= + X Y - b Z           \label{eq_lorenz3}
 \end{array} 
 \ ,
\end{equation}
which is a system of three coupled first-order differential equations,
with $X$ the circulatory fluid flow velocity, $Y$ the temperature
difference between ascending and descending fluid elements,
$Z$ the deviations of the vertial temperature profile from its
equilibrium value, and $r$ is the control parameter 
measuring the magnitude of the temperature difference $\Delta T$.
The Lorenz model can describe the transition from heat 
conduction to convection rolls, where Lorenz discovered the 
transition from deterministic to chaotic system dynamics.

Thus, the Lorenz model demonstrates that a temperature gradient
(for instance below the photosphere) transforms (a possibly turbulent)
random motion into a highly-organized rolling motion (due to the
Rayleigh-B\'enard instability) and this way organizes the plasma
into nearly equi-sized convection rolls that have a specific
size (such as $w \approx 1500$ km for solar granules). 
The self-organization process thus creates order (of granules
with a specific size) out of randomness (of the initial turbulent
spectrum). 

Since convection is the main energy transport process inside
the Sun down to $0.7 R_{\odot}$, larger convection rolls than
the granulation pattern can be expected. Krishan (1991, 1992) 
argues that the Kolmogorov turbulence spectrum $N(k) \propto k^{)-5/3}$
extends to larger scales and possibly can explain the observed
hierarchy of structures (granules, mesogranules, supergranules, 
and giant cells) by the same self-organization process.

At smaller scales, a subpopulation of mini-granular structures
has been discovered, in the range of $w \approx 100-600$ km,
(Fig.~\ref{f_minigranules}), 
predominantly confined to the wide dark lanes between regular
granules, often forming chains and clusters, but being different
from magnetic bright points (Abramenko et al.~2012).
A set of TiO images of solar granulation acquired with the 
1.6 meter New Solar Telescope at Big Bear Solar Observatory 
was utilized. The high-contrast speckle-reconstructed images of 
quiet-sun granulation (Fig.~\ref{f_minigranules}), 
allowed to detect, besides the regular-size granules, the small 
granular-like features in dark inter-granular lanes, named as 
{\sl mini-granules}. Mini-granules are very mobile and short-lived.  
They are predominantly located in places of 
enhanced turbulence and close to strong magnetic fields in 
inter-granular lanes. The equivalent size of detected granules 
was estimated from the circular diameter of the granula's area. 
The resulting {\sl probability density functions (PDF)} for 36 
independent snapshots 
are shown in gray on the left frame of Fig.~(\ref{f_abramenko}). 
The average PDF (the red histogram) changes its slope in 
the scale range of $\approx 600-1300$ km. This varying power law 
PDF is suggestive that the observed ensemble of granules may 
consist of two populations with distinct properties: regular 
granules and mini-granules. A decomposition of the observed PDF 
showed that the best fit is achieved with a combination of a 
power law function (for mini-granules) and a Gaussian 
function (for granules). Their sum fits the 
observational data. Mini-granules do not display any characteristic 
(``dominant'') scale. This non-Gaussian distribution of sizes 
implies that a more sophisticated mechanism with more 
degrees of freedom may be at work, where any small fluctuation 
in density, pressure, velocity and magnetic field may have 
significant impact and affect the resulting dynamics.
It is worth to note that a recent direct numerical simulation 
attempt (Van Kooten and Cranmer 2017) produced the PDF of 
granular size in agreement with the observed one in 
Fig.~(\ref{f_abramenko}). The authors concluded that the population 
of mini-granules is intrinsically related to non-linear turbulent 
phenomena, whereas Gaussian-distributed regular granules originate 
from near-surface convection. 

{\sl \underbar{Critical Assessment:} The size distribution of
granulation cells in the solar photosphere does not form a
power law distribution, but clearly shows a preferred spatial
scale of $\approx 1000$ km, which renders a regular spatial
pattern (S), rather than a scale-free distribution. However, 
a power law distribution has been found for the newly discovered
``mini-granules'' in a size range of 100-600 km, which contradicts
a self-organizing convective process that creates bubbles of equal
sizes. The physical process of convection that is driven by
a temperature gradient and the Rayleigh-B\'enard instability (I) 
is well-understood and known as the Lorenz model. A caveat is
how much the magnetic field plays a role in the solar
convection zone, requiring a model with magneto-convection and 
hydromagnetic (Parker and Kruskal-Schwarzschild) instabilities.
Anyway, a self-organization process is warranted based on the
preferred scale of convective rolls (Table 1: qualifiers I,S).}

\subsection{	Magnetic Field Self-Organization 		}

How is the solar magnetic field organized and how does the
resulting magnetic field self-organize into stable 
structures? It is said that sunspots and pores represent
the basic stable structures that are visible in the photosphere, 
but their sub-photospheric formation (driven by the solar dynamo) 
and stability are long-standing problems. In the following we
discuss a few papers that explicitly use the term ``self-organization'' 
in this context.

Takamaru and Sato (1997) propose a self-organization system
that evolves intermittently and undergoes self-adaptively 
local maxima and minima of energy states. The nonlinear 
interactions of twisting multiple flux tubes lead to local
helical kink instabilities, resulting in the formation of a
knotted structure. Intermittent reconnection with neighbored
flux tubes in the knotted structure releases energy and
restores the original configuration, a process that exhibits
self-organization in an open complex nonlinear system where 
energy is externally and continuously supplied.

Vlahos and Georgoulis (2004) state that non-critical
self-organization appears to be essential for the formation 
and evolution of solar active regions, since it regulates the 
emergence and evolution of solar active regions, perhaps 
characterized by a percolation process (Schatten 2007; 2009), while 
the energy release process is governed by self-organized criticality.
Georgoulis (2005, 2012) explores various (scaling and multi-scaling,
fractal and multi-fractal) image-processing techniques to measure 
the expected self-organization of turbulence in solar magnetic fields.
However, no difference was found in the turbulence spectrum between 
flaring and non-flaring active regions.

Chumak (2007) proposes a dynamic self-organization model of the
active region evolution in terms of a diffuse aggregation process 
of magnetic flux tubes in the upper levels of the solar convection zone. 
The physical model is governed by hydrodynamics, magnetic forces, and 
additional random forces. 

Kitiashvili et al.~(2010) describes the process of magnetic field 
generation as a self-organization process:
{\sl The simulations reveal two basic steps in the process
of spontaneous formation of stable structures that are the key for
understanding the magnetic self-organization of the Sun and the
formation of pores and sunspots: (1) formation of small-scale
filamentary magnetic structures associated with concentrations of
vorticity and whirlpool-type motions, and (2) merging of these
structures due to the vortex attraction, caused by converging
downdrafts around magnetic concentration below the surface}, 
reaching magnetic field strengths of $B \approx 1500$ G at the 
surface and $B \approx 6000$ G in the interior. The structure 
was found to remain stable for at least several hours. 
Examples of the simulated formation and evolution of magnetic 
structures are shown in Fig.~(\ref{f_kitiashvili}).

Although the term ``self-organization'' is not explicitly mentioned
in recent (realistic) radiative 3-D MHD simulations of
Abbett (2007), Cheung et al.~(2007), Martinez-Sykora et 
al.~(2008, 2009, 2011), Tortosa-Andreu and Moreno-Insertis (2009),
Stein et al.~(2011), Stein (2012), and Rempel and Cheung (2014), 
we can interpret the generation of stable coherent magnetic 
structures in the turbulent convection zone as a manifestation
of a self-organizing process. Basically, these global MHD dynamo
models generate coherent flux ropes that rise towards the
solar surface (Fig.~\ref{f_rempel_cheung}). 
There is no need to insert sub-photospheric 
flux ropes in the simulation box as done earlier, because 
recent 3-D MHD simulations added the evolution of realistic
magneto-convection as a time-dependent boundary to drive 
the flux emergence process (Cheung and Isobe 2014; 
Cheung et al.~2017). The fact that susnpots always appear
within a time scale comparable to the flux emergence time
of an active region, providing magnetic flux to the sunspot,
indicates that coherent magnetic structures self-organize
deep in the convection zone. There the Rossby number is
less than unity and convection is constrained by differential
rotation and meridional flows. 

As a disclaimer, we have to
be aware that these 3-D MHD simulations capture a local box
only, rather than being global. A self-consistent
generation of magnetic flux, simulated on a global scale
that includes the entire spherical convection zone of the Sun,
is presented in Miesch et al.~(2000), which produces laminar
and turbulent states, driven by the differential solar rotation.  
Related work describes convection and dynamo action in rapidly 
rotating suns (Brown et al.~2010), or in large-scale dynamos 
with turbulent convection and shear (K\"apyl\"a et al.~2012). 
In order to understand the basic mechanism of the formation of 
magnetic flux concentrations, numerical 3-D MHD simulations were
performed that study the turbulence contributions to the mean
magnetic pressure in a strongly stratified isothermal layer 
with a large plasma beta (Brandenburg et al.~2012). By applying
a weak uniform horizontal mean magnetic field, the 
{\sl negative effective magnetic pressure instability (NEMPI)}
is activated, which reduces the turbulence and thus the
turbulent pressure. If this reduction is more than the
magnetic pressure, then the weakly magnetized region will have
a reduced total pressure, which leads to a collapse of the
field into a stronger tube. Since this mechanism generates
order from turbulence, it can be considered to be a self-organziation
process (Robert Cameron, private communication). In the global
3-D MHD simulations of Hotta et al.~(2015), an efficient small-scale
dynamo generates the magnetic field, which has a feedback on the
poleward meridional flows, and thus displays the characteristic feedback
feature of a self-organizing process. A simulation of the convective dynamo
in the solar convective envelope has been conducted by Fan and Fang (2014),
which is driven by the solar radiative diffusive heat flux, exhibiting
irregular cyclic behavior with oscillation time scales ranging
from about 5 to 15 yr and undergoes irregular polarity reversals,
as it is typical for self-organizing limit cycles far off a
stationary equilibrium.

{\sl \underbar{Critical Assessment:} Ideas of applying 
self-organization processes to generate the magnetic field in the 
solar convection zone or in the solar corona are mentioned only
briefly in the reviewed papers (or not at all), but no quantitative 
models or measurements are presented that would allow us to discriminate
which magnetic structures have a random pattern and which ones
exhibit some ordered pattern. The magnetic flux on the solar 
surface was found to have a power law size distribution 
(Parnell et al.~2009), which is rather consistent with a 
self-organized criticality process. The envisioned feedback 
mechanisms inculde the kink instability, the NEMPI instability,
percolation, diffuse aggregation, and vortex attraction, but none 
of these processes has been characterized in emerging flux 
simulation in terms of self-organization. So, we can 
observe spatial patterns of photospheric magnetic flux patches (S), 
but are not sure which instability (I) enacts self-organization
(Table 1: qualifiers I(?),S).}

\subsection{	The Hale Cycle 				}

The global magnetic field of the Sun undergoes a cyclic
transition from a global poloidal field to a highly-stressed
toroidal field in 11 years, switching the magnetic polarity
during this process, so that the original polarity is restored
after two cycles, yielding a 22-yr cycle that is called
the (magnetic) ``Hale cycle''. There exist over 2000 publications
about the solar magnetic activity cycle. A recent review
can be found in Hathaway (2015).

A physical model of the Hale Cycle is the Babcock-Leighton
dynamo model (Babcock 1961; for a review see Charbonneau 2014), 
which explains the winding-up of the highly-stressed toroidal 
field as a consequence of the differential rotation (during 
the rise phase of the cycle), and is followed by a gradual decay
with decreasing sunspot number and meridional diffusion 
of the magnetic field, leading to a relaxed poloidal field 
during the solar cycle minimum. An example of a 3-D MHD simulation
of the solar convection zone is shown in Fig.~(\ref{f_dynamo}).
The observed variation of the sunspot number between 
the years 1870 and 2017 is shown in Fig.~(\ref{f_butterfly}).

The variability of the solar cycle can be understood in terms
of a weakly nonlinear limit cycle affected by random noise
(Cameron and Sch\"ussler 2017), quantified in normal form 
in terms of the Hopf bifurcation (Fig.~\ref{f_cameron}, 
\ref{f_limitcycle}, Appendix C). 
The presence of a limit cycle is a common property in
coupled nonlinear dissipative systems, which is most
easily understood in terms of the Lotka-Volterra equation
system (Haken 1983), known as the predator-prey equation in
ecology (Fig.~\ref{f_limitcycle} bottom; Appendix D),
\begin{equation}
 \begin{array}{ll}
  \dot X &=   k_1 X - k_2 X Y \\
  \dot Y &= - k_3 Y + k_2 X Y \\
 \end{array}
 \label{eq_volterra}
\end{equation}
This equation system has a periodic solution, which is
called the limit cycle. Critical points occur when
$dX/dt=0$ and $dY/dt=0$, which yields a stationary point
in phase space at $X=k_3/k_2$ and $Y=k_1/k_2$. 
Applying the Lotka-Volterra equation
system to the solar cycle, $X$ represents the poloidal field
and $Y$ the toroidal field, $k_1$ the growth rate of the
poloidal field, $k_3$ the growth rate of the toroidal field,
and $(k_2)$ a nonlinear interaction term between the two
field components. The Lotka-Volterra equations describe
the emergence and sustained oscillation in an open system
far from equilibrium, as well as emergence of spontaneous
self-organization (Demirel 2007). An application of the
Lotka-Volterra system to the complex system of the solar cycle
is discussed in Consolini et al.~(2009), where a double dynamo
mechanism is envisioned, one at the base of the convection
zone (tachocline), and a shallow subsurface dynamo.
The deeper dynamo dominates the poloidal field,
while the shallower dynamo controls the toroidal field. 
In summary, the limit cycle represents a highly-ordered
self-organizing 11-year (22-year) pattern of the solar magnetic 
activity, which cannot be explained with a random process.

A chaotically modulated stellar dynamo was modeled also based on 
bifurcation theory, where modulation of the basic magnetic
cycle and chaos occur as a natural consequence of a star
that is in transition from a non-magnetic state to one with
periodically reversing fields (Tobias et al.~1995). 

{\sl \underbar{Critical Assessment:} The solar cycle is a very
periodic phenomenon with little variation in each cycle, which
is a classic example of a nonlinear dissipative system with 
limit-cycle behavior (LC), such as the Hopf bifurcation
(Appendix C) or Lotka-Volterra equation system (Appendix D).
The limit cycle produces a regular temporal pattern (T), and
the cycle variation modulates the magentic flux and area 
on the solar surface like-wise (S). The physics of the solar
cycle is also well-understood in terms of the Babcock-Leighton
model, where the differential solar rotation is the driver,
and a twisted magnetic field relaxation mechanism acts as the
feedback mechanism. The underlying instability still needs
to be identified and may depend on both the shallow dynamo
or the deep dynamo in the tachocline at the bottom of the
convection zone (Table 1: qualifiers LC,I[?],S,T).}

\subsection{	Evaporation-Condensation Cycles		}

Solar observations show that coronal loops routinely harbor
flows that result from the complex physics of the solar transition
region (e.g., Peter et al.~2006). Upflows can generally be
understood as the result of heated plasma from the
chromosphere ascending into coronal loops (chromospheric
evaporation), as modeled from EUV, soft X-ray, and hard X-ray
observations. These upflows frequently happen
during solar flares, but equally occur as a consequence
of other coronal heating mechanisms also, in active regions,
in Quiet Sun regions (explosive events, EUV brightenings), 
and even in coronal 
holes (plumes, jets). At the same time there is numerous
evidence for downflows, also called {\sl ``coronal rain''}
or {\sl ``coronal condensation''}, mostly observed in H$\alpha$ 
(first reported by Leroy 1972)
and UV lines of cooler temperatures (Schrijver et al.~2001;
De Groof et al.~2005). The combined pattern of
upflows and downflows is also referred to as
{\sl ``evaporation-condensation cycle''}
(Krall and Antiochos 1980), which we consider
under the aspect of a self-organization process here. 

The earliest physical interpretation of evaporation-condensation
cycles has been modeled in terms of the thermal instability,
which constitutes a chromosphere-corona coupling or feedback 
mechanism between the heating rate and the cooling rate in a 
coronal loop (Kuin and Martens 1982). Such a system can exhibit
a stable static equilibrium if the coupling between the 
chromosphere and the corona is sufficiently strong, but for
typical coronal loop conditions the system is expected not
to be stable, resulting into a cyclic solution that corresponds
to the limit cycle of a coupled nonlinear system.
The physical model predicts that a temporal excess of heating
leads to an excess conductive flow at the loop base, which results
into chromospheric evaporation with increasing pressure and density, 
and in turn amplifies the radiative loss, leading to a thermal
(or radiative) instability with subsequent condensation or
downflow of cool material. Kuin and Martens (1982) use the following
form of the hydrodynamic equations for a 1D loop,
\begin{eqnarray}
  {\partial n \over \partial t} &=& - {\partial \over \partial z} (nv) \label{eq_hd1} \\
  {dv \over dt} &=& - {2 \over m_H n} {\partial p \over \partial z} 
	- g_{\parallel}                                                \label{eq_hd2}\\
  {3 \over 2} {dp \over dt} &=& - {5 \over 2} p {\partial v \over \partial z}
	  - {\partial \over \partial z} 
	    \left[ \kappa_0 T^{5/2} {\partial T \over \partial z} \right] 
	  + E_H - n^2 \Psi(T)                                          \label{eq_hd3}\\
  p 	&=& n k_B T = \mu m_H n c_s^2	                               \label{eq_hd4}
\end{eqnarray}
where $p$ is the pressure, $n$ the particle density, $v$ the
plasma velocity, $T$ the electron temperature, $t$ the time,
$k_B$ the Boltzmann constant, $m_H$ the hydrogen mass,
$g_\parallel$ the gravitational acceleration along the loop,
$c_s$ the isothermal sound speed, $\mu=0.5$ the molecular
weight, $\kappa_0$ the Spitzer conductivity, $E_H$ the
heating rate (assumed to be spatially constant), and
$\Psi (T)$ is the radiative loss function (approximated
with a power law $\Psi (T) = \Psi_0 T^{-\gamma}$). 
Kuin and Martens (1982) find static solutions for some
parameters of the loop length $L$ and heating rates $E_H$.
The time-dependent solutions can be approximated by the
following coupled equation system for the dimensionless
temperature $X=T/T_0$ and density $Y=n_e/n_0$ parameters,
\begin{eqnarray}
  {d X \over d t} &=& {1 \over Y} [ 1 - Y^2 \Psi(X) - \alpha (X - 1)] \label{eq_hd5} \\
  {d Y \over d t} &=& f \alpha (1 - X^{-1})                           \label{eq_hd6}
\end{eqnarray}
Similar to the Lotka-Volterra equation system (Eq.~\ref{eq_volterra}),
this rate equation system has a limit cycle at the critical point
$dX/dt=0$ and $dY/dt=0$, requiring $f \alpha (1 - X^{-1}) = 0$ and
$[ 1 - Y^2 \Psi(X) - \alpha (X - 1)]/Y =0$, which yields the solution 
$X=1$ and $Y=1/\sqrt{\Psi(X=1)}$ for the limit cycle
at the attractor point.  The X-Y phase diagram of some
quasi-stationary solutions is shown in Fig.~(\ref{f_kuin}). 

Numerical 1-D hydrodynamic simulations of the condensation of
plasma in loops of wide ranges of lengths and temperatures
(10 Mm $\le$ L $\le$ 300 Mm; 0.2 MK $\le$ T $\le$ 2 MK) reproduce
the cyclic pattern, starting with chromospheric evaporation,
followed by coronal condensation, then motion of the condensation region 
to either side of the loop, and finally loop reheating with a period 
of 1h to 4 days (M\"uller et al.~2003, 2004, 2005).
It is found that the radiatively-driven thermal instability occurs about an
order of magnitude faster than the Rayleigh-Taylor instability, which can
occur in a loop with a density inversion at its apex also 
(M\"uller et al.~2003).
Simulations with different heating functions reveal that the process of
catastrophic cooling is not initiated by a drastic decrease of the total loop
heating rate, but rather results from a loss of equilibrium at the loop apex as
a natural consequence of quasi-steady footpoint heating (M\"uller et al.~2004;
Peter et al.~2012). The same effect of a loss of equilibrium can occur
in the case of repetitive impulsive heating (e.g., Mendoza-Briceno et al.~2005;
Cargill et al.~2013). 

EUV intensity pulsations with periods from 2 to 16 hrs have been 
discovered to be quite common in the solar corona and especially in 
coronal loops (Auch\`ere et al.~2014; Froment et al.~2015). The three 
loop events shown in Fig.~(\ref{f_froment}), studied in detail by 
Froment et al.~(2015),
have time periods of 3.8, 5.0 and 9.0 hrs and are lasting over several 
days. They were interpreted in terms of thermal non-equilibrium evaporation 
and condensation cycles (Froment et al.~2015, 2017). In Fig.~(\ref{f_froment}) 
the temperature and the total emission measure are shown, extracted from a DEM 
analysis using the method developed by Guennou et al.~(2012,a,b; 2013). 
The temperature corresponds to the peak temperature of the DEM, and the total 
emission measure is proportional to the squared density along the 
line-of-sight.

Uzdensky (2007a,b) proposes a similar self-organization process for 
coronal heating. This self-regulating process keeps the coronal
plasma roughly marginally collisionless. The driver of the
self-organization process is the magnetic reconnection in the
collisional Sweet-Parker regime. The feedback mechanism is the
inhibition of magnetic reconnection triggered by density increases
due to chromospheric evaporation. After some time, the conductive
and radiative cooling lowers the density again below the
critical value and fast reconnection sets in again. Thus,
the self-organization process is made of repeating cycles
of fast reconnection, evaporation, plasma cooling, and 
re-building of magnetic stress. A similar self-regulation
mechanism controlled by marginal collisionality in magnetic 
reconnection is explored in Cassak et al.~(2008) and 
Imada and Zweibel (2012). The cyclic behavior has been
simulated with a 1-D hydrodynamic model that is driven
by gravity and the density dependence of the heating function
(Imada and Zweibel 2012).

{\sl \underbar{Critical Assessment:} 
The evaporation-condensation scenario of coronal loops predicts 
a quasi-periodic time pattern, but not much is known about the
degree of periodicity, and whether this corresponds to a 
quasi-periodic self-organizing limit cycle. The quasi-periodic 
patterns discovered by Froment et al.~(2015, 2017), which exhibit
phase-shifted oscillations between the emission measures and
temperatures in active regions, reveal large fluctuations in the
emission measure versus temperature diagram (Fig.~\ref{f_froment}),
which may indicate strong nonlinearities near the limit cycle
or inadequate background subtraction in the differential emission
measure analysis. Although the physics of the evaporation-condensation 
cycle is well understood, to deduce the time evolution of the heating rate,
electron density, and temperature from observational data,
adequate background subtraction needs to be performed in order 
to establish whether the observations are well described by a
limit-cycle system (Table 1: qualifiers I,LC[?],T[?]).}

\subsection{	Quasi-Periodic Radio Bursts		}

We identify more than 150 publications that report or model
periodic (oscillatory) or quasi-periodic solar radio bursts. 
Many of these quasi-periodic solar radio emissions are believed
to be generated by various plasma instabilities (Benz 1993).
The degree of periodicity was found to vary from random to
strictly periodic (e.g., Aschwanden et al.~1993).
In an early review, solar radio pulsations were classified 
into three different models: (1) MHD oscillation eigenmodes;
(2) cyclic self-organizing systems; and (3) modulation of
magnetic reconnection, particle injection, or acceleration
(Aschwanden 1987). Here we discuss only the second group
in terms of self-organization mechanisms, which 
observationally can be easily distinguished from the first
group: MHD oscillations are strictly periodic, while 
limit cycles in self-organizing systems produce less
regular quasi-periodic pulse patterns. We have also to be aware
that the periodicity of solar radio bursts can only be
inferred from time profiles (Figs.~\ref{f_radio_osc}, 
\ref{f_zurich}), while spatial fine structures mostly cannot 
be resolved by {\sl remote-sensing} observations 
with current radio instruments.

Self-organizing systems with limit cycles were initially
applied to loss-cone instabilities occurring in the aurora,
where two types of waves (electrostatic and upper hybrid 
waves) exchange energy in a limit cycle, driven by 
the loss-cone instability (Trakhtengerts 1968), a concept
that was then applied to solar radio pulsations also
(Zaitsev 1971; Zaitsev and Stepanov 1975; Kuijpers 1978;
Bardakov and Stepanov 1979; Aschwanden and Benz 1988).  
The two-component nonlinear systems of self-organization 
are controlled either by wave-wave interactions, or by
wave-particle interactions (also called a {\sl quasi-linear
diffusion} process). 

The process starts with the
development of a nonthermal particle distribution 
(such as electron beams, loss-cones, pancakes, or rings),
which then become unstable and transform kinetic energy
into various waves (such as whistler waves, upper-hybrid waves,
Langmuir waves, or electron-cyclotron maser emission), 
relaxing the unstable particle distribution then (in form
of a plateau for beams or a filled loss-cone). After this
feedback, when new particles arrive, the relaxed particle distribution 
becomes unstable again and the entire nonlinear cycle starts over. 
For the case of electron-cyclotron emission, for instance, the 
dynamics of the wave-particle interaction can be described by the
following system of coupled equations (e.g., Aschwanden and Benz 1988),
\begin{eqnarray}
 {\partial N({\bf k},t) \over \partial t}
 + {\bf v}_g({\bf k}) {\partial N({\bf k},t) \over \partial {\bf r}}
 &=& \Gamma( {\bf p}, {\bf k},f) N({\bf k}, t)
 - \gamma({\bf p}, {\bf k}, f) N({\bf k}, t) \ , 
\label{eq_maser1} \\
 {\partial f({\bf p},t) \over \partial t}
 + {\bf v}({\bf p}) {\partial f({\bf p},t) \over \partial {\bf p_j}}
 &=& {\partial \over \partial j}
 \hat D_{ji}( {\bf p}, {\bf k}, N) {\partial f({\bf p}, t) \over \partial p_i}
 + {\partial f({\bf p}, f) \over \partial t}|_{source}
 + {\partial f({\bf p}, f) \over \partial t}|_{loss} 
\label{eq_maser2} \ ,
\end{eqnarray}
where the waves are represented by the photon number density
$N({\bf k}, t)$ in {\bf k}-space, the particle system is
described by its density distribution $f ({\bf p}, t)$ in
momentum space, $\Gamma ( {\bf p}, {\bf k}, f )$ is the
wave growth rate, $\gamma({\bf p}, {\bf k}, f )$ is the
wave damping rate, ${\bf v}_g({\bf k})$ is the group
velocity of the emitted waves, and $ \hat D_{ji}({\bf p}, {\bf k}, N)$
is the quasi-linear diffusion tensor. 
Eq.~(\ref{eq_maser1}) is the wave equation that
describes the balance between emission and growth and damping
rate, while Eq.~(\ref{eq_maser2}) describes the evolution of the particle
distribution. The interaction between waves and particles is
expressed by the quasi-linear diffusion tensor.
In addition, there is a source term of particles 
(with large pitch angles), as well as
a loss term (which quantifies the precipitating particles with
small pitch angles out of the loss-cone.) 

A complete analytical 
solution of this coupled integro-differential equation system
is not available,
but a limit-cycle solution applied to the case of electron-cyclotron
maser emission has been calculated (Aschwanden and Benz 1988).
The pulse period $\tau_{lc}$ of the limit cycle has been found to be the
geometric mean of the wave growth time $\tau_g$ and the
particle diffusion time $\tau_d$,
\begin{equation}
	\tau_{lc} = 2 \pi \sqrt{\tau_g \ \tau_d} \ , \label{eq_maser3}
\end{equation}
which is a close analogy to the limit cycle of the Lotka-Volterra
equation system (Eq.~\ref{eq_volterra} and Appendix D) or a coupled differential
equation system (Appendix B). 
In summary, such a self-organizing system
is driven by coherent wave growth stimulated by an unstable
(loss-cone) particle distribution via the relativistic (cyclotron
or gyromagnetic Doppler resonance 
condition), while the feedback mechanism represents the back reaction
that flattens the unstable particle distribution (via quasi-linear diffusion), 
which in turn quenches coherent wave growth until the loss-cone
is filled again with new particles and the cyclic wave-particle
interaction starts over. The result is a stationary
quasi-periodic pattern of coherent radio emission,
which is strictly periodic in the limit cycle only,
but becomes aperiodic depending on the inhomogeneity, anisotropy, 
time-dependence, and noise of the control parameters.  
Dabrowski and Benz (2009) find generally a good correlation
between decimetric pulsations and hard X-rays.

{\sl \underbar{Critical Assessment:} The quasi-periodicity of 
solar radio bursts as observed in dynamic spectra is the most
convincing signature of a self-organizing process, in contrast to
a time series with random time intervals, as it would be expected
for self-organized criticality models. The spatial counterpart (S)
of the quasi-periodic temporal scales (T) is generally not observed
due to the lack of radio images with high spatial resolation.
Nevertheless, quasi-periodic time intervals are consistent with
a limit cycle (LC) of a nonlinear dissipative system, but there
are many plasma instabilities that can operate as a positive
feedback mechanism, either in terms of wave-particle interactions
(e.g., loss-cone or beam instabilities),
or wave-wave interactions. Thus, more data modeling, possibly with
high-resolution imagery and magnetic field modeling is required to
identify the relevant instabilities that control a self-organizing
process in the generation of quasi-periodic radio bursts 
(Table 1: qualifiers T, LC, I[?]).}

\subsection{	Zebra Radio Bursts		}

While the existence of self-organizing systems observed in solar 
radio bursts is mostly inferred from the quasi-periodicity of observed
temporal patterns, there exists another category of solar radio
bursts that exhibits very regular periodic patterns in the frequency domain. 
The most striking example is the so-called {\sl zebra burst} 
(Fig.~\ref{f_zebra}), 
which reveals drifting parallel bands of quasi-stationary radio 
emission with harmonic frequency ratios.  Theoretical interpretations 
include (i) models with interactions between electrostatic 
waves and whistlers, and (ii) radio emission at the double-plasma
resonance (Kuijpers 1975, 1980; Zheleznykov and Zlotnik 
1975a,b; Mollwo 1983; Winglee and Dulk 1986; Chernov 2006; 
Chen et al.~2011), 
\begin{equation}
	\omega_{UH} = \left( \omega^2_{Pe} + \omega^2_{Be} \right)^{1/2}
		= s \ \omega_{Be} \ ,  \label{eq_zebra}
\end{equation}
where ${\omega}_{UH}$ are upper hybrid waves,
${\omega}_{Pe}$ is the electron plasma frequency,
${\omega}_{Be}$ is the electron cyclotron frequency,
and $s$ is the integer harmonic number, which introduces
a periodic pattern in the resonance frequency. If the
magnetic field structure $B(h)$ with altitude $h$ is known, 
the harmonic frequencies can be mapped onto a periodic
spatial pattern (Fig.~\ref{f_zebra}, right panel).  Either way, harmonic 
resonances (of the gyrofrequency) create order out of randomness 
for this type of radio bursts, in analogy to mechanical resonances 
that produce harmonic patterns of planet orbits.    

In recent models, the double-plasma resonance mechanism faces a number of 
difficulties in explaining the dynamics of zebra stripes (i.e., sharp changes of 
the frequency-drift rate, a large number of stripes, frequency splitting of 
stripes, super-fine millisecond structure), because the magnetic field and 
density cannot change as rapidly. Improved models are in progress 
(Karlicky et al.~2001; LaBelle et al.~2003; Kuznetsov and Tsap 2007; 
Karlicky and Yasnov 2015). New calculations concern the increments of the 
upper-hybrid waves under double-plasma resonance conditions, the ring distribution 
of high-speed electrons with relativistic corrections, different temperatures 
of the background plasma, and optimum wave numbers 
(Benacek et al.~2017). It has been shown that the optimum 
increment for electron velocities is $v \approx 0.1$ c, with a 
narrow dispersion. If the speed is $\approx 0.2$ c, the increment sharply 
decreases and the flux maxima are washed out in the continuum for several 
cyclotron harmonic numbers $s$. Thus, these calculations show the inefficiency 
of the double-plasma resonance mechanism. Under such conditions it becomes clear, 
that the double-plasma resonance mechanism cannot explain the majority of 
zebra stripes. An additional complication is the simultaneous occurrence of 
decimetric millisecond spikes.

In the whistler model, all the aforementioned properties of zebra burst stripes 
have been explained by physical processes that occur during the coalescence of 
Langmuir waves $(l)$ with whistler waves $(w)$, producing transverse waves,
$l + w \mapsto t$ (Kuijpers, 1975; Chernov 1976; 1990; 2006; 2011). 
Langmuir waves and whistlers can be generated by the same fast particles trapped 
in magnetic islands (Berney and Benz, 1978). 

The spatial structure of zebra radio bursts is believed to originate in magnetic 
islands after coronal mass ejections. Therefore the close connection of zebra bursts 
with fiber bursts is simply explained by the acceleration of fast particles 
in magnetic reconnection regions in the lower or upper part of magnetic islands.

A wavelike or saw-tooth frequency drift of stripes was explained by the 
switching of the whistler instability from the normal Doppler-cyclotron resonance 
into the anomalous one (Fig.~2b in Chernov 1990). Such switching should lead 
to a synchronous change of the frequency drift of stripes and spatial drift 
of the radio source, since whistlers generated at normal and anomalous
resonances move in opposite directions. New injections of fast particles 
cause sharp changes in the frequency drift rate and oscillation pattern
of zebra stripes. Low frequency absorption (i.e., black stripes of zebra bursts) 
are explained by quenching of the plasma wave instability due to diffusion 
of fast particles by whistler waves.

The superfine structure is generated by a pulsating regime of the whistler 
instability with ion-sound waves (Chernov et al.~2003). Rope-like chains 
of fiber bursts are explained by a periodic whistler instability between two 
fast shock fronts in a magnetic reconnection region (Chernov 2006). In the
whistler model, zebra-stripes can be converted into fiber bursts and back
(Fig.~\ref{f_chaos_zebra}), which exhibits morphological changes from
chaos to order, and in reverse direction. 
A comparative discussion of observations of zebra and fiber bursts and 
different theoretical models can be found in the reviews of Chernov (2012; 2016).

{\sl \underbar{Critical Assessment:} The most striking pattern that
hints to a self-organization process is the periodic appearance of
bands in dynamic spectra of some solar radio bursts, which is
interpreted in terms of gyroharmonic resonances (R). In principle,
a periodic pattern in radio frequency can be mapped to a periodic
pattern in spatial structures (S), using the plasma frequency 
relationship $f_p \propto \sqrt{n_e}$ and a density model $n_e(h)$ 
as a function of the altitude $h$, while there is no obvious
periodic pattern of temporal structures (T) expected. The driver
mechanism that produces zebra bursts is likely to be a population 
of nonthermal particles, while the counter-acting feedback mechanism
has been modeled in terms of electrostatic waves, whistler waves,
or the double-plasma resonance. The observational verification of
any of these wave types is still very challenging with remote-sensing
techniques (Table 1: qualifiers R,S,I[?] ). --- Note that the spatial
pattern of a zebra skin has also been classified as a self-organization
process in biology (e.g., Camazine et al.~2001), where the light and
dark pigmentation is created by the diffusive interaction of chemical 
activation (driver) and inhibition (feedback) during the embryonic 
development.}

\clearpage

\section{       STELLAR PHYSICS                 }

\subsection{    Star formation          }

The spatial distribution of star formation in a galaxy is not uniform
but is concentrated in a number of small localized areas in galaxies.
Young stellar associations with their H II regions and molecular
clouds (Fig.~\ref{f_nebula_pillars}) 
are manifestations of the ordered distribution of matter
participating in the star formation processes, governed by
self-organization in a nonequilibrium system (Bodifee 1986).
A star formation region can be modeled by the following system of
coupled equations (Bodifee 1986):
\begin{eqnarray}
        {dA \over dt} &=& K_1 S + K_2 S - K_3 M^2 A   \label{eq_star1}\\
        {dM \over dt} &=& K_3 M^2 S - K_4 S M^n       \label{eq_star2}\\
        {dS \over dt} &=& K_4 S M^n S - K_1 S - K_2 S \label{eq_star3}\\
        {dR \over dt} &=& K_1 S \ ,                   \label{eq_star4}
\end{eqnarray}
where $A$ is the mass of the interstellar atomic gas,
$M$ is the mass of the interstellar molecular gas (with dust),
$S$ is the mass of the stellar material (young stars with
their associated H II regions), $R$ represents the total mass
of ``old'' stars (stellar remnants and low-mass main-sequence stars),
and $n$ is a stability parameter. (Stability is granted for
$n \ge 2$ for any $[k_1, k_2]$ pair, see Fig.~3 in Bodifee 1986).
A graphic representation of the star formation process is
depicted in Fig.~(\ref{f_bodifee}).
This coupled system of differential equations, after elimination
of $S$ and the introduction of dimensionless variables, can be
simplified to
\begin{eqnarray}
	{d a \over d \tau} &=& 1 - a - m - k_1 m^2 a   \label{eq_star5} \\
	{d m \over d \tau} &=& k_1 m^2 a - k_2 m^n 
			+ k_s m^n a + k_2 m^{n+1} \ ,  \label{eq_star6}
\end{eqnarray}
where $a$ is the fractional mass of atomic gas,
$m$ is the fractional mass of molecular gas,
$\tau = (K_1 + K) t$ is a dimensionless time variable,
$k_1$ is the the efficiency of production of molecules,
and $k_2$ is the efficiency of triggered star formation.
It is found that this coupled equation system has
three stationary states, two of them trivial (all mass
contained in either atomic or molecular gas), and
a non-trivial stationary solution (Bodifee 1986). 
The latter solution is not necessarily stable against
perturbations, but can evolve into a limit-cycle 
oscillation, constrained by the conditions
$da/d\tau=0$ and $dm/d\tau=0$ in Eqs.~(\ref{eq_star5}-\ref{eq_star6}).
Near the limit cycle, the oscillation is
maintained without an external periodic driving force,
producing repetitive violent bursts of star formation,
separated by long quiescent periods. Similar
limit-cycle solutions were found by Ikeuchi and Tormita
(1983), where supernova remnants control the hot,
warm, and cold gas, and a diffusion transport term
is added. The main spatial manifestation of this
self-organizing mechanism is the spiral structure
of galaxies, in analogy to rotating spiral vortices
formed in chemical auto-catalytic oscillations
(Zaikin and Zhabotinsky 1970; Winfree 1972, 1973;
Cox et al.~1985). 

Subsequent simulations of interstellar turbulence and
star formation include isothermal models of molecular 
clouds and larger-scale multi-phase models to simulate the
formulation of molecular clouds. They {\sl show how
self-organization in highly compressible magnetized
turbulence in the multi-phase interstellar medium can
be exploited to generate realistic initial conditions
for star formation} (Kritsuk et al.~2011).
Multiple states of star-forming clouds have been identified
in 3-D MHD simulations: gravity splits the clouds into
two populations, one low-density turbulent state, and one
high-density collapse state (Collins et al.~2012).

{\sl \underbar{Critical Assessment:} The evidence for
self-organization in the star formation process is the
morphological change from an initial randomized molecular
cloud to concentrations in a number of small localized
(H II) zones in galaxies, which represent spatially
ordered structures (S), in contrast to the initially
uniform randomness of the interstellar gas. 
A size distribution of ordered structures, however,
has not been quantified yet.
The corresponding ordered time structures (T) are produced
by repetitive violent bursts of star formation. The
physical process of self-organization in star formation
is modeled in terms of highly compressible magnetized
turbulence, which can trigger instabilities ending with
high-density collapses (I) (Table 1: qualifiers I,S,T).}

\subsection{    Stellar and other Quasi-Periodic Oscillations   }

The origins of quasi-periodic oscillations (QPO) observed from various 
stellar sources (pulsars, cataclysmic variable stars, neutron stars, binary 
stars, active galactic nuclei, etc.) are largely not understood. 
Interpretations have focused on attributing the overall variability to 
accretion fluctuations, with the QPO produced by modulation of the accretion, 
for example by resonance-like interactions between natural rotational and 
orbital frequencies or -- more germane to this article -- nonlinear chaotic 
dynamics perhaps with modulated limit cycles, which point to self-organizing 
systems.

A number of attempts to detect and characterize deterministic chaos from 
astronomical time series data have been made.  For example, the irregular 
X-ray variability of the neutron star Her X-1 has been analyzed with the 
method of Procaccia (1985) and detection of a low-dimensional attractor 
(D $\approx$ 2.3) and some higher-dimensional chaos was inferred for the 
accretion disk (Voges et al. 1987). The light curves of three long-period 
cataclysmic variable stars have been analyzed with the technique of 
Grassberger and Procaccia (1983a,b) in the search of an attractor 
dimension, but the light curves could be modeled with a periodic and 
a superimposed random component (Cannizzo et al. 1990). Evidence for a 
low-dimensional attractor with a dimension of D $\approx$ 1.5 was found 
in the Vela pulsar with a correlation sum technique (Harding et al.~1990).

However, much of this earlier work has proved to be questionable.  For 
example the result found by Voges et al. (1987) was disputed by Norris 
and Matilsky (1989), who concluded that the insufficient signal-to-noise 
ratio does not allow to distinguish from an ordinary attractor 
contaminated with noise. Since the attractor dimension is equivalent 
to the number of coupled differential equations, we would expect a 
lowest attractor dimension of D = 2 for the Lotka-Volterra equation 
system, or D = 3 for the Lorenz model.  Based on a careful simulation 
study of non-chaotic random data Harding et al.~(1990) questioned the 
significance of their own result quoted above. They concluded 
``It appears that the correlation sum estimator for dimension is 
unable to distinguish between chaotic and random processes.''  

This important cautionary remark is reinforced by at least two key 
theoretical results. Eckmann and Ruelle (1992) presented an elementary 
proof that the correlation dimension estimated with the 
Grassberger-Procaccia algorithm cannot exceed the value $2\ log_{10}(N)$, 
where $N$ is the number of points in the time series.  (One finds in 
the astronomical literature a number of dimension estimates approximating 
this value, suggesting that they are entirely spurious.)  
Osborne and Ruelle (1992) disproved several then traditional views in 
this context, showing that essentially any correlation dimension can 
be found for entirely random (i.e. lacking any ``deterministic chaos'') 
colored random data simply by choosing the appropriate index for the 
power-law power spectrum of the data. They concluded ``These results have 
implications on the experimental study of deterministic chaos as they 
indicate that the sole observation of a finite fractal dimension from 
the analysis of a time series is not sufficient to infer the presence 
of a strange attractor in the system dynamics.''

All in all, these theoretical limits and the realities of signal-to-noise 
and length of available time series cast a pall on the quest for evidence 
of nonlinear dynamics in astronomical systems that continues to some 
extent today.  However the landscape of time domain astronomy is 
improving with respect to time coverage, sampling cadence and 
signal-to-noise, and perhaps prospects for more definitive characterization 
of underlying dynamics of variable objects will improve accordingly.

Based on a more elaborate and physically motivated approach, a time 
series from the R Scuti star, a RV Tau type star, was found to exhibit 
deterministic chaos (with an embedding dimension of 4), because it 
was not multi-periodic and could not be generated by a linear 
stochastic process (Buchler et al.~1996). The quasi-periodic light 
curve is shown in Fig.~\ref{f_buchler} (top panel), along with a 
synthetic light curve generated with a corresponding low-dimensional 
(strange) attractor (Fig.~\ref{f_buchler} middle and bottom). However,
Mannattil et al.~(2016) offer a detailed criticism of this methodology, 
albeit in the different context of X-ray variability.

On the other side of the theory-observation coin, two simple physical 
metaphors incorporating self-organization ideas have inspired 
independent quasi-stochastic models reproducing some features of the 
observed variability of accretion sources.  The dripping handrail 
(Scargle et al. 1993) evokes an analogy between astrophysical accretion 
on the one hand and the accumulation, flow, and dripping of moisture 
on a stairway's handrail on the other hand.  These authors quantified 
the quasi-periodic oscillations (QPO) of the low-mass X-ray binary star 
(LMXB) Scorpius X-1 with a wavelet based power spectrum that was found 
to be consistent with the spectrum computed for a dripping handrail 
accretion model, a simple dynamical system that exhibits transient 
chaos (Scargle 1993; Young and Scargle 1996). This highly oversimplified 
picture nevertheless explains the 1/f and QPO features --- typically 
though to be separate phenomena ---  as two aspects of a single physical 
process, notably ascribing the variability as quasi-random due to 
non-linear dynamics in a constant external accretion flow (and not 
due to a postulated random accretion).

The sandpile metaphor independently inspired a self organization model 
(Mineshige et al.~1994a,b; Mineshige and Negoro 1999),
similar to the dripping handrail, physically somewhat more realistic 
in that its 2D geometry allowed treatment of angular momentum transport 
within the accretion disk. On the other hand these authors postulated
randomness for accretion, although quasi-randomness is generated
automatically by nonlinearities in their model even with steady accretion,
for the same reasons as with the dripping handrail model.

The fluctuation power spectra of accreting black holes, neutron stars,
and white dwarfs that are accreting gas from a stellar companion 
sometimes exhibit peaks at certain frequencies (Remillard and McClintock 2006; 
van der Klis 2006). These are also seen from some supermassive black holes 
powering active galactic nuclei (Smith et al. 2018). These peaks are called 
{\sl ``quasi-periodic oscillations'' (QPOs)}, since they are usually not very 
narrow.  The frequencies observed in the neutron star and white dwarf 
sources are time-dependent, 
usually being positively correlated with the luminosity (proportional 
to the mass-accretion rate). The black hole sources exhibit two classes 
of QPOs, separated in frequency by a factor of at least 30. Only the 
{\sl high-frequency quasi-periodic oscillations (HFQPOs)} 
have a fixed frequency, which is slightly below that
 of the innermost stable orbit in the accretion disk (and therefore 
inversely proportional to the mass of the black hole). However, these HFQPOs 
have a relatively small duty cycle. A mysterious property of many of them is 
the 3:2 ratio of the two highest frequencies (Wagoner 2008). There is no 
complete physical theory that can explain this fact, and no numerical 
simulations reproduce the observed QPOs. The accretion disks are very 
turbulent, driven by the conversion of the differential rotational energy 
via the magneto-rotational instability. In addition, they are subject to 
viscous and thermal instabilities, on time scales greater than the orbital 
period at that radius. The hotter ``corona'' of the neutron star and 
black hole disks appears to up scatter the cooler thermal photons from 
the disk into X-rays, without seriously demodulating them.

{\sl \underbar{Critical Assessment:} A number of oscillatory light curves 
from various types of stars have been recorded, which clearly establish
the presence of non-random ordered time structures (T). The spatial
counterparts (S), of course, cannot be resolved in stellar distances.
The time evolution is generally quasi-periodic, which is typical for
nonlinear dissipative systems with limit cycles (LC), and low-dimensional
attractors have been identified from those time series. The physical
mechanism or instability (I) that is responsible for stellar quasi-periodic 
oscillations is less clear, but an accretion model (i.e., the dripping 
handrail model) has been proposed (Table 1: qualifiers T, LC, I[?]).}   

\subsection{    Pulsar Superfluid Unpinning           }

In the crust of a neutron star or pulsar, the neutron 
superfluid coexists with a lattice of nuclei (Fig.~\ref{f_neutron_star}).
The rotation in a superfluid occurs along quantized
vortex lines only, which must be able to move outward
freely, in order that the superfluid can follow the
observed, electromagnetically driven braking of the
pulsar's rotation. However, there are
pinning centers in the neutron star crust that inhibit
free vortex motion. Therefore the vortex lines could
be pinned to the nuclei in certain layers of the crust
(Anderson et al.~1981).
An angular velocity lag builds up between the
crust and superfluid as a consequence. When the lag
exceeds a threshold, vortex lines unpin catastrophically
and move outward, transferring angular momentum from the
superfluid to the crust and producing an observable
impulsive spin-up of the star, known as a ``pulsar glitch''.

Pulsar glitches are generally interpreted in terms
of the self-organized criticality model, due to the
scale-invariant, power law-like distributions of sizes
and exponential waiting time distributions
(Melatos et al.~2008; Espinoza et al.~2011). In this
scenario, superfluid vortices pin metastably in
macroscopic domains and unpin collectively via
nearest-neighbor avalanches. 
Recent quantum mechanical simulations, in which the
evolution of the pinned, decelerating superfluid is
described by the time-dependent Gross-Pitaevskii equation,
have identified two knock-on processes responsible for
mediating vortex avalanches: local, hydrodynamic,
nearest-neighbor repulsion and nonlocal, acoustic-wave
unpinning (Warszawski et al.~2012). The simulations
also reproduce the size and waiting-time statistics
in observational data, albeit over a relatively small
dynamic range because computational limitations restrict
the simulated system to $<200$ vortex lines at present
(Warszawski and Melatos 2011; Melatos et al.~2015).
Alternatively, 
Melatos and Warszawski (2009) propose a noncritical
self-organization process (which they call ``coherent
noise'' according to Sneppen and Newman 1997), 
where the global Magnus force acts uniformly
on vortices trapped in a range of pinning potentials
and undergoing thermal creep. In this scenario,
Melatos and Warszawski (2009) find that vortices again
unpin collectively, without nearest-neighbor
avalanches, but still produce a scale-free size
distribution as observed. 
The microscopic self-organization processes of
nuclear matter in neutron star crusts has also
been simulated in crystalline lattices, {\sl where
the system organizes itself into exotic structures}
(Sebille et al. 2011; Caplan and Horowitz 2017). 
When the magnetic field in the superconducting
stellar interior is included, the vortex lines
and magnetic flux tubes self-organize into a
turbulent, reconnecting tangle, sustained by stellar
braking (Drummond and Melatos 2017). The flux tubes
act as pinning centers as well, widening the scope
for vortex line avalanches to occur.

In the pulsar glitch self-organization process, the driver
is the electromagnetic braking of the star, while the
feedback mechanism is the local interplay between the
superfluid Magnus force and the pinning potentials, which
regulate semi-coherent unpinning (Cheng et al.~1988).
An alternative yet analogous scenario involving elastic
stresses (star quakes) has also been proposed
(Middleditch et al.~2006). A promising theoretical
framework that is applicable to a wide variety of
self-orgnaizing systems of this kind is the mean-field model 
of a state-dependent Poisson process, indroduced originally
in the context of forest fires (Daly and Porporato 2006),
and solar flares (Wheatland 2008), and generalized
recently to neutron stars (Fulgenzi et al.~2017) and biological
applications (Miles and Keener 2017). The model makes
quantitative predictions of size and waiting time distributions
and size-waiting time correlations as a function of the driving
rate, independent of the detailed microphysics.

{\sl \underbar{Critical Assessment:} One manifestation
of self-organization is the lattice grid of nuclei in
the neutron superfluid zone of a neutron star, which is
a highly ordered spatial (S) structure (like a crystal), 
opposed to a random-like thermodynamic fluid in normal
stars. The physical model involves the rapid rotation
of a pulsar (driver), and a feedback mechanism is 
given by the inhibition of vortex motion in the superfluid 
unpinning potentials (I). The feedback mechanism
maintains a semi-coherent unpinning, which represents a
spatial ordered structure (S) also. A caveat of this model
is that observations of neutron star glitches exhibit
scale-free power law distributions, which are typical
for self-organized criticality models (Table 1:
qualifiers S[?], I[?]).} 

\clearpage

\section{       GALACTIC PHYSICS                        }

Observable galaxies arise when gas (``baryons'') 
flows into concentrations (``halos'') of dark matter and 
forms stars, which themselves radiate and excite residual 
gas to radiate in a number of ways. The morphologies of the 
galaxies we now see is a time-slice of ongoing processes 
that include: (i) build-up of halos massive enough to retain 
gas within a framework of intersecting cell-walls and 
filaments; (ii) continuing gas inflow (only about half of the 
baryons are currently within star-forming halos); 
(iii) outflow of gas (and some 
recycling) driven by winds and jets from bursts of star 
formation, supernovae, and central supermassive black holes; 
(iv) gravitationally driven encounters between halos, 
described as major mergers (when the masses are comparable), 
producing spiral arms and disks), and minor mergers or captures 
of little galaxies by large ones (which can make star streams, 
rings, disks, and central bulges, and in the process initiate 
driving of spiral arms). The starting point is a random 
distribution of small ($\approx 10^{-5}$) density fluctuations 
in the distribution of dark matter through the universe, with 
the spectrum of those fluctuations described by $N(\delta \rho) 
\propto (\delta \rho)^{-1}$ (the Harrison-Zeldovich spectrum). 
A constraint throughout is that the mass of central black holes 
is close to $\approx 0.8 \ 10^{-3}$ of the mass of stars 
through much of the cosmic history.

Over several decades now, many groups have modeled this scenario 
of a universe with N-body simulations (with N gradually increasing 
from $10^6$ to $10^{10}$ and more), and a brief summary of the 
results is {\sl ``Any correct description of our universe must 
look very much like $\Lambda$CDM on large scales, seeded by a 
nearly scale-invariant fluctuations spectrum that is dominated 
by dark energy''} (Bullock and Boylan-Kolchin 2017). That is, 
theory and observations agree well for length scales of a 
megaparsec and more. The situation on small scales is very much 
less satisfactory (Bullock and Boylan-Kolchin 2017; 
Naab and Ostriker 2017; Freeman 2017; Concelice 2014). 
One approach has been the ``zoom simulation'' 
(Springel et al.~2005) that switches 
from large scale considerations to something like the size of 
a galaxy and includes gas processes, star formation, and dust 
attenuation within either an adaptive mesh refinement or a 
smoothed particle hydrodynamic code. Naab and Ostriker (2017) 
show results (2011-2015) from six groups. Each resembles 
some real galaxy (e.g., see figures in Conselice 2014), but 
resemble all to the scenario called ``flocculent''
(Elmegreen and Elmegreen 1987), rather than the ``grand design''
(Elmegreen 2011). In addition, it is a general principle that 
a theoretical process that mimics the real world, does not 
proof its correctness (an argument often used in discussions 
of biological evolution).

The next question is what physical model can produce spirals 
(at least numerically calculated), and which scenario can be 
described in terms of self-organization? 
Binney and Merrifield (1998) state that the arms nearly always 
trail; they are bright because the young stars have formed 
there, but there is some enhanced density of old stars as well 
($\approx 40\%$). Gas is needed to sustain the 
process, so galaxies of the Hubble type S0 generally do not 
show arms, and bars and companions can be drivers. However 
Binney and Merrifield (1998) consider spiral arms, although 
appearing as prominent features, not to be very important in the 
great scheme.

First, it is necessary to understand how some galaxies 
develop disks. Jeans (1915), starting with methods due to 
Boltzmann and treating stars as the gas particles, concluded 
that a non-spherical system with stellar motions describable 
as gas streams could not be static. 
Bertil Lindblad (1926, 1927, 1925) recognized that the Milky Way, 
or at least the parts of it he could study, is rotating, that 
spiral arms seen in the newly-recognized extragalactic nebulae 
would wind up fairly quickly, and that a pattern of higher 
density in the arms, rotating more slowly than matter, could 
be more stable. Lindblad describes his mathematical methods 
as deriving from work by Poincare, also applied to gases.

Sufficient angular momentum produces disks. Lindblad thought 
the rotation might arise from galaxy encounters or mergers, 
though we now associate mergers with destruction of disks and 
formation of ellipticals. Instabilities tend to form warps 
and bars (which the Milky Way has both) 
(Bland-Hawthorn and Gerhard 2016), and it is true that while
2/3 of S-type galaxies now have bars, they were rare at $z=1$, 
but disks still survive. 
Ostriker and Peebles (1973) proposed in their highly-cited
paper that an extended, dark, spheroidal halo would permit 
survival. Bland-Hawthornie and Gerhard (2016) 
show NGC 3 spirals rectified to face-on might be confused 
with the Milky Way if we could see it face-on from outside. 
None has two dominant arms of the type of M51 galaxy, but 
none shows the complexity as the products of ``Zoom-in 
simulations'' (Springel et al.~2005). 
It is perhaps significant that the Milky Way 
also belongs to the rare ``green valley'' category of 
galaxies that are neither blue, vigorous star formers, 
nor red and dead.

Comments specific to the Milky Way include that it reached its 
mostly 2-armed state about 9 Gyrs ago 
(Francis and Anderson 2009), that its present conditions was 
probably triggered by a first encounter with the Sagitarius 
dwarf spheroidal galaxy (Purcell and Bullock 2011). 
M31 incidentally also has a ``driving companion'' and both 
galaxies have their dwarf spheroidal companions largely 
organized in a planar thin structure that is also not understood.

According to Freeman (2017) true bulges come from mergers, while 
instabilities in disks provide bars and pseudo bulges as in the 
Milky Way. Our thick disk (which does not have arms, nor do other 
thick disks (they are nearly but not quite ubiquitous) formed 
11-12 Gyrs ago, at $z=2-2.5$ equivalent time, which is very 
close to the peak of the star formation rate for the local 
universe as a whole (Madau and Dickinson 2014) and probably 
also for the Milky Way (though we expect to know more about 
this topic when the Gaia data are fully in and analyzed), 
but it is likely that the Milky Way spiral pattern requires 
more than one mechanism. The oldest spiral reported so far 
(in a sea of dwarf irregular structures imaged by HST in its 
deep field) is labeled as Q2343-BX442 (Law et al.~2015) 
and is of the ``companion driven'' variety, being seen at a 
redshift that corresponds to an age near 11 Gyrs, and 
evolution tends to move galaxies from {\sl flocculent} to {\sl grand 
design} (Francis and Anderson 2009).

There are also many galaxy-evolution issues that do not obviously 
interact with spiral structure, for instance the correlation 
of stellar masses with central {\sl Super Massive Black Hole} masses 
(Heckman and Best 2014) and deciding whether they radiate enough 
ultraviolet to re-ionize the universe at $z \approx 6$ (Stark 2016).
Locally first, if not temporally first, come perturbations exerted 
on disks by a companion galaxy or bar. The former, with a swing 
amplifier, undoubtedly makes things to start as tidal distortions 
and end up looking like grand design spirals (Toomre 1981). 
And an ensemble of molecular clouds in the 
potential of a barred galaxy comes to trace out a spiral type S 
with their star formation (energy dissipation) 
(Combes and Gerin 1985). But see Toomre (1977) for how far the 
purely gravitational processes had come 40 years ago. 
Woltjer (1965), on the other hand, was certain (and has been 
since at least Woltjer 1959) that magnetic fields had to be part of the 
answer, since the energy densities in the Galactic plane in field, 
cosmic rays, and random gas motions are roughly equal, and both 
field lines and gas motions seemed to be at least partially 
along the arms.

This brings us to the density wave theory of Lin and Shu (1964) 
and Shu (2016). They decided to tackle the problem for the middle, 
that is, to impose pitched arms on the disk density structure 
and gravitational potential and see what happened. The short 
answer was ``Not much''; That is the pattern persisted, 
soliton-like, for several times the rotation period of the 
model galaxy, and the pitch-angles of real galaxies look like 
the calculated ones (Pour-Imani and Kennefick et al.~2016). 
The modern version naturally looks a good deal more complex 
than the 1964 original, but also has some applicability to 
the rings of Saturn and hot-Jupiter formation around other 
stars (Shu 2016).

The chief competing theory for some years, applicable 
particularly to spiral in the past and to flocculent ones in 
{\sl Stochastic Self-Propagating Star Formation (SSPSF)}, but 
forward by Mueller and Arnett (1976) and further developed 
by Gerola and Seiden (1978). The idea is that a random 
fluctuation in gas density yields a small burst or star 
formation; winds from the stars and supernovae move outward, 
compressing gas, which, in turn, forms stars. Differential 
rotation in the disk, which will eventually wind up and 
spoil arms, in the short term ($\approx 10^8$ years), so 
the galactic rotation period stretches out those regions 
of propagating star formation into arc-shaped features. 
Thus SSPSF is a possible ``starter'' to establish S-shaped 
perturbations to the the Lin and Shu (1964) process started. 
Auer (1999) combined the 
processes in roughly this fashion and regarded the result 
as a good way of looking at the initiation, development, 
and eventual washing out of spiral structure. Since all 
processes (Naab and Ostriker gas outflows, Lin and Shu 
magnetic confinement, SSPSF, and event companion driving) 
require the presence of gas to form new stars and make 
the arms visible spiral arms necessarily transient, 
on time scale from $10^8$ to $10^{10}$ years.

In summary, the structure of our Milky Way was triggered 
when the Sagitarius dwarf first passed through the disk 
about 9 Gyrs ago (Purcell et al.~2011). The arms have 
been preserved by a density wave (Lin and Shu 1964) with 
a swing amplifier (Toomre 1981, 1977), but an image 
reconstructed from HI data and starcluster information 
suggests with the bar as a likely additional driver 
(Combes and Gerin 1985), but an image reconstructed from HI 
and starclusters data (Bland-Hawthorn and Gerhard 2016; 
Elmegreen 2011) that we also have transient, flocculent spurs 
and other structures (Elmegreen and Elmegreen 1987). 
Gas inflow, which continues along the filaments that connect 
galaxies (Faucher-Giguere and Angles-Alcazar 2017), tends 
to make the disk larger and less dense and capable of 
continuing star formation (Nabb and Ostriker 2017).

We turn now to the four cases of spiral formation or 
preservation that appear to be most closely connected with 
self-organization.

The spiral pattern of differentially rotating galactic disks
represents a self-organization process (Fig.~\ref{f_galaxy_hubble}). 
Nozakura and Ikeuchi (1988)
model irregular and regular spiral patterns of the interstellar
medium with a reaction-diffusion process (Fig.~\ref{f_galaxy_spiral}), 
which is a self-organization process known in chemistry (e.g., 
Bray, W.C. 1921; Cox et al.~1985).      

The physical model of a differentially rotating galactic disk
of Nozakura and Ikeuchi (1988) contains the following assumptions:
(i) The interstellar medium has two components $\rho_1({\bf r},t)$
and $\rho_2({\bf r},t)$; (ii) The interstellar medium exhibits a 
limit-cycle behavior around a steady point $(\rho_{10}, \rho_{20})$
in the $\rho_1-\rho_2$ phase space; (iii) the galactic disk
is 2-dimensional and infinitely extended; (iv) the rotation curve of
the galactic disk is flat, $V({\bf r})=const=V_0$ throughout the
disk $(0 \le r < \infty)$; and (v) the propagation of the local
interstellar medium is expressed by the scalar
diffusion matrix ${\bf D} = diag(D,D)$, where $D>0$ is a
diffusion coefficient. The model with the following 
reaction-diffusion equations with advective terms satisfies 
these 5 conditions:
\begin{equation}
	{\partial \over \partial t} \left( {\rho_1' \over \rho_2'} \right)
	+ {V_0 \over r} {\partial \over \partial \theta} 
	\left( {\rho_1' \over \rho_2'} \right)
	= \left[ 
           \begin{array}{cc}
		\lambda(\rho') & -\omega(\rho') \\
		\omega(\rho')  & -\lambda(\rho') 
	   \end{array}
         \right]
	 \left( {\rho_1' \over \rho_2'} \right)
	+ {\bf D} \nabla_r^2 
	 \left( {\rho_1' \over \rho_2'} \right) \ ,
	\label{eq_nozakura}
\end{equation}
where $\rho_i'({\bf r},t)=\rho_i({\bf r},t)-\rho_{10}, i=1,2$  
are the deviations of $\rho_1$ and $\rho_2$ from their steady
values $\rho_{10}$ and $\rho_{20}$,
$\lambda(\rho')$ and $\omega(\rho')$ are the nonlinear
reaction terms concerning mainly with the stability and the
oscillation frequency of the system,
and ${\bf D}$ is the diffusion matrix.
Nozakura and Ikeuchi (1988) found nonlinear rigidly-rotating
spiral wave solutions for this analytical model, which is
designed to have limit-cycle solutions. In this model,
rigidly-rotating spiral structures are a consequence of the
balance between the winding effect of differential rotation
and the straightening effect of diffusive propagation.
 
More complex galactic models with chaotic orbits and
massive central masses (possibly attributed to a central
black hole) were investigated by Kalapotharakos et al.~(2004). 
Small central masses with a ratio of $m < 0.005$ were found
to organize chaotic orbits with Lyapunov exponents too small to
develop chaotic diffusion during a Hubble time. Large
central masses $(m \gapprox 0.004)$, produce about the
same amount of chaotic orbits, but the Lyapunov exponents 
are larger, so that the secular evolution evolves into a
new equilibrium. The underlying self-organization mechanism
converts chaotic orbits into ordered orbits of the 
{\sl Short Axis Tube} type (Kalapotharakos et al.~2004). 

A spatio-temporal self-organization in galaxy formation 
has been found from a relationship between the number 
of star formation peaks (per unit time) and the size of 
the temporal smoothing window function 
(used to define the peaks), holding over 
a range of $\Delta t = 10-1000$ Myr (Cen 2014).
{\sl This finding reveals that the superficially chaotic
process of galaxy formation is underlined by temporal
self-organization up to at least one Gyr} (Cen 2014). 

The observed hierarchy of galactic structures, from giant
cellular voids to enormous superclusters, with galaxies
distributed within intricate networks of arcs and cells,
clearly indicates some self-organizing process that is not 
consistent with a random distribution (Krishan 1991, 1992).
Two different scenarios are usually considered: (i) the
hot dark matter (HDM) scenario with initial large-scale 
structures that fragment into smaller ones, and (ii) the 
cold dark matter (CDM) scenario where the smaller structures
form first, coalescing then to larger galactic structures.
However, besides the self-organization of structures seen in 
luminous matter, the existence of dark matter (Trimble 1987)
may have its own ``dark self-organization''.
Related self-organization processes may drive gravitational 
clustering and/or turbulent cascading (Krishan 1991, 1992). 

{\sl \underbar{Critical Assessment:} According to the Hubble
galaxy classification, galaxies can be formed in different
morphologies, from ellipticals (type E0-E7) to normal spirals
(type Sa-Sc) and barred spirals (SBa-SBc), which all represent
a spatial pattern (S) observed in our present time-slice.
The evolution from an initial random-like state to a
well-ordered spatial structure (S) with a spiral pattern reveals
the action of a self-organization process. 
Physical models of galaxy formation include at least three 
scenarios that are more or less consistent with the $\Lambda$CDM
cosmology: (i) Interaction with a nearby companion, (ii)
the Lin-Shu density wave theory, or (iii) flocculent bursts
of star formation that get dragged out by differential rotation.
Thus, a combination of gravity and rotation is a most likely
driving force, while the feedback force or self-organizing
instability is still open and is currently investigated with
large-scale numerical N-body simulations.
(Table 1: qualifiers S, I[?]).}

\clearpage

\section{	COSMOLOGY				}

Self-organization inherently involves regulated change.  
That is, it must be dynamical, involving driver forces and positive
feedback mechanisms. Each of these features has an associated scale. 
When looking for the self-organizational aspects of cosmology, 
relevant scale sizes run from (at least) as small as the Planck scale 
($\approx 10^{-33}$ cm) to the size of the observable Universe 
($\approx 10^{28}$ cm), spanning over 60 orders of magnitude. 
Perhaps even more challenging, modern science has yet to reveal exactly 
whether and how one may correctly separate our understanding of 
particles and fields 
from that of the space-time in which they exist, especially at scales 
near to and less than the Planck scale.  For the purposes of this review, 
we will simply focus on the organizational aspects of mainstream cosmology.  

\subsection{	Einstein-de Sitter and $\Lambda$CDM Models		}

The first goal of cosmology is to understand how space changes with time. 
Einstein ``interpreted gravity as a manifestation of geometry''
(Misner et al.~1973). He showed us that a ``4-D space-time'' 
formed by merging 3-D space and the time dimension into 
one continuum (space-time), can respond to any form of energy. 
Einstein's equation couples the curvature of 
space-time, i.e. gravity, to the stress-energy inside it. Therefore, 
placing energy sources inside a well-chosen unified 4-D space-time 
geometry is the best way to quantify how they coevolve.  
To examine how order developed, we recognize that the expansion 
dynamics of space-time sensitively depends upon the amount(s) of 
each energy component inside it, and the evolving organization of 
those energy components depends upon the expansion dynamics. Such 
coupling or ``feedback'' can lead to an evolution that ``self-organizes''.
Thus, when space-time contains matter and/or vacuum energy, 
interesting processes can emerge. 

Einstein's first solution to his 
equation assumed that space-time contained uniformly distributed 
normal matter, but was static; however, quickly realizing that 
space-time containing only matter would collapse, unless another 
component was included to resist it, he added a positive 
``cosmological constant term''. This is the equivalent of a uniform 
vacuum energy that counter-balances the curvature-producing
effect of matter.   
Soon thereafter de Sitter produced a model envisioning a maximally 
symmetric space whose ``metric'' (curvature) is the same 
{\sl at all times and all places}, which also included both uniformly 
distributed matter and a cosmological term to balance it.  For such a 
space-time (de Sitter model 1), one may chose a time coordinate and 
its associated family of space like coordinates (``slicings'') that 
correspond to specific values of that time coordinate, to thereby 
represent geometrically flat (Euclidean), positively-curved, or 
negatively-curved 3-D spaces.  However, the de Sitter (model 1) space 
does not restrict this choice, so it does not select a specific 
cosmology, {\sl per se}.  In other words, maximally symmetric de 
Sitter (model 1) space may just rest! However, {\sl maximally symmetric
space-times are ... not reasonable models of the real world}
(Carroll 2004). Then, de Sitter found that in a space-time with 
only a cosmological constant and no matter, test particles would 
accelerate away from one another! It is this second version 
(de Sitter model 2) with accelerating expansion that is normally 
associated with the ``de Sitter space''. This was progress.  

In the 1920s, Alexander Friedman and Georges LeMaitre further studied 
how the inclusion of matter in Einstein's Equation could affect things, 
but this time their independently derived solutions narrowed down the 
space-time symmetries to yield a cosmological model of a universe that 
can undergo smooth expansion (Fig.~\ref{f_cosmology}). More 
importantly, if one assumes that space-time contains a homogeneous and 
isotropic cosmic fluid composed of given matter, radiation and/or 
vacuum energy densities, the Friedman-LeMaitre solutions will: (i) limit 
its geometrical possibilities by selecting slicings from among the flat, 
positively, and negative curvature options for the 3-D spatial part of 
the metric, and (ii) determine its expansion dynamics! 
Matter and radiation resist or slow down the expansion rate, while
vacuum energy does the opposite. In addition, initial densities of
radiation and matter decrease with volume, while vacuum energy
normally stays constant, all the while remaining isotropic and
homogeneous. Their model is reflected in the following two equations 
known as the ``Friedman Equations'':
\begin{equation}
	\left({ \dot{a} \over a } \right)^2 = {8 \pi G \over 3} \rho
		- {k \over a^2} \ , \label{eq_cosmo1}
\end{equation}
\begin{equation}
	\left({ \ddot{a} \over a } \right) = - {4 \pi G \over 3} 
		(\rho + 3 p ) \ ,   \label{eq_cosmo2}
\end{equation}
where changes in the scale factor $a$ are related to total energy density 
$\rho$ (which can include matter, radiation, and vacuum), the overall 
geometrical curvature $k$ (can be positive, negative, or zero), 
and pressure p.  Note that the term in parentheses in Eq.~\ref{eq_cosmo1}  
is the ``Hubble parameter''.
These equations were derived by inserting 
the ``Robertson-Walker metric'' into Einstein's Equation, 
which then expresses how the scale factor (size) 
of 3-D space changes with time. This combination successfully related 
the scale factor to the evolving stress-energy of the Universe.

It was eventually recognized that our actual Universe (i) is geometrically 
flat $(k=0)$, (ii) has been expanding for almost 14 billion years, (iii) its 
cosmic fluid has passed through stages where its dominant component was 
radiation, then matter, and then vacuum energy, and (iv) we have been in 
a quasi-de Sitter accelerating expansion stage for the last 6 billion years! 
(Fig.~\ref{f_cosmology}).  
There is compelling evidence that the cosmic fluid is made up of normal 
matter and radiation, a form of matter that does not emit or absorb 
electromagnetic radiation ({\sl Cold Dark Matter or ``CDM''}), and vacuum 
energy ({\sl ``Dark Energy''}).  It is then no surprise that the current 
cosmological model (``$\Lambda$CDM'') is based upon the Friedman Equations,
although it is still undecided whether the vacuum energy component has a
constant value $(\Lambda)$ or is changing with time 
(Fig.~\ref{f_cosmology}). 

\subsection{ 	Evolution of Matterless Space-Time	}
 
The self-organizational concepts have been applied to the creation of de 
Sitter (model 2) space-time (only vacuum energy) and to the more relevant 
Friedman-LeMaitre space-time that obeys the Friedman equation. Creation 
of de Sitter (model 2) space-time from quantum fluctuations, combining 
causality and gravity with quantum theory, was discussed by Ambjorn et al.~2008. 
Viewed as a self-organization process, many microscopic 
constituents exhibit a collective behavior and give rise to a unified, 
smooth space-time macrostructure in this model. However, one must keep in mind that 
this pure de Sitter (model 2) space-time, without matter, is maximally 
symmetric and too broad to reflect the real Universe. 

In another approach quantum gravity is described as a network that 
self-organizes into a discrete 4-D universe, in analogy to the 
ferro-magnetic Ising model for space-time vertices with an 
anti-ferromagnetic Ising model for the links. The ground state 
self-organizes as a new type of low-clustering graph with finite 
Hausdorff dimension 4 (Trugenberger 2015). Once again, this work 
does not appear to directly lead to the $\Lambda$CDM model.

\subsection{	Evolution of Space-Time with Matter and Radiation	}

The largest order out of random process in astrophysics today 
is the production of the observed large scale structure of 
galaxies and clusters of galaxies throughout the cosmos. The $\Lambda$CDM 
model, together with the theory of cosmic inflation, lay the foundation 
(i) for generating the initial conditions for structure formation, 
(ii) for creating matter and radiation, and (iii) for the subsequent 
hierarchical growth of the structure of matter via gravitational 
instability. 
	
Because Einstein's Equation relates space-time curvature (i.e., gravity) 
to the stress-energy of its contents, fluctuations of energy density will 
generate fluctuations of curvature. ``Inflation theory'' envisions a very 
early burst of quasi-de Sitter (model 2) expansion during near-Planck scale 
stochastic quantum fluctuations of the inflation-driving scalar energy 
field (``inflaton''), and generates the corresponding space-time curvature 
fluctuations that expand superluminally to semi-classical scales. 
As inflation ends, the inflaton transfers its remaining energy into 
radiation and particles that, during the first few minutes of the Big Bang, 
evolve through a nucleosynthesis stage into a plasma of ``normal matter'' 
(comprised primarily of hydrogen and helium atoms and electrons) and 
gravitationally-interacting-only Dark Matter. As the continuing expansion 
further cools the cosmic fluid (plasma) further, the theory goes, 
it is attracted by the curvature fluctuations (gravitational potentials) 
originating from the inflaton field, and ultimately collapse into the 
structure we see today.

In a very recent paper, Ge and Wang (2017) set forth an approach
to derive cosmological dynamics starting with the physics of
quantum entaglement. Building upon earlier ideas that space-time
geometry could be the result of the entanglement of macroscopic
quantum states, together with recent work by Jakobson hypothesizing
a relationship between Einstein's equation for gravity and vacuum
entanglement of quantum states, Ge and Wang (2017) were able to
derive the flat-Universe Friedman equations 
(\ref{eq_cosmo1}) and (\ref{eq_cosmo2}) above.
It will be interesting to see if further work exploiting the apparent
deep connection between quantum information theory and the emergence
of space-time will successfully be applied to the entire $\Lambda$CDM
Universe paradigm with its inflaton field, dark matter, and dark energy.

\subsection{	The Cosmic Microwave Background 	}

Measurements of the {\sl Cosmic Microwave Background (CMB)} over the past 25 
years strongly support the idea that at about 400,000 years into the 
Big Bang the temperature of the H/He plasma dropped to around 3000 
degrees K, allowing the electrically charged free electrons and 
nuclei to then combine into neutral atoms. At this point, known as 
``recombination'', electromagnetic radiation (photons) that had 
previously enabled the plasma to resist, gravitational collapse was 
released, carrying the image of the last surface from which it 
scattered.  Continued expansion of the Universe then caused the 
wavelengths of the released photons to stretch from visible to 
microwave values. 

The revolutionary and Nobel Prize winning (Smoot and Mather)
{\sl Cosmic Background Explorere (COBE)} satellite work 
(Mather et al.~1991; Boggess et al.~1992), 
followed by the {\sl Wilkinson Microwave Anisotropy Probe (WMAP)} 
(Bennett et al.~2013; Hinshaw et al.~2013),
and Planck satellites (Planck Collaboration 2016), with increasing 
sensitivity and resolution, precisely mapped that 13.8-billion-year-old 
microwave image of the celestial sphere of the cosmos 
which encoded much of the physics of the early Universe  
(Fig.~\ref{f_cobe}).

The CMB revealed that the H/He plasma had begun to form slight 
over densities and under densities (anisotropies), i.e., clouds, 
of a range of sizes that reflected a Gaussian random distribution, 
just as predicted by inflation theory.  The CMB analysis also 
revealed the relative amounts of normal matter, dark matter, and 
vacuum energy contained in the ``cosmic fluid'' at that time, and 
therefore the geometry and age of the currently observable Universe. 
This information combined with the distance-marking capability, 
demonstrated for Type Ia supernovae in Nobel Prize winning work
(Riess et al.~1998; Perlmutter 1999), together with
other measures of the expansion history of the Universe,
provided the initial and continuing conditions 
that determined large scale structure. Fig.~(\ref{f_inflation})
depicts the overall evolution.

Interestingly, when the inflaton field transfers its energy into 
radiation and particles and the Universe then re-heats to thermal 
equilibrium to continue the hot Big Bang, the cosmic fluid may be 
treated as a perfect fluid undergoing an adiabatic expansion.  
As the temperature drops with expansion and matter is created, 
the big picture is still isotropic and homogeneous. 
At the time of the CMB 
release, the curvature fluctuations had caused anisotropies in 
the matter distribution on the order of one part in 100,000.  
Long before resolution of the CMB image, astronomers knew from
observation that anisotropies at least this large were needed
to ``make structure on time''.
Thereafter these wispy but critically important over-densities 
of matter underwent rapid local amplification via linear 
fluctuations, and 
then collapsed by non-linear gravitational forces into structures. 
So here, the inflaton field served to generate both the driving force of 
the process as well as the primordial quantum fluctuations that 
seeded the gravitational feedback needed for structure formation 
from the thermally equilibrated and cooling cosmic fluid. Recent 
galaxy surveys clearly show a distribution pattern
(Fig.~\ref{f_blanton}).

\subsection{	Formation of Large Scale Structure	}

N-body simulations of the post-CMB evolution of dark matter into 
large scale structure were carried out by the Virgo Consortium 
(Millenium Simulation or MS) in 2005 (Springel et al.~2005). 
The basic simulation, including only dark matter, successfully 
reproduced the ``cosmic web'' topology (Libeskind et al.~2017)
of a $\Lambda$CDM Universe, 
with its known clusters and filaments of size $\approx 100$ Mpc 
and its essentially isotropic homogenous matter distribution on 
larger scales (Fig.~\ref{f_millenium_sim} bottom to top).  The formation 
of galaxies and quasars was then separately accounted for by 
adding semi-analytic modeling to test the importance of 
``baryonic effects'', such as gas cooling, star formation, 
feedback, etc., within the simulated dark matter 
substructures. Then, the group examined galaxy clustering, 
luminosities and colors (star formation rate and age) and 
compared these with observations.      

More recently, the Illustris Simulation (IS) carried out a 
series of large-scale simulations of galaxy formation that 
included both gravity and hydrodynamics to directly account 
for the baryonic component (gas, stars, supermassive black 
holes, etc.). The $\Lambda$CDM model and recent CMB derived 
cosmological parameters were again used to set the initial 
conditions of the simulation, which began 12 million years 
after the Big Bang and ran forward for about 14 billion years.  
As seen in Fig.~(\ref{f_unisim3} from bottom to top), the IS 
depicts the evolution of dark and baryonic matter from the 
linear through the non-linear collapse stages, and beautifully 
reproduces the growth of structure that favorably compares 
with observations.  Using short time steps, the IS was also 
able to show the time evolution of baryon parameters (gas 
temperature, density and metallicity) associated with 
simulated Active Galactic Nuclei explosions 
(Fig.~\ref{f_unisim2} from left to right). 

So how do the MS and IS relate to the theme of this review? 
They show that the $\Lambda$CDM model, including inflation, 
well describes the physical evolution of the large-scale 
structure of the Universe to the present time.  The most 
interesting feature of the underlying cosmological model 
is that a period of exponential inflation can temporarily 
remove all disorder within the causal speck of space-time 
that then grows into our current observable Universe.  
At the end of inflation, the matter content of the cosmic 
fluid is created from the remaining inflaton energy and 
re-heats to a temperature at or above that envisioned by 
grand unified particle theories.  Then, the ``Big Bang'' 
ensues.  This model ensures that the post-inflation 
evolution of the Universe is basically isotropic and 
homogenous, as the Friedman Equation assumes.  However, 
it is most fortunate for us that the large-scale structure 
is not perfectly isotropic and homogeneous, but exhibits 
a definite ordering into filaments, cluster nodes and voids 
on scales smaller than $\approx$100 Mpc where galaxies and 
stars, heavier elements, and life came to be.  

\subsection{ Self-Organization and Logistic Growth 	}

The topology of the large-scale structures of our Universe
can be characterized by fractal geometry (Murdzek and Iftimie 2008).
Using the recently completed redshift surveys (e.g., Fig.~\ref{f_blanton}),
which provide galactic right ascension $(l)$, declination $(b)$, and
redshift $(z)$, one can transform the redshift $z$ into
a distance $r$,
\begin{equation}
	r = {c \over H} \int_0^z 
	{dz \over \sqrt{\Omega_M(1+z)^3} } \ , \label{eq_cosmo3}
\end{equation}
where $c$ is the speed of light, 
$H$ is the Hubble constant, $H=70$ km s$^{-1}$ Mpc$^{-1}$,
and $\Omega_M=1$ is the the Einstein-de Sitter model, which then
yields the 3-D space coordinates $(l,b,r)$, from which the fractal
(Haussdorf) dimension $D$ can be obtained, which defines a
fractal volume $V_{fractal} \propto r^D$ that is smaller than
the Euclidean volume $V \le r^3$, since $D \le 3$.
Murdzek and Iftimie (2008) 
find this way a lowest fractal dimension of $D \approx 1.3$ for
nearby galaxies ($r = 25$ Mpc), which monotonously grows and
saturates at a value of $D \lapprox 2.0$ at the largest distances
($r \gapprox 250$ Mpc). They interpret the distance-dependent
fractal dimension function $D(r)$ as a radial (or temporal) 
evolution that can be modeled with a logistic curve (or Verhulst 
equation),
\begin{equation}
	{dD(r) \over dr} = \Gamma D(r) 
	\left( 1 - {D(r) \over D_{\infty}} \right) \ , \label{eq_cosmo4}
\end{equation}
where $\Gamma$ is the exponential growth rate, and 
$D_{\infty}=D(r=r_{\infty})$ is the asymptotic limit at an infinite
distance, also known as {\sl carrying capacity} (or 
maximum amount of ressources) in ecological models. 
The authors argue that the logistic growth model
(Verhulst law), which describes nonlinear growth phenomena 
in a closed system with a limited total ressource quantity $(D_{\infty})$, 
is a concept that agrees with the nonlinear theory of structure formation,
and thus indicates a self-organized universe. The self-organizational
aspect is the predicted feedback that the growth rate $dD(r)/dr$ of the
fractal dimension $D$ is decreasing to zero (in the asymptotic
limit) when the scale is increased $r \mapsto \infty$ (in Eq.~\ref{eq_cosmo4}). 
Interestingly, this
model predicts an almost finite universe, where the mass or energy
asymptotically vanishes at large distances ($r \gapprox 250$ Mpc).
It also predicts 2-D galactic structures at large distances, 
and 1-D structures (filaments or curvi-linear threads) at nearby 
galactic distances of $r \lapprox 25$ Mpc.

\subsection{ Self-Organization of Interacting Cosmic Fluid Components	}

Self-organization of components of the cosmic fluid into stars 
and galaxies was covered earlier in this review where dark and 
baryonic matter are assumed to only interact gravitationally. 
However, non-gravitational interactions of dark matter and 
dark energy have also been studied in a cosmological model 
with diffusion (Szydlowski and Stachowski 2016).  The state 
variables of the density parameter for matter (dark and visible) 
and of the rate of growth of energy transfer between the dark 
sectors can be coupled using the Lotka-Volterra framework, 
from which it was demonstrated that the de Sitter solution is 
a global attractor for all trajectories in the phase space 
(Szydlowski and Stachowski 2016).  In a related approach, 
called the ``Jungle Universe'', the dynamics of homogeneous 
and isotropic Friedman-Lemaitre universes are considered as a 
special case of a generalized Lotka-Volterra system, where 
the competitive species are the barotropic fluids that fill 
the universe (Perez et al. 2014). 

{\sl \underbar{Critical Assessment:}  
The large-scale structure of the Universe 
is seen to have resulted from a combination of 
quantum-fluctuation-seeded gravitational collapse and more 
complex particle physics, all the way back to the Big Bang. 
Self-organization concepts applied to cosmology are extremely
scanty in literature (amounting to a few sentences
in a few cosmology papers) and appear not to relate to the
$\Lambda$CDM model. Quantitative measurements of spatial 
(S) or temporal structures (T) that discriminate against random 
patterns, identification of nonlinear dissipative systems with
driver and positive feedback mechanisms, critical instabilities (I),
and possible limit-cycle (LC) equilibria need to be identified.
In conclusion, there is a lot of room for modeling of cosmological 
models in terms of self-organization 
(Table 1: qualifiers S[?], T[?], [I?], LC[?]).}

\clearpage

\section{	DISCUSSION					}

In this interdisciplinary review we aim to point out some
universal properties of nonlinear systems governed by
self-organization. We discussed 6 cases in planetary
physics, 6 cases in solar physics, 3 cases in
stellar physics, one case in galactic physics, and
some tentative ideas in cosmology, amounting to 17 systems in
the field of astronomy and astrophysics (Table 1). 
Self-organizing systems, however, have been found
in many more scientific disciplines, such as in 
ionospheric physics, magnetospheric physics, plasma
physics, physics, chemistry, biology, social science,
and computer science, as the 51 examples compiled
in Table 2 demonstrate. 

In Table 4 we juxtapose the system characteristics of
self-organizing systems to random systems, as well as
to self-organized criticality systems. The three different
dynamic system types are visualized also for the same 
medium, such as sand, in Fig.~(\ref{f_sand}). 
In the following we characterize each system in turn.

\subsection{	Characteristics of Random Processes 	  }

Random or stochastic processes can be characterized
with the statistics of 
independent events. The mathematical distribution of 
independent events can be derived from rolling dices, 
which leads to a binomial distribution and can be 
approximated by a Gaussian function 
(also called normal distribution)
in the limit of an infinite number of dices, or with a
Poisson distribution or exponential distribution in the
limit of rare events. A time series of random events
consists of irregular, intermittent events and the 
resulting power spectrum is characterized by white noise 
(i.e., a flat power spectrum $f(\nu)=const$). We can
consider random processes in time or in space. If spatial
structures are produced by a random process, their size
distribution is theoretically a Gaussian function, 
with a well-defined
mean and standard deviation, where the mean defines a specific
preferred spatial scale. From the thermodynamic or information 
(theory) point of view, the entropy is increasing with time
in random processes.  Examples of random processes are
Brownian motion of gas molecules, diffusion processes,
the detected photons from a star, electrical 
current fluctuations due to thermal noise, or patterns
of sand at the beach (Fig.~\ref{f_sand}a).
(For a concise summary of the statistics of random processes
see Section 4 in Aschwanden 2011). 

\subsection{	Characteristics of Self-Organized Criticality 	}

Self-organized criticality systems (Bak et al.~1987; Pruessner 2012;
Aschwanden et al.~2016) 
are completely different from random processes, which is 
experimentally and observationally demonstrated by the 
appearance of scale-free power law distributions
of spatial and temporal sizes. Avalanches in self-organized
criticality systems represent coherent structures in the time 
domain (1/f-noise),
in contrast to incoherent noise in random systems.
Quantitatively, a size distribution
of avalanches in a self-organized criticality system can be
simulated from chain reactions of nearest-neighbor interactions
in a lattice grid, where a critical threshold of the gradient 
(or curvature radius) between next-neighbor interactions has to 
be exceeded, before an avalanche can start. The avalanches
occur intermittently in such a complex system, and the
intervening time intervals (waiting times) obey an exponential
random distribution function. The spatial structure of the avalanches
is fractal (or multi-fractal), which corresponds to a power law
distribution also. The reason why such a dissipative nonlinear
system is called ``self-organizing'', 
(e.g., the critical slope of a sandpile) is the fact that the
system automatically maintains the critical state of avalanching
without external control, as long as the energy input into this
open system is steady and stochastic. The microscopic structure
of a self-organizing system is maintained in the time average, 
and thus the entropy of the system is invariant when averaged over
many avalanches. Examples of self-organizing systems are sand piles 
(Fig.~\ref{f_sand}c), earthquakes, solar flares, forest fires, 
stock market fluctuations, etc. (see a representative list of 
phenomena in Section 1 of Aschwanden 2011).

\subsection{	Characteristics of Self-Organization 	  }

Although the two terms ``self-organization'' and ``self-organized
criticality'' sound confusingly similar, they characterize two
completely different types of dissipative nonlinear systems,
to which terminology we will adhere for historical reasons. 
The only commonality between the two systems is that both are open
nonlinear dissipative systems with external energy input, and that both 
maintain some system property in an automated way without external
control. A self-organizing system has always a primary
driving force, and a secondary counter-acting force that acts
as a stabilizing positive feedback reaction to the driving force.
A system can only be called to be a self-organization process,
when the feedback mechanism leads to a quasi-stationary
stabilization of the combined system. Otherwise, a system
with a negative feedback (or none) will evolve away from a 
stationary state and end in a catastrophic way. 

What sets a self-organizing system apart from a random system
is the ability to create {\sl ``order out of chaos''}, or
better ``order out of randomness'', since the term ``chaos''
is already used in nonlinear physics to characterize a particular 
type of nonlinear system behavior that is non-deterministic . 
Therefore, the property of self-organization is also called
spontaneous order, a process where some form of overall order
arises from local interactions between parts of an initially
disordered system. In principle, every ordered structure that is 
significantly different from randomized distributions requires
an ordering mechanism, or a self-organizing process. We have
seen in this review that every self-organizing mechanism 
observed in astrophysics can be modeled by a system of
coupled differential equations, among which the type of 
a Lotka-Volterra system is most prominent. These differential equations
describe the interaction between a driving force and a positive
feedback force, which generally have a limit-cycle solution,
and therefore can sustain a quasi-stationary oscillation 
near the limit cycle, which is also called attractor
(or strange attractor if it has a fractal structure).
The quasi-stationary, quasi-periodic system dynamics near a 
limit cycle is the essential characteristic of self-organization
processes. 

What are the size distributions of a self-organizing
system? Since the limit cycle represents a fixed value of
a time period (which often corresponds also to a fixed value
of a spatial structure), the size distribution is expected to be
peaked or quantized at this particular value. For instance, the solar
granulation exhibits a fixed spatial scale of $w \approx 1500$ km
for convection cells (granules), and a temporal scale (or life time)
of $\tau \approx 8-10$ min. In the case of planetary systems, each
planet has its own attractor and limit-cycle dynamics, which are
moreover weakly coupled by harmonic ratios in a N-body system.
Since the limit-cycle solution often
contains some random noise, the size distribution is generally
not sharply quantized like a delta-function, but rather broadened
to a single or multiple Gaussian functions, see Gaussian distribution
of granules in Fig.~(\ref{f_abramenko}, right panel). 

Since self-organizing systems create ``order out of randomness'',
the entropy is decreasing during the evolution from an
initially disordered system to a self-organized limit-cycle
behavior. It is also said that self-organizing systems evolve
into a dynamics far away from thermal equilibrium. 
A limit cycle is defined by a critical point (in phase space)
around which the quasi-stationary oscillation dynamics occurs, 
where the amplitude of the oscillation is a measure how far off
the system evolves from a system equilibrium solution.  
The thermal equilibrium therefore corresponds to the asymptotic 
limit of a vanishing limit-cycle amplitude, where every
dynamic system variable becomes constant. 

In Fig.~\ref{f_sand}b) we show the pattern of sand dunes,
which self-organized in an interplay between gravity and wind,
forming ripple patterns with a fixed ripple separation scale, 
which is distinctly different from sand piles generated by 
self-organized criticality avalanching (Fig.~\ref{f_sand}c),
or from a sand beach shaped by random processes.

\subsection{	The Physics of Self-Organization Systems 	}		

It should be clear by know that the term ``self-organization''
merely expresses a category of a nonlinear dissipative system
behavior, which is a general property of complex system behavior,
but does not define a specific physical model for an observed
phenomenon. The choice of a particular physical model that is
applied to an observed phenomenon is a matter of interpretation.
We can study the dynamic system behavior with purely mathematical
models (of coupled differential equation systems) without
specifying a physical application (e.g., see examples in textbook 
by Strogatz 1994, Chapter 7). However, since we are most interested in
obtaining physical insights from the observed phenomena, we
identified the underlying driving forces and positive feedback 
mechanisms for each of the 17 studied astrophysical phenomena
(Table 1).

A summary of observed astrophysical phenomena and the
self-organizing driver forces and feedback mechanisms is given
in Table 1, based on the concepts offered in the reviewed
publications. Drivers can be the gravitational force,
the centrifugal force (from rotation), differential rotation,
solar radiation, temperature gradients, convection, magnetic stressing,
plasma evaporation, acceleration of nonthermal 
particles, or the cosmic expansion. Feedback mechanisms
involve mostly instabilities (i.e., the magneto-rotational
or Balbus-Hawley instability, the Rayleigh-B\'enard 
instability, turbulence, vortex attraction,
magnetic reconnection, plasma condensation, loss-cone
instability), but also resonances (mechanical orbit
resonance, double plasma resonance). While instabilities
mostly evolve into limit-cycle behavior, which constrains
one specific time scale, resonances can produce ordered
structures at multiple quantized values (such as 
harmonic orbit resonances of planets, or magnetic 
harmonics in upper hybrid waves of solar radio bursts).

We described the underlying physical models with systems
of coupled differential equations in this review.
Very few equation systems can be analytically solved, if at all.
For instance, even the basic Lotka-Volterra equation system
has a transcedental (implicit) solution (Appendix D).
Consequently, the more complex cases that involve
a hydrodynamic approach (photospheric granulation,
chromospheric evaporation, star formation, galaxy
formation), an MHD approach (protoplanetary disks, 
solar magnetic fields, the Hale cycle), or N-body problems
(planetary spacing, planetary rings and moons),
have to be studied by numerical simulations.
Numerical solutions of coupled differential equation
systems can now easily be obtained with numerical
minimization algorithms (in form of time profiles 
$X(t), Y(t)$, or phase diagrams $Y(X)$, see 
Fig.~(\ref{f_limitcycle}).

\section{	CONCLUSIONS 					}

In this multi-disciplinary review we provide, for the first time, 
a compilation of 17 astrophysical phenomena that have
been associated with self-organization mechanisms. 
The conclusions of this study are:

\begin{enumerate}

\item{Self-organization is a very multi-disciplinary
	subject that has been applied in
	planetary physics, solar physics, stellar physics,
	galactic physics, cosmology, ionospheric physics,
	magnetospheric physics, laboratory plasma physics,
	condensed matter physics, chemistry, biology, 
	social science, and computer science.}

\item{Self-organizing systems in astrophysics create 
	spontaneous {\sl order out of randomness}, during
	the evolution from an initially disordered system 
	to an ordered and more regular quasi-stationary system, via: 
	(i) quasi-periodic limit-cycle dynamics, and/or  
	(ii) resonances (i.e., harmonic mechanical resonances, 
	or harmonics of the gyrofrequency).}   

\item{Self-organizing processes are not controlled 
	from outside, but are driven by global forces (inside an open
	dissipative system), such as gravity, rotation,
	thermal pressure, or acceleration of nonthermal particles,
	in the case of astrophysical applications.}
 
\item{The limit-cycle behavior of astrophysical self-organization
	processes occurs due to a positive feedback mechanism that
	couples with the primary driver. This feedback mechanism
	is often an instability, such as the magneto-rotational
	instability, the Rayleigh-B\'enard convection instability, 
	turbulence, vortex attraction, magnetic reconnection, 
	plasma condensation, or a loss-cone instability.} 

\item{Physical models of an astrophysical self-organization process
	require a hydrodynamic approach (photospheric granulation,
	chromospheric evaporation, star formation, galaxy formation), 
	an MHD approach (protoplanetary disks, solar magnetic field, 
	Hale cycle), or N-body simulations (planetary spacing, 
	planetary rings and moons).}

\item{The entropy in self-organization processes is decreasing 
	during the evolution from an initially disordered system 
	to a self-organized limit-cycle behavior, in contrast to
	random processes where the entropy increases, or to
	self-organized criticality systems where the entropy
	remains invariant in the long-term time average.}

\item{The Lotka-Volterra equation system represents a useful tool
	to study the dynamical behavior of nonlinear dissipative
	systems, which are likely to evolve into a limit-cycle
	behavior for long-lived quasi-stationary phenomena.} 

\end{enumerate}

While the modeling of systems with self-organization was severely
hampered in the past, due to the mathematical difficulty of finding
analytical solutions for coupled integro-differential equation systems,
it is expected that the use of numerical computer simulations will
be capabable to produce realistic models in the future, for 
hydrodynamic, MHD, and N-body problems.

\clearpage

\section*{ APPENDIX A: Keplerian Orbits 		}

\begin{figure}
\centerline{\includegraphics[width=1.0\textwidth,angle=90]{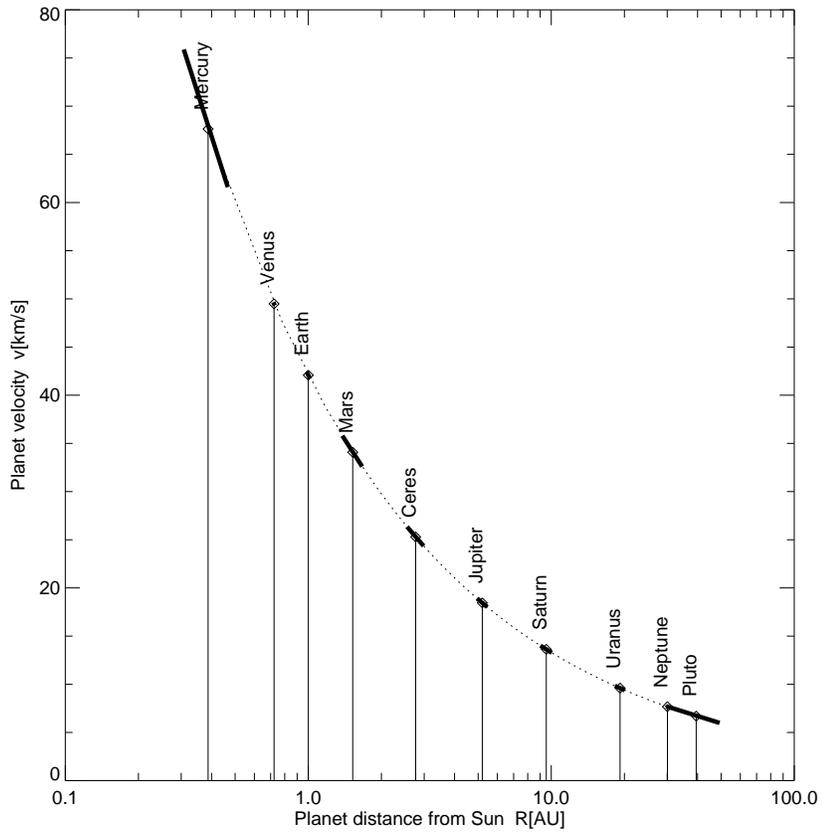}}
\caption{Phase-space diagram of planet distances from Sun, $r$, and
planet velocity, $v$, for Keplerian orbits. The Keplerian orbit is
marked with a dotted curve, the planet positions in phase space with
diamonds, and the planet motions in phase space with thick curve 
segments. Note that Mercury and Pluto, the two planets with the
largest orbit eccentricity, have the largest trajectories in
phase space. The mean planet locations (diamonds) in phase space
represent attractors of a nonlinear system with limit cycles.}
\label{f_kepler}
\end{figure}

\begin{figure}
\centerline{\includegraphics[width=0.8\textwidth,angle=270]{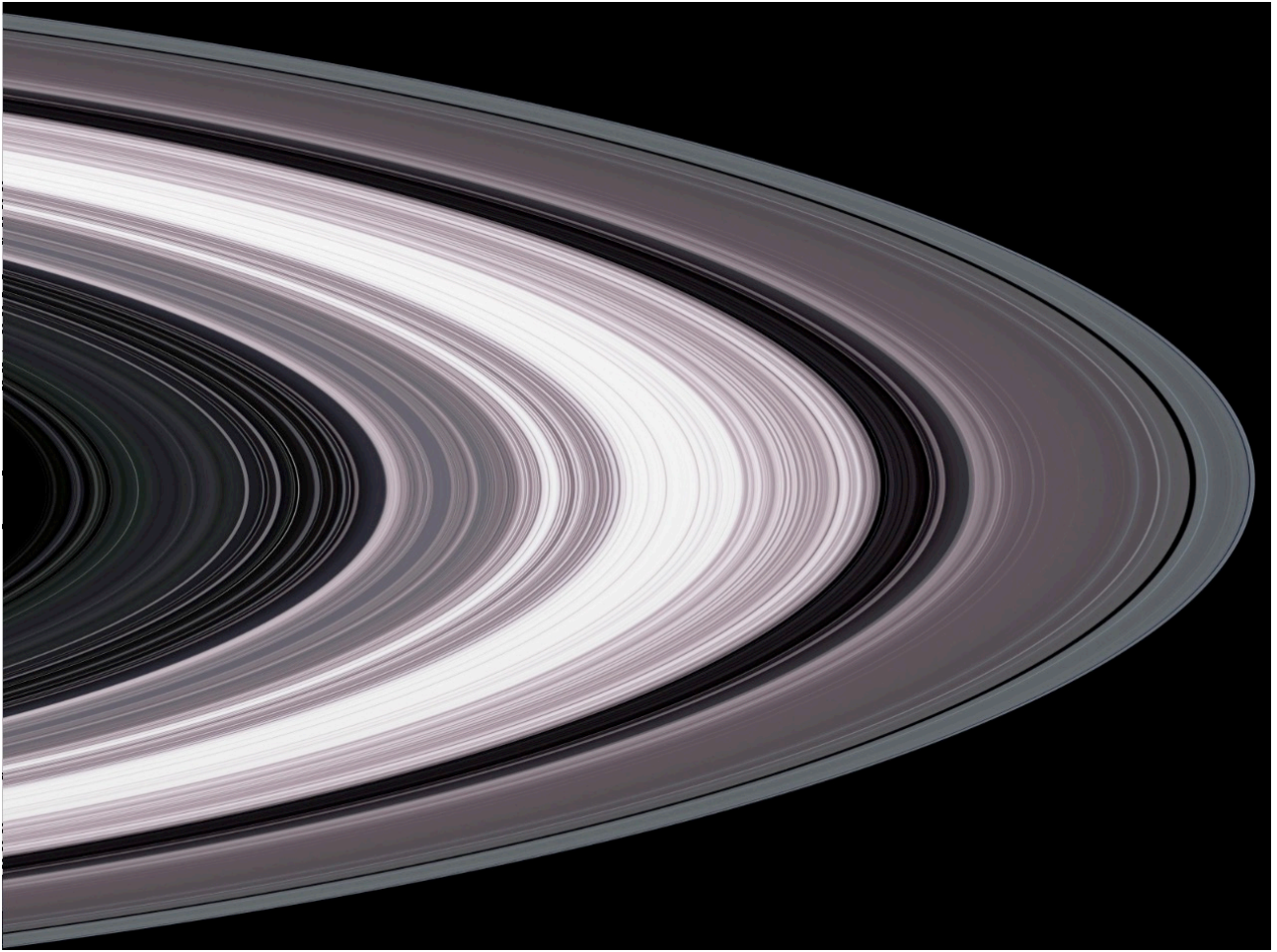}}
\caption{First radio occultation observation of Saturn's rings on 2005 May 3
with the Cassini spacecraft, using the radio bands of 0.94, 3.6, 13 cm.
The spatial resolution is $\approx 10$ km. The largest gap is the Cassini 
division, and the last outer (spatially resolved) gap is the Encke division 
[Credit: Cassini-Huygens mission, NASA].}
\label{f_saturn}
\end{figure}

\begin{figure}
\centerline{\includegraphics[width=0.8\textwidth,angle=90]{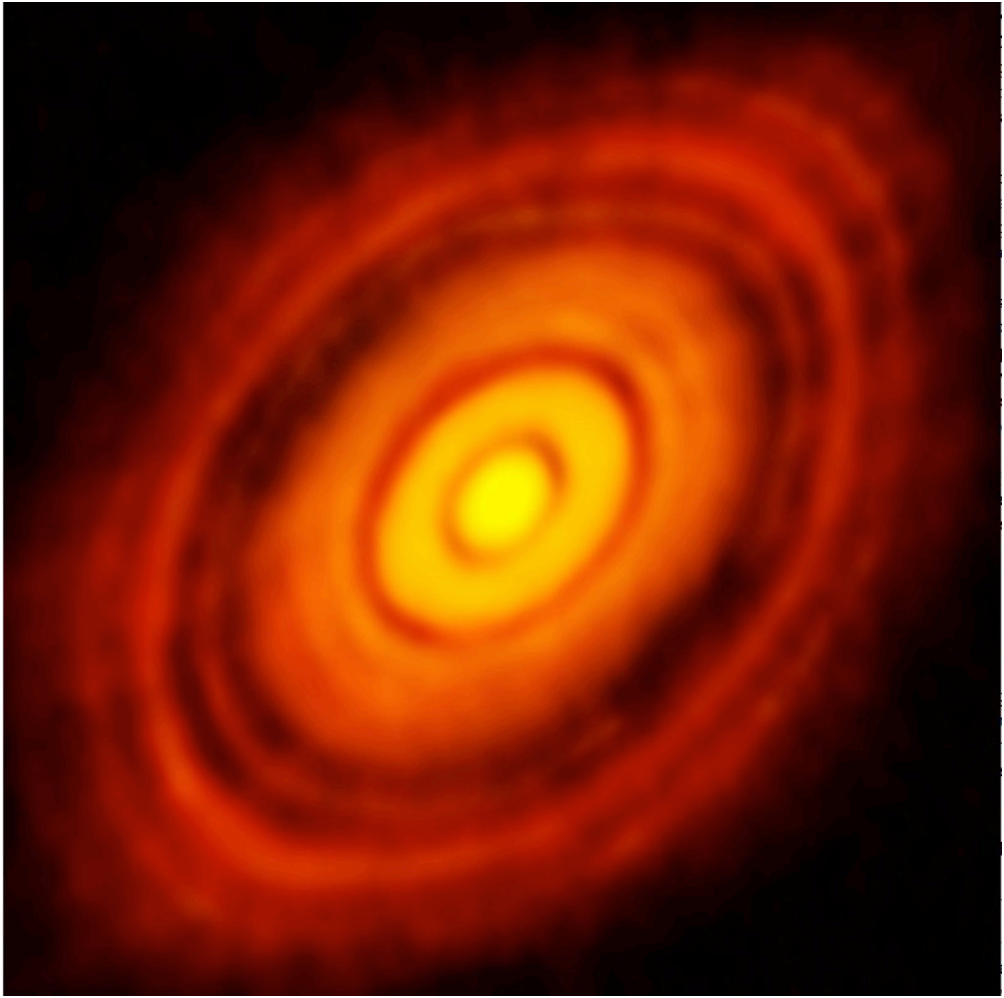}}
\caption{Thermal emission from milimeter-sized dust grains settled in the 
midplane of the HL Tauri disk, featuring a series of axisymmetric rings 
and gaps [ALMA Partnership 2015].}
\label{f_hltauri}
\end{figure}

\begin{figure}
\centerline{\includegraphics[width=1.0\textwidth,angle=270]{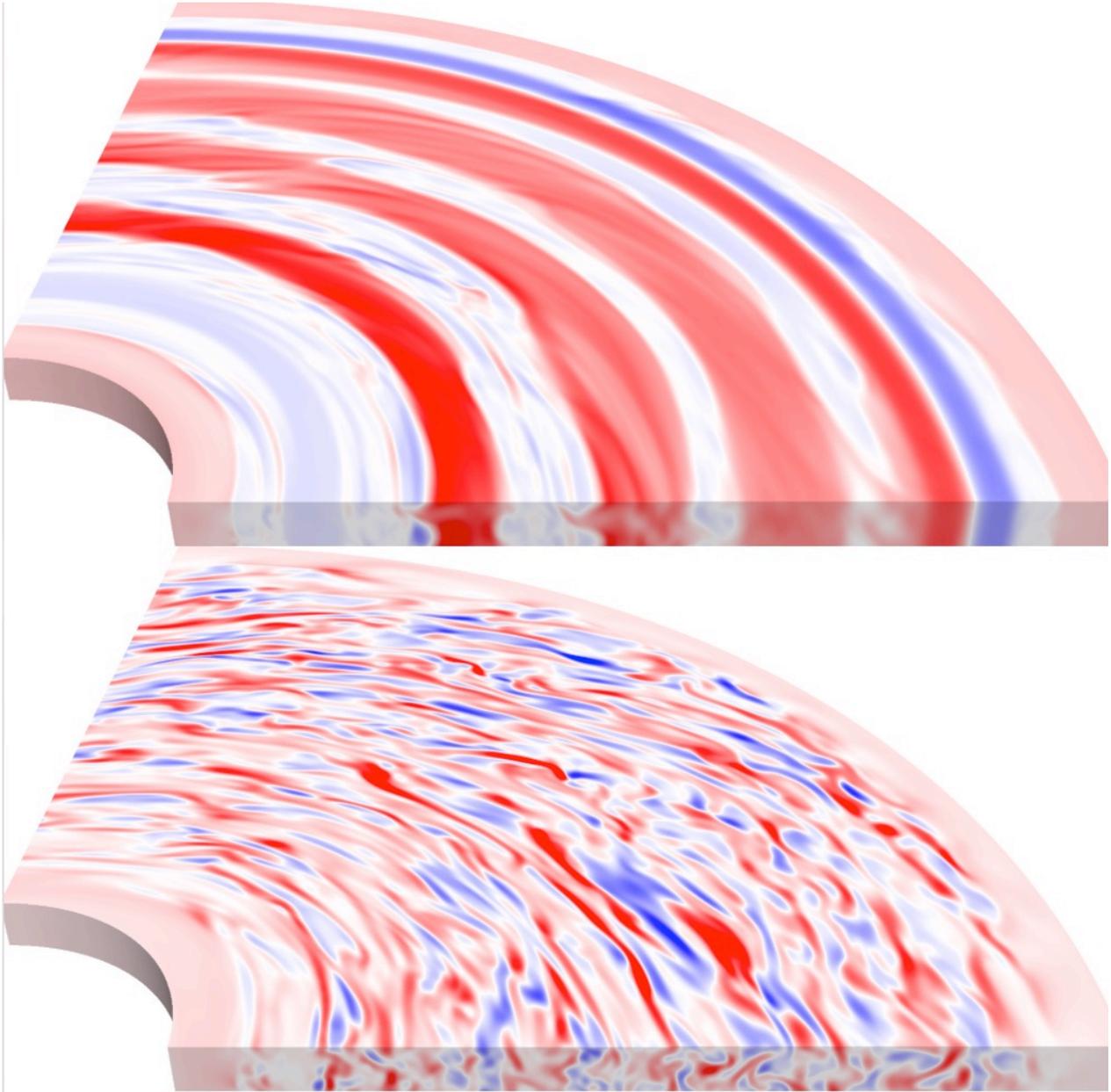}}
\caption{Axial magnetic field in two non-stratified simulations of 
protoplanetary disks. \emph{Bottom}: MRI-driven turbulence in ideal MHD; 
\emph{Top}: Ordered phase displaying axisymmetric rings of magnetic 
flux in Hall-MHD [B\'ethune et al.~2016].} 
\label{f_bethune}
\end{figure}

\begin{figure}
\centerline{\includegraphics[width=1.0\textwidth,angle=270]{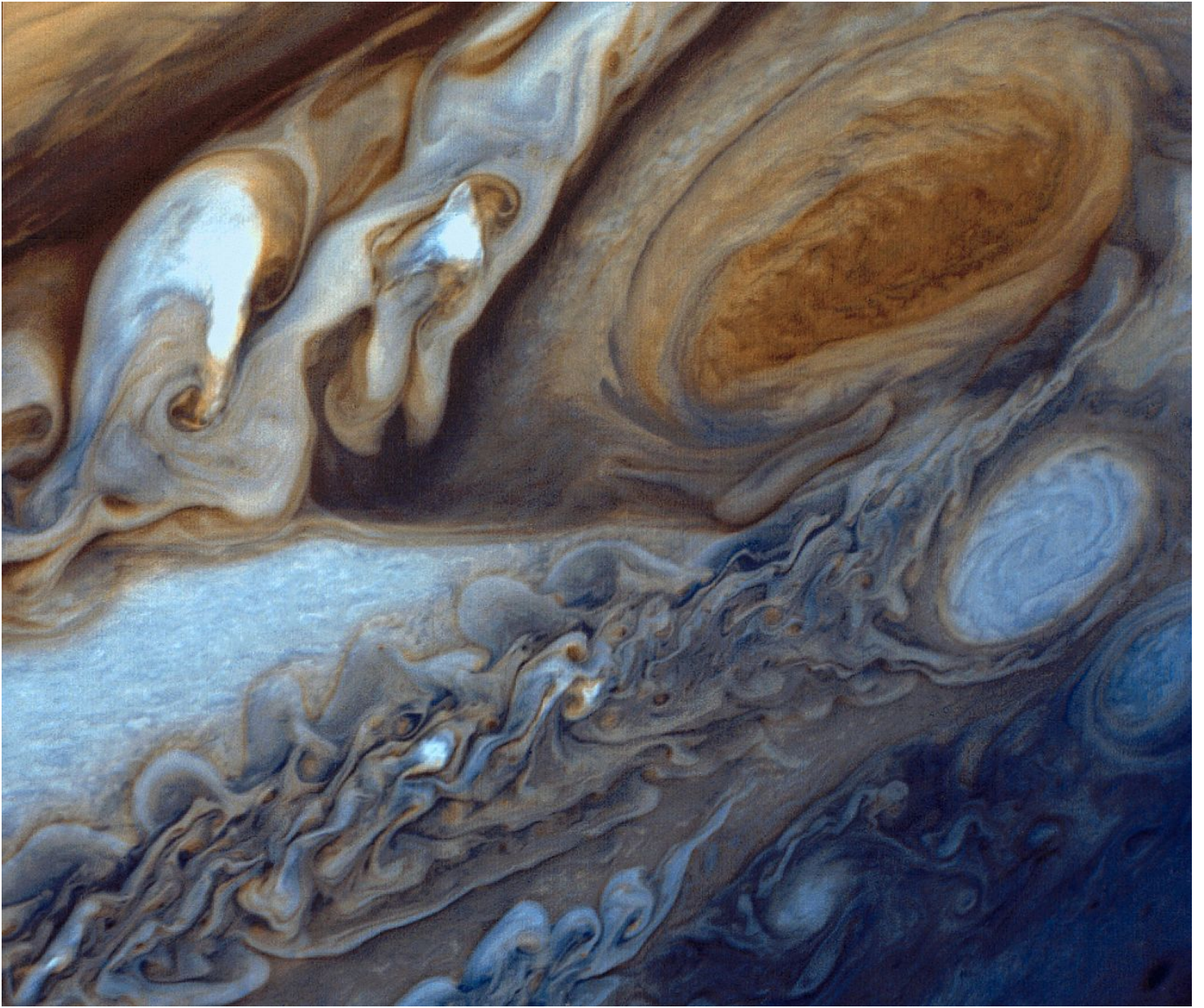}}
\caption{False-color image of the Great Red Spot of Jupiter,
observed with Voyager [credit: NASA, Caltech/JPL - 
http://www.jpl.nasa.gov/releases/2002/release$\_$2002$\_$166.html].}
\label{f_jupiter}
\end{figure}

\begin{figure}
\centerline{\includegraphics[width=0.55\textwidth,angle=270]{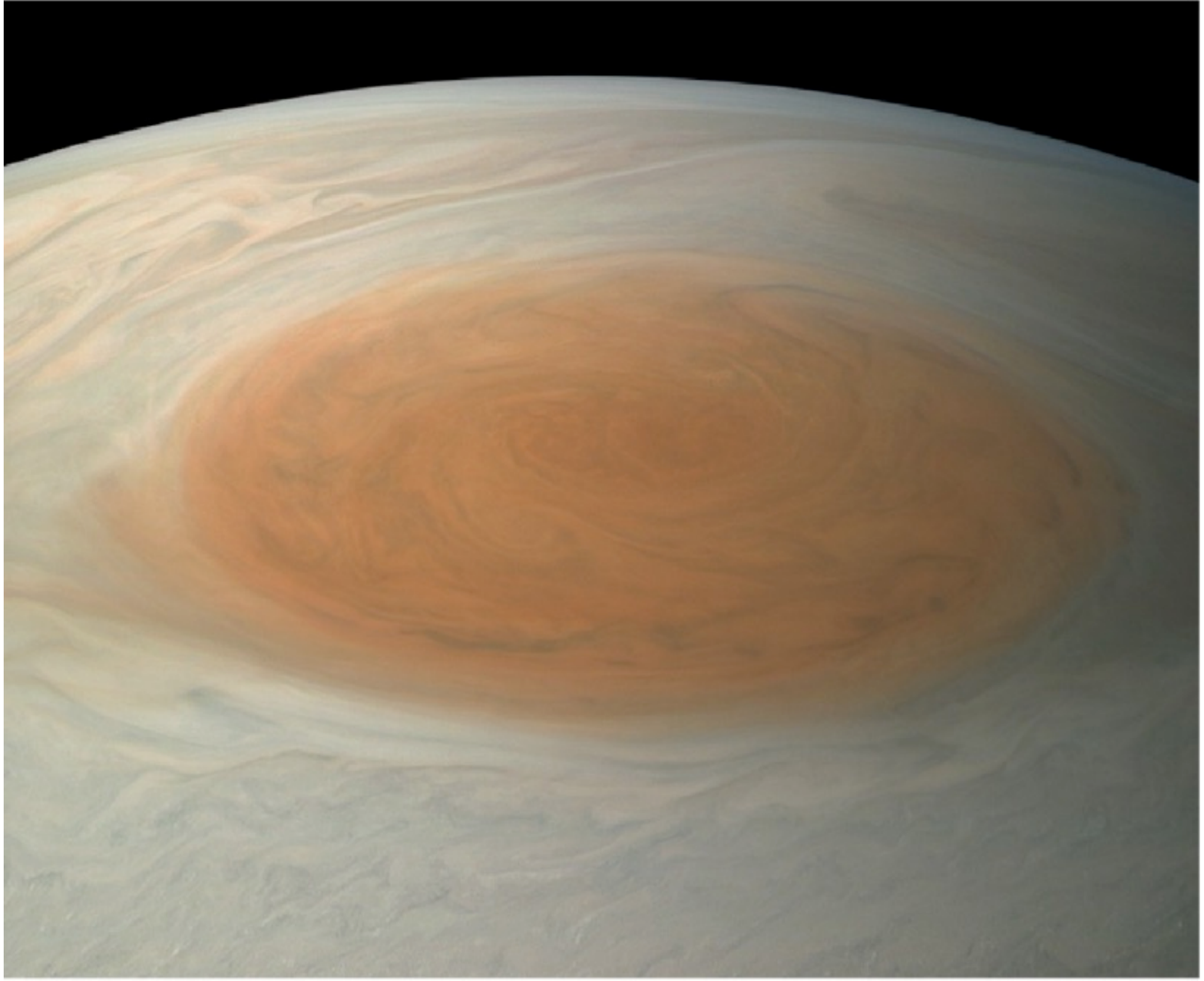}}
\centerline{\includegraphics[width=0.60\textwidth,angle=270]{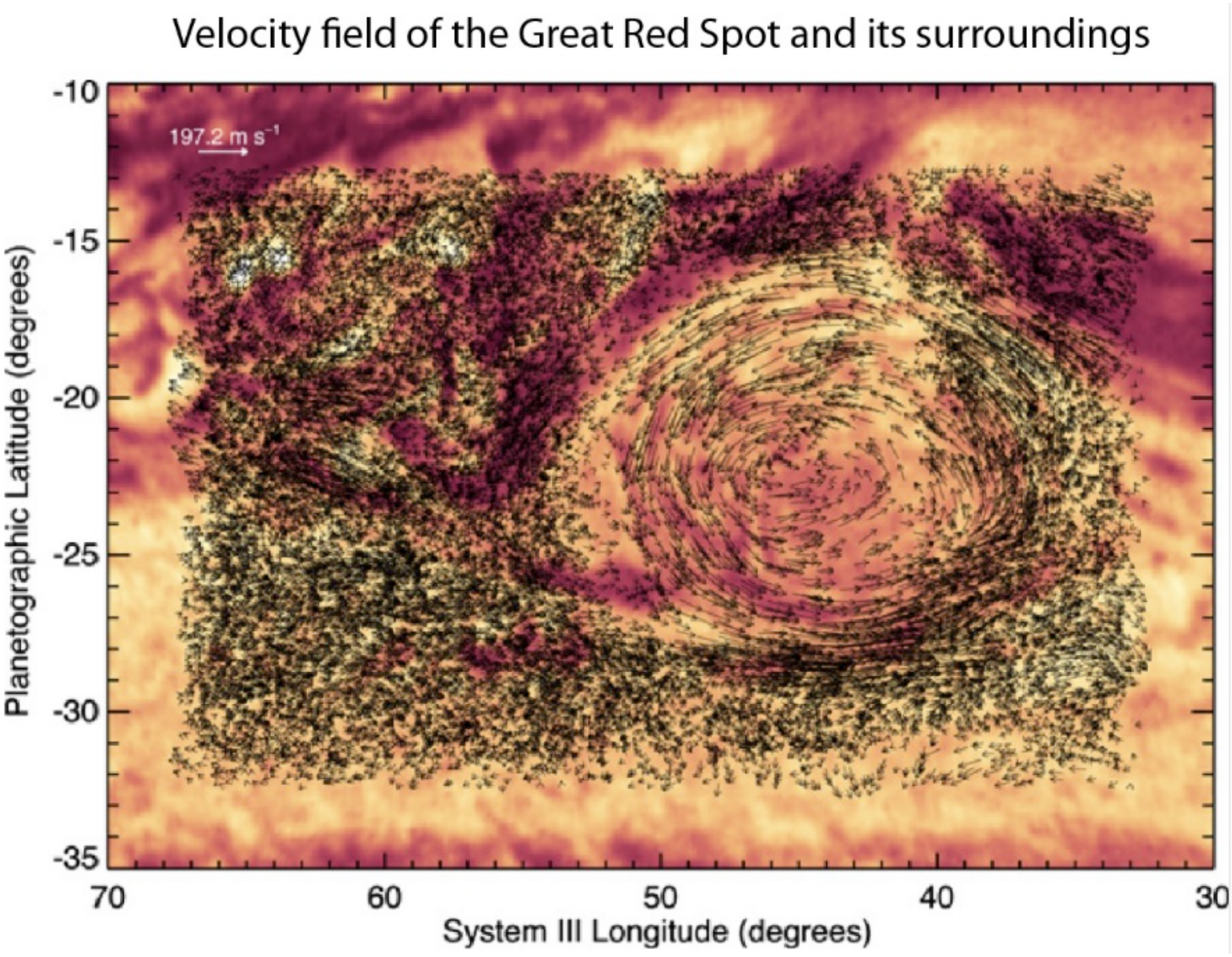}}
\caption{{\sl Top:} Great Red Spot rendered in true colors recently 
obtained by the Juno spacecraft in 2017  
[Image credit: NASA/JPL-Caltech/Space Science Institute, SwRI/MSSS,
Bj\"orn Jonsson]. 
{\sl Right:} Visualization of the velocity field [Simon et al.~2014].}
\label{f_jupiter_vortex}
\end{figure}

\begin{figure}
\centerline{\includegraphics[width=0.95\textwidth,angle=0]{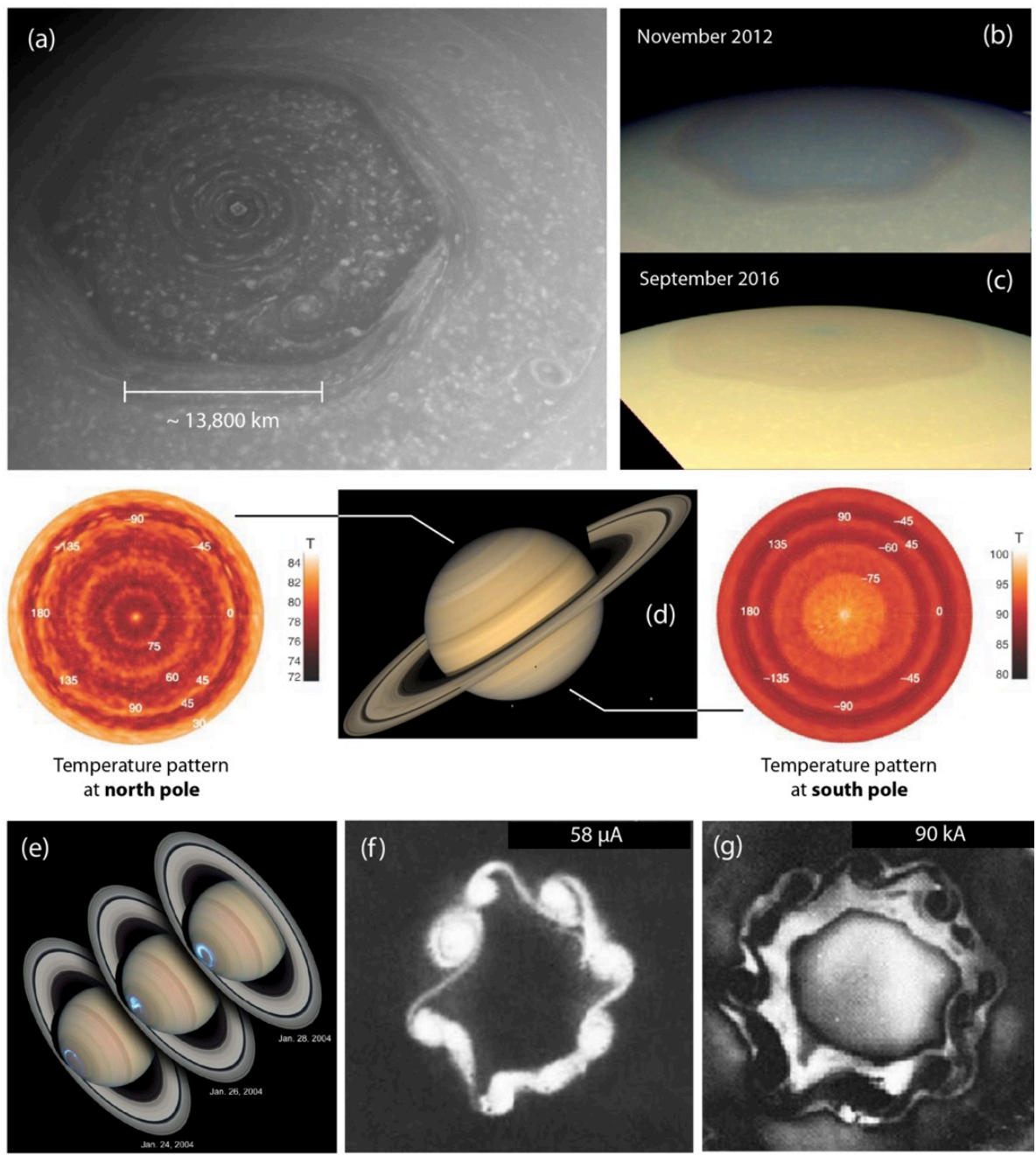}}
\caption{(a) Picture of Saturn's hexagon at the noth pole (view: 53$^\circ$ 
above the ringplane). Image taken by the Cassini spacecraft, 2013. 
[Image credit: NASA/JPL-Caltech/Space Science Institute]. 
(b) and (c): Images of the hexagon taken by Cassini in 2012 and 2016. 
[Image credit: NASA/JPL-Caltech/Space Science Institute.] 
(d) Temperature distribution at the north and south plane (in the 
troposphere at 100 mbar) according to Fletcher et al.~(2008). 
(e) Aurora on Saturn [Image credit: NASA]. 
(f) and (g) show discharge structures of electron beam 
interactions with a fluorescence screen at 58 $\mu A$ 
(f) or with a steel witness plate at 90 kA (g) (Parett et al.~2007).}
\label{f_saturn_hexagon}
\end{figure}

\begin{figure}
\centerline{\includegraphics[width=0.8\textwidth,angle=270]{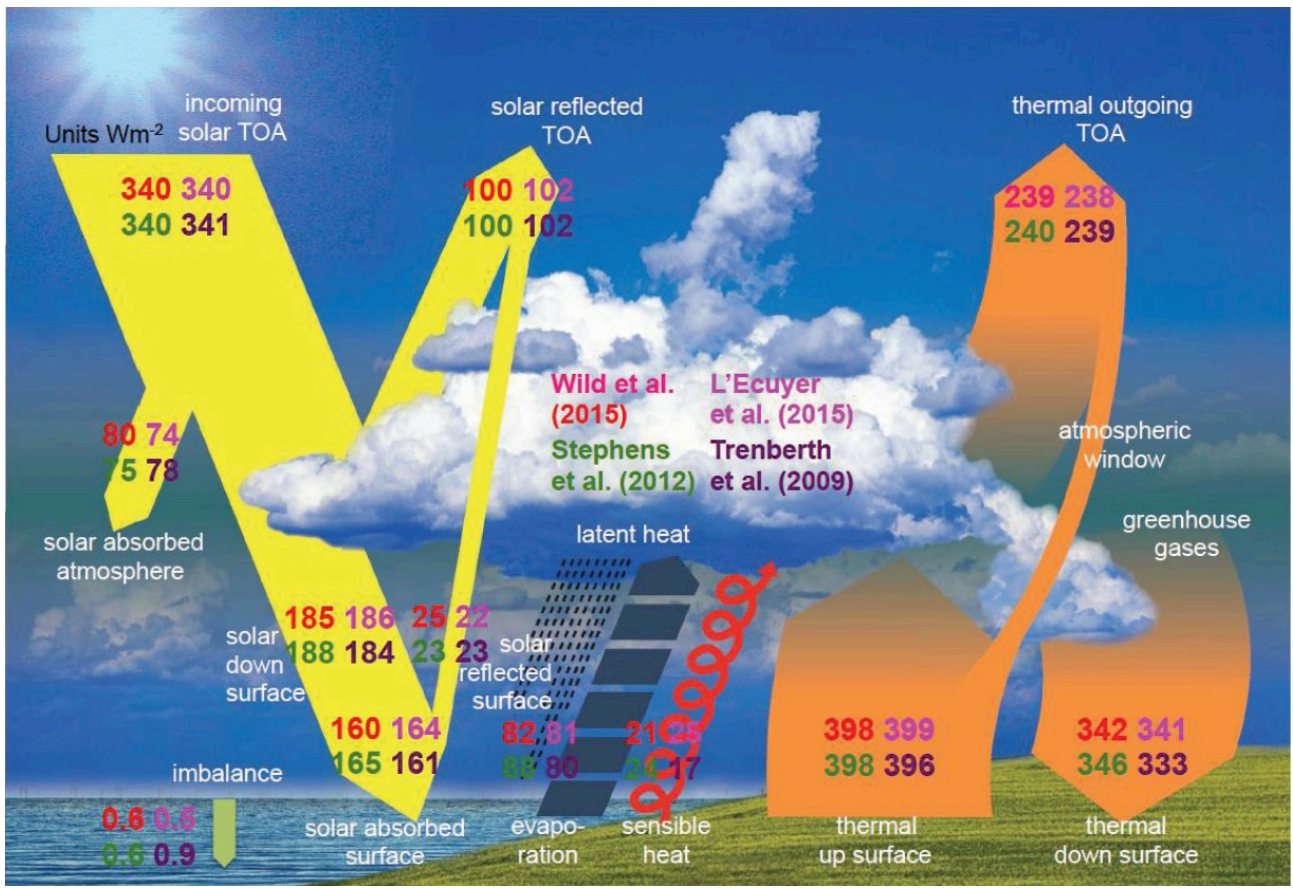}}
\caption{Comparison of different recent global annual mean energy balance
estimates for present day conditions as published by Wild et al.~(2015)
(upper left red values), L'Ecuyer et al.~(2012) (upper right pink values),
Stephens et al.~(2012) (lower left green values), and 
Trenberth et al.~(2009) (lower right purple values). Units in W m$^{-2}$
[Credit: Figure adapted from Wild 2017].}
\label{f_climate}
\end{figure}

\begin{figure}
\centerline{\includegraphics[width=0.7\textwidth,angle=0]{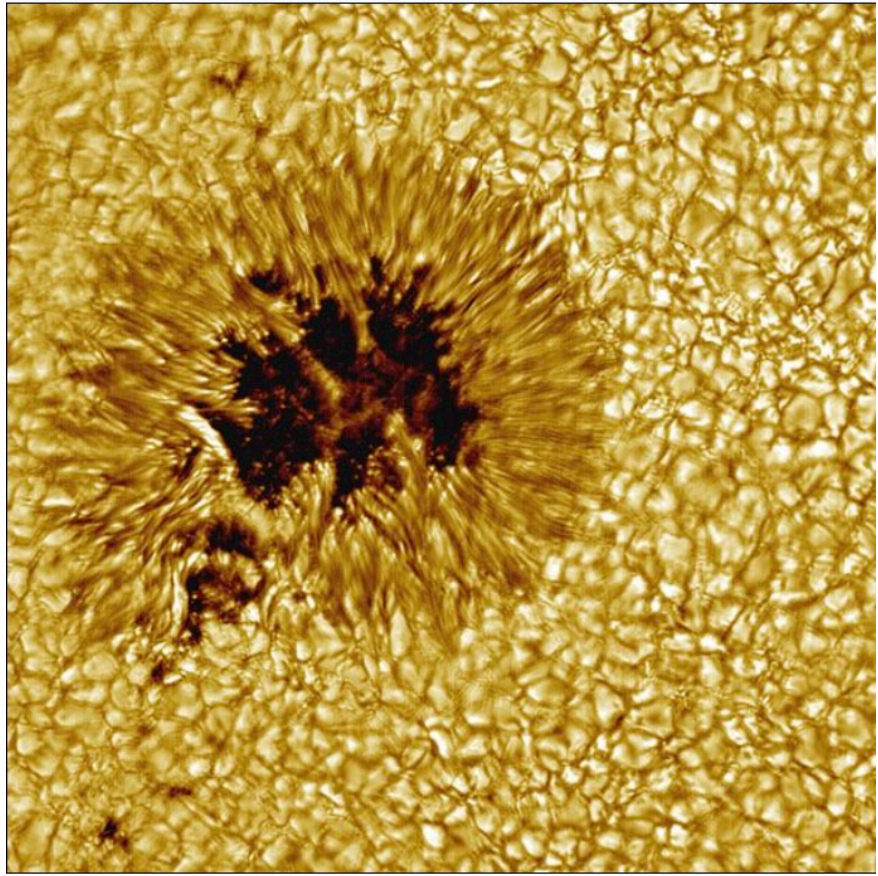}}
\caption{Closeup of photospheric granulation pattern and a sunspot 
near the center of the Sun. Note that the average size of granules
has a typical size of $w \approx 1500$ km [credit: NSO, NOAO, 
https://apod.nasa.gov/apod/ap051106.html].}
\label{f_granulation}
\end{figure}

\begin{figure}
\centerline{\includegraphics[width=0.8\textwidth,angle=0]{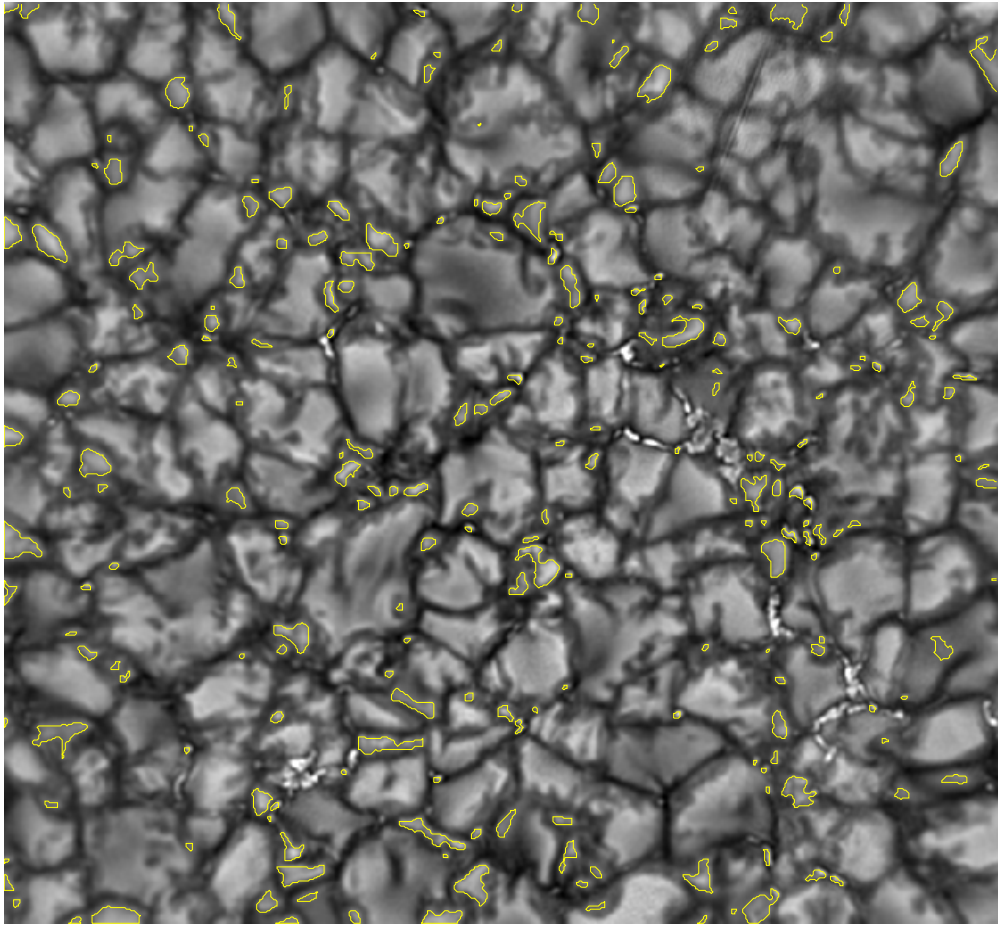}}
\caption{A TiO image of the solar surface is shown, containing 
normal granules and mini-granules in a Quiet Sun region, 
observed with the New Solar
Telescope (NST). Mini-granules are outlined with yellow contours,
which show granular-like features of sizes below 600 km located
in dark intergranular lanes. Note that the mini-granules do not
coincide with magnetic bright points [Abramenko et al.~2012].}
\label{f_minigranules}
\end{figure}

\begin{figure}
\centerline{\includegraphics[width=1.0\textwidth,angle=0]{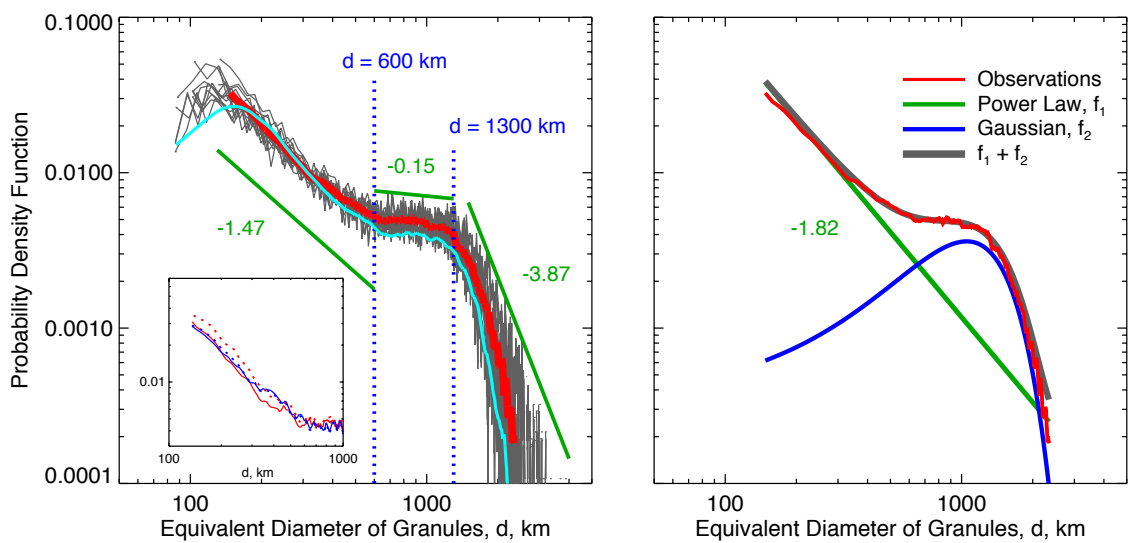}}
\caption{The probability density function of the equivalent diameter 
of granules (in units of km) is shown, observed in Quiet Sun regions
with the New Solar Telescope (NST). The regular granules have a 
size of $w \approx 500-2000$ km, while the range of $w \approx 100-500$ 
km exhibits the new phenomenon of ``mini-granules'' [Abramenko et al.~2012].}
\label{f_abramenko}
\end{figure}

\begin{figure}
\centerline{\includegraphics[width=1.0\textwidth,angle=270]{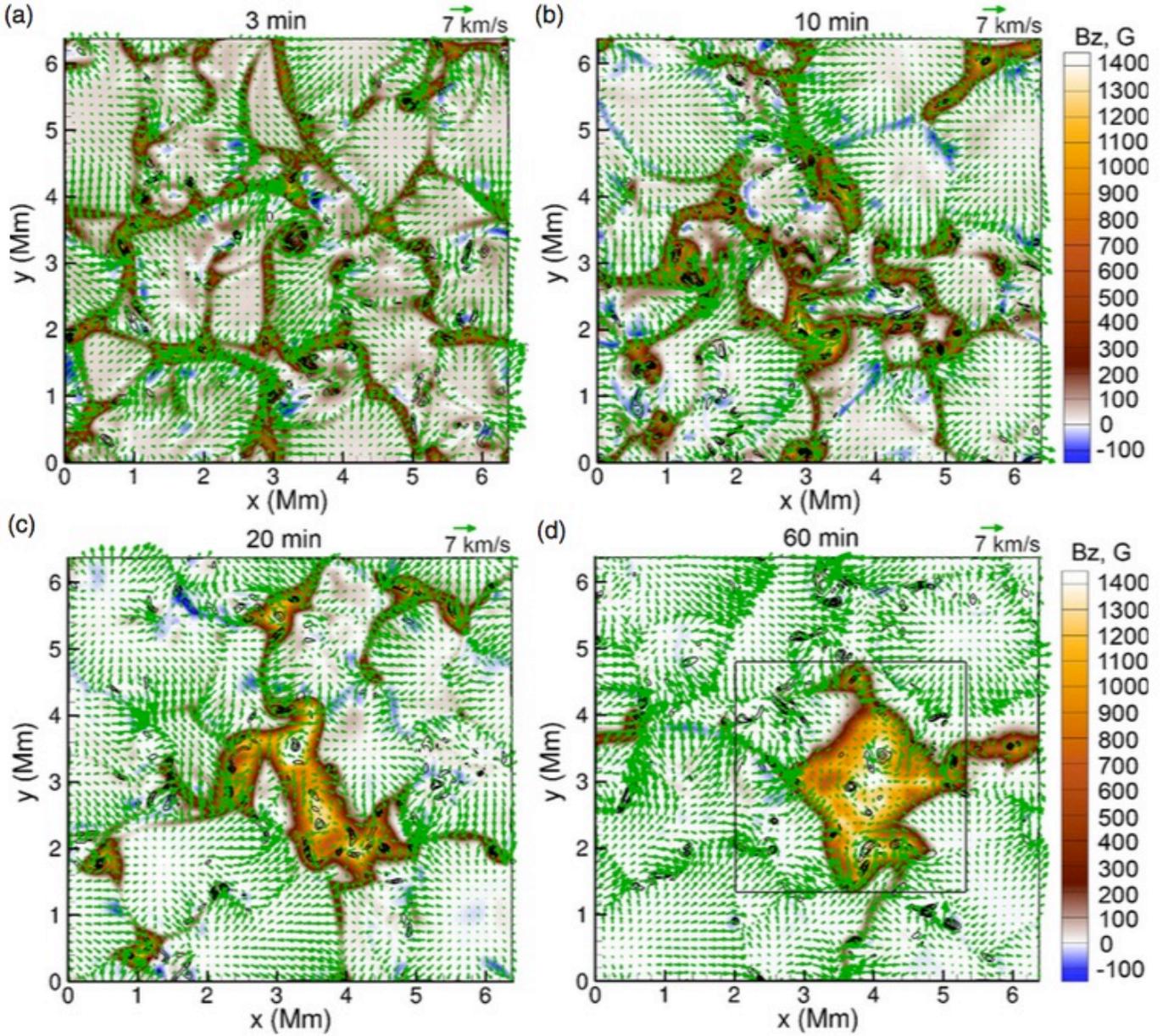}}
\caption{Four snapshots of the simulated formation of magnetic field
structures (pores and sunspots), showing the surface distribution of
the vertical magnetic field (color background), the horizontal flows
(arrows), and the vorticity magnitude (black contour lines), at 4 times
(3, 10, 20, 60 min) from the moment of initiation of a uniform
magnetic field ($B_z=100$ G) [Kitiashvili et al.~2010].}
\label{f_kitiashvili}
\end{figure}

\begin{figure}
\centerline{\includegraphics[width=1.0\textwidth,angle=0]{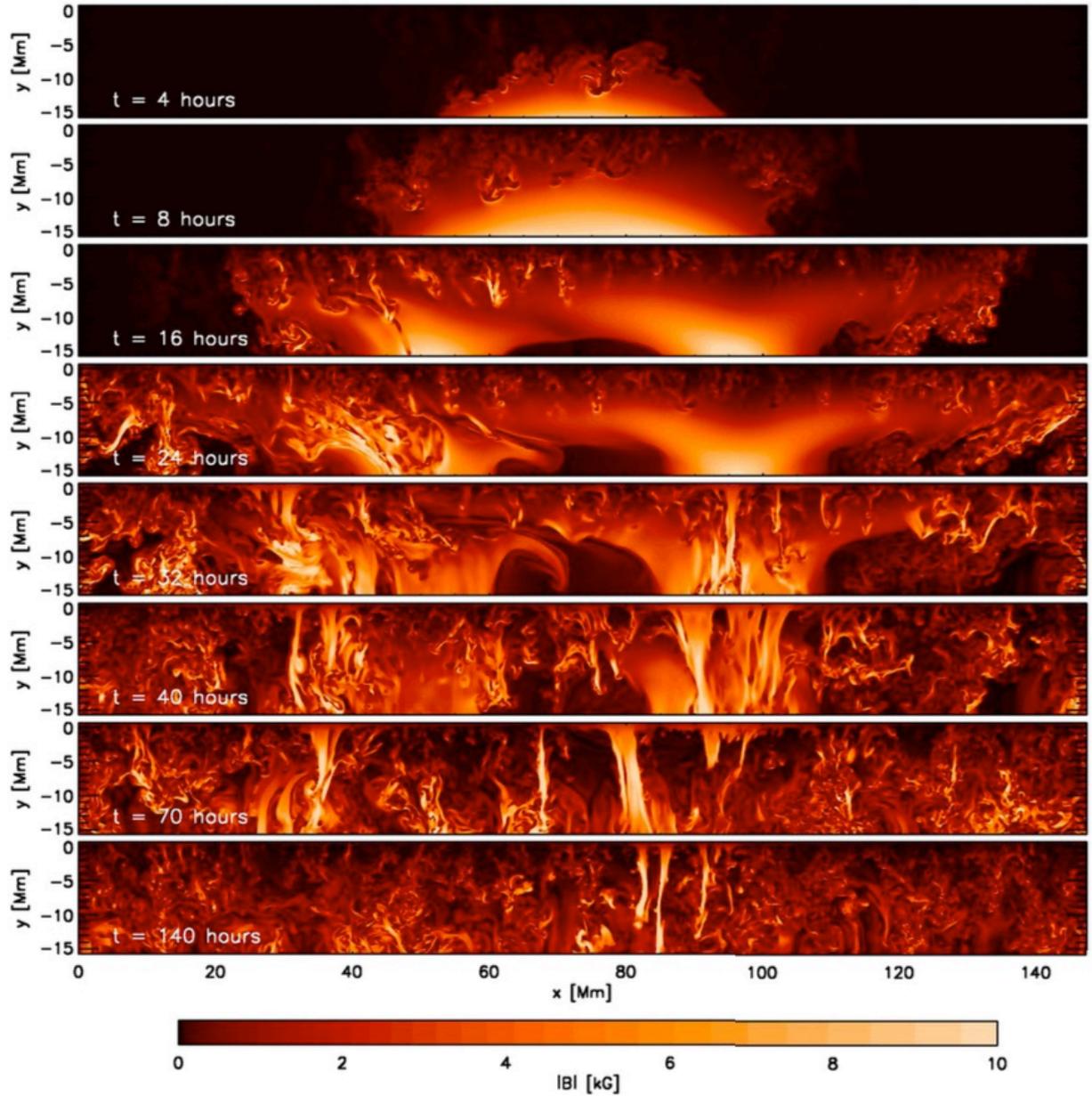}}
\caption{Time evolution of the magnetic field strength $|B|$ 
during magnetic flux emergence on a vertical
cut through the center of the domain along the x-axis. The first two 
snapshots show the subsurface field evolution prior to the appearance
of flux in the photosphere, the remaining six snapshots correspond to the
photospheric magnetograms [Rempel and Cheung 2014].}
\label{f_rempel_cheung}
\end{figure}

\begin{figure}
\centerline{\includegraphics[width=1.0\textwidth,angle=0]{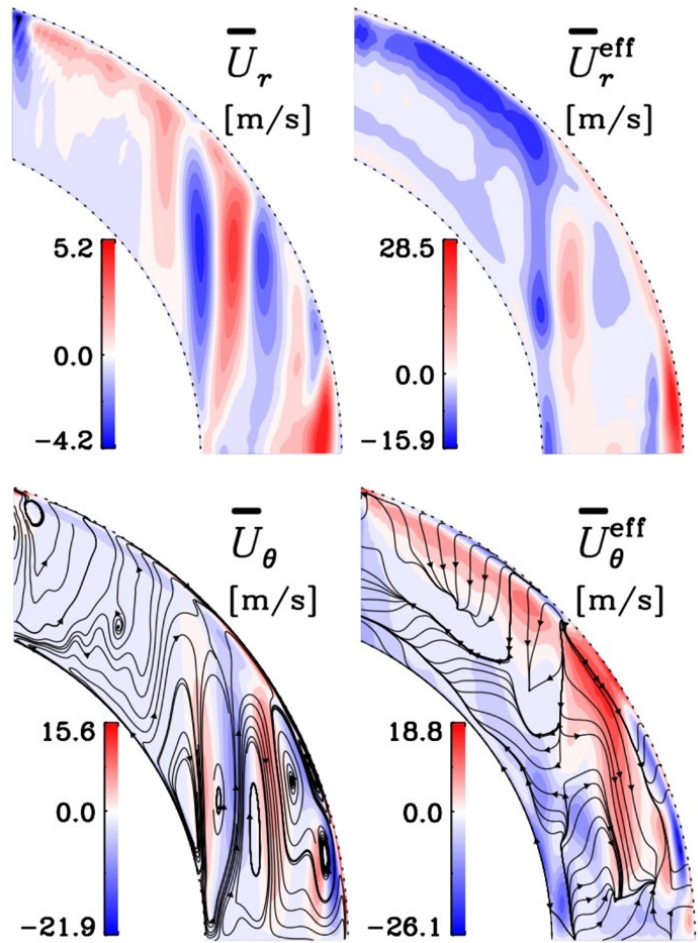}}
\caption{The solar dynamo action obtained form a 3-D MHD simulation
is depicted in the convection zone, showing the time-averaged radial
and effective radial flow (top panels), and the latitudinal
and effective latitutdinal flow [Warnecke et al.~2017].}
\label{f_dynamo}
\end{figure}

\begin{figure}
\centerline{\includegraphics[width=1.0\textwidth,angle=270]{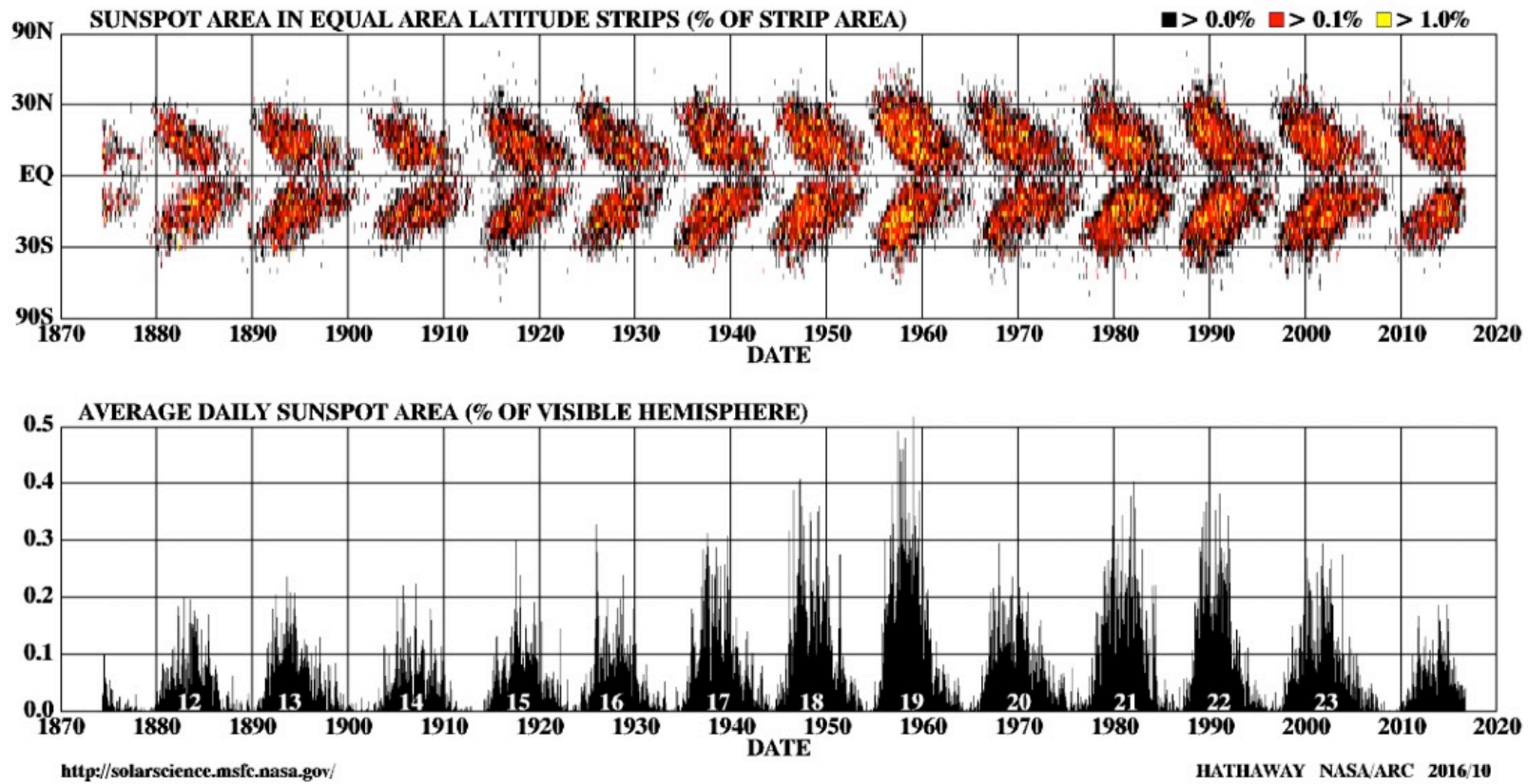}}
\caption{The variation of the sunspot number from 1870 to 2020, showing the
11-year periodicity in the average daily sunspot area (bottom panel)
and in the latitude distribution (butterfly diagram in top panel).
[Credit: http://solarscience.msfc.nasa.gov/, David Hathaway, NASA/ARC].}
\label{f_butterfly}
\end{figure}

\begin{figure}
\centerline{\includegraphics[width=1.0\textwidth,angle=270]{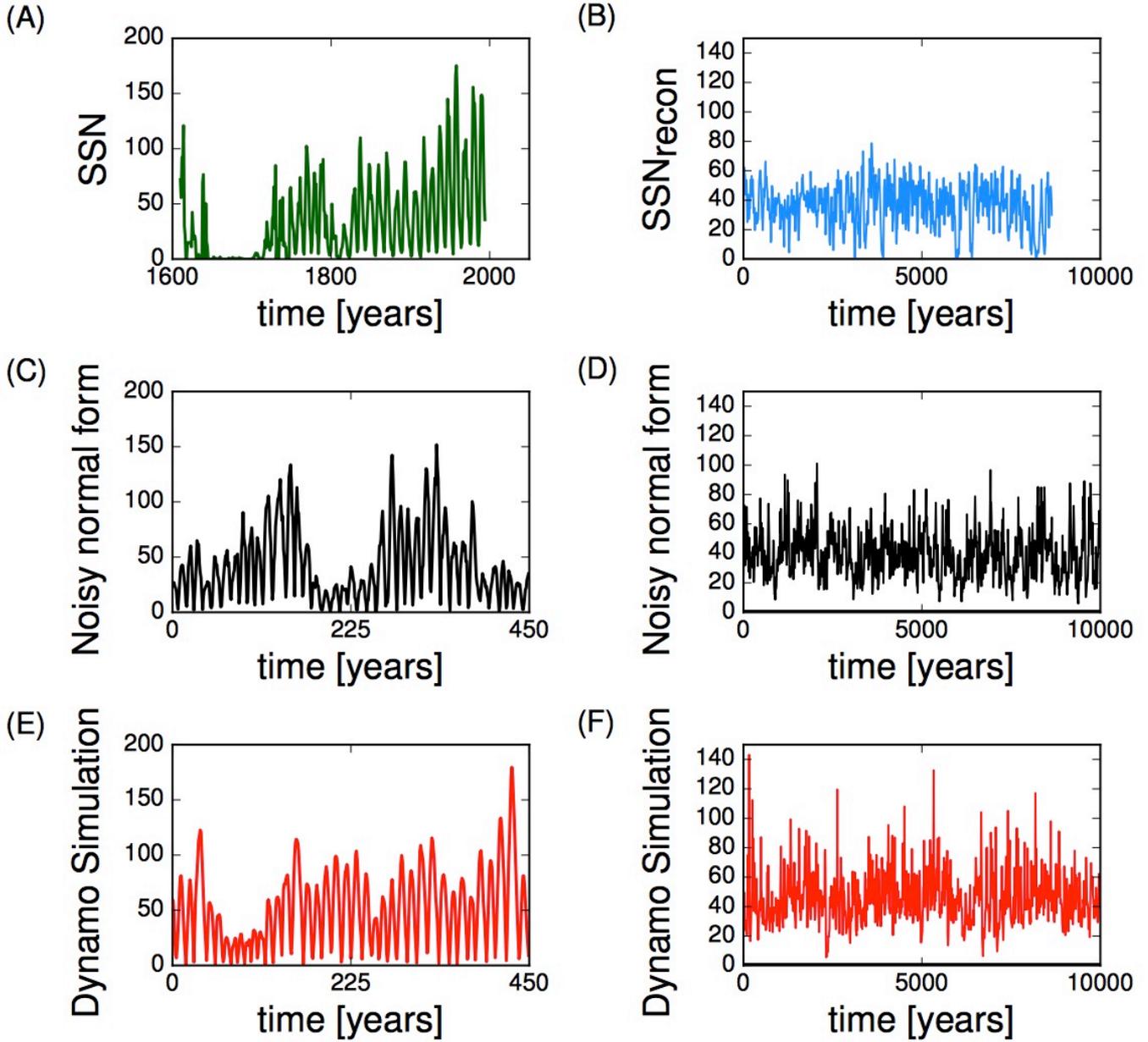}}
\caption{(A) Time series of the observed sunspot numbers (SSN);
(B) Sunspot number reconstructed from cosmogenic isotopes (SSN$_{recon}$);
(C,D) Monte-Carlo simulations of a weakly nonlinear, noisy limit cycle
(Hopf bifurcation normal-form model); (E,F) Results from Babcock-Leighton
dynamo model with fluctuating sources [Cameron and Sch{\"u}ssler 2017].} 
\label{f_cameron}
\end{figure}

In principle, the Keplerian orbits of planets can be understood
as a limit cycle of a self-organizing system. The dynamics of
two planets can be written in a Hamiltonian form (Nesvorny and
Vokrouhlick 2016),
$$
	H = H_K + H_{per} = \sum_{j=1}^2
	\left( {p_j^2 \over 2 \mu_j} - G {\mu_j M_j \over r_j} \right) 
	+ H_{per} \ ,
	\eqno(A1)
$$
where $H$ is the total Hamiltonian, $H_K$ is the Keplerian part,
$H_{per}$ is the perturbation part, $M_*$ is the solar mass,
$M_j = m_j + M_*$ and $\mu_j=m_j M_*/M_j$ the reduced masses. 
If we neglect the perturbations, the Hamiltonian form is equivalent
to the conservation of kinetic and gravitational potential energy,
$E_{kin}+E_{grav}=(1/2) m v^2 - G M_* m/r = 0$, from which the
following relationship between the distance $r$ and velocity $v$
results,
$$
	v(r) = \sqrt{ {2 G M \over r} }
	\eqno(A2)
$$
We show this relationship between the planet distance $r$ and
the planet velocity $v(r)$ in the phase space $[r,v]$ 
in Fig.~(\ref{f_kepler} dotted curve). For each of the 10 planets
(including Ceres and Pluto), there is a fixed point
$[r_i, v_i], i=1,...,10$ that represents an attractor for the
planet motion in phase space. 
For planets that have a circular
orbit with no eccentricity, the Keplerian planet motion is 
confined to a constant single fixed point in phase space
($r_i=const$, $v_i=const$). However, all planet orbits move
on ellipses with some eccentricity $e$, in the cartesian space 
$[x,y]$. The minimum $r_{min}$ and maximum distance $r_{max}$
are given by the eccentricity $e=c/a$,
$$
	\begin{array}{ll}
	r_{min} =& a ( 1 - e ) \\
	r_{max} =& a ( 1 + e ) 
	\end{array}
	\ ,
	\eqno(A3)
$$
where $a$ is the major axis of the ellipse, $b$ is the minor axis,
$c=\sqrt{a^2 - b^2}$ is the distance of one ellipse focal point (where
the Sun is) from the center of the ellipse, and $e=c/a$ is the
eccentricity. Defining the mean distance with $a$ and the orbital
period with $T$, we retrieve Kepler's third law, $a^3 \propto T^2$,
using $v = 2 \pi R / T$ and Eq.~(A2).

We show the planet motion in phase space $[r_i, v_i]$
in Fig.~(\ref{f_kepler} thick curve segments), which cover the range of
$r_{min} \le r \le r_{max}$ for each planet. We see that only
Mercury and Pluto cover an appreciable distance in phase space,
because they have the largest eccentricities of $e_{mercury}=0.2056$
and $e_{pluto}=0.2488$. Therefore, every planet performs an
oscillatory motion around their mean distance (from the Sun) with
an orbital period $T$ that can be considered as a limit cycle.
Gravitational perturbations will alter the Keplerian orbits slightly,
which we neglected here. The orbital period $T$ marks the temporal 
scale that is self-organized in this system. In addition, the system
self-organizes low harmonic ratios between adjacent planet pairs,
which is necessary to warrant long term stability of the planet system.
These quantized harmonic ratios mark the spatial scales. Therefore,
a planet system self-organizes both temperal and spatial
scales.

\section*{ APPENDIX B: Coupled Differential Equations	 }

A basic analytical system that exhibits oscillatory limit-cycle behavior
can be described by the following non-homogeneous linear 
differential equation system (Aschwanden and Benz 1988),  
$$
 	\begin{array}{ll}
  		{dX / dt} &=   b_1 + a_{11} X + a_{12} Y \\
  		{dY / dt} &=   b_2 + a_{21} X + a_{22} Y 
 	\end{array}
 	\ .
 	\eqno(B1) 
$$
This equation system contains linear terms only (in $X(t)$ and $Y(t)$), 
which is appropriate for the dynamic system behavior under the influence 
of small perturbations. The general (complex) solution depends on the 
value of the discriminant $D$ and trace $S$ of the non-homogeneous equation,
$$
 	\begin{array}{ll}
  		S  & =  {1 \over 2} (a_{11} + a_{22}) \\
  		D  & =  (a_{11} a_{22} - a_{12} a_{21})   
 	\end{array}
 	\ .
 	\eqno(B2) 
$$
The general complex solution of the non-homogeneous differential
equation system for $(S^2 - D) \neq 0$ is given by,
$$
 	\begin{array}{ll}
  		{X(t)} &=   X_1 \exp^{z_i t} + X_0  \\
  		{Y(t)} &=   Y_1 \exp^{z_i t} + Y_0  
 	\end{array}
 	\ .
 	\eqno(B3) 
$$
where the exponential coefficient has two solutions,
$$
 	\begin{array}{ll}
  		{z_1} & =   S + [(S^2 - D)]^{1/2} \\
  		{z_2} & =   S - [(S^2 - D)]^{1/2} 
 	\end{array}
 	\ .
 	\eqno(B4) 
$$
For the linearly dependent case of $(S^2-D)=0$ the general
solution is given by
$$
 	\begin{array}{ll}
  		{X(t)} &=  (X_2 t + X_1) \exp^{S t} + X_0  \\
  		{Y(t)} &=  (Y_2 t + Y_1) \exp^{S t} + Y_0  
 	\end{array}
 	\ .
 	\eqno(B5) 
$$
The non-homogeneous coefficients are,
$$
 	\begin{array}{ll}
  		X_0  & =  {(b_2 a_{12} - b_1 a_{22}) / D} \\
  		Y_0  & =  {(b_1 a_{21} - b_2 a_{11}) / D} 
 	\end{array}
 	\ .
 	\eqno(B6) 
$$
Using $X_1$ and $X_2$ as free parameters, the coefficients
$Y_1$ and $Y_2$ are,
$$
 	\begin{array}{ll}
  	Y_1  & = {1 \over a_{12}} \left( {a_{22} - a_{11} \over 2} \right) X_1 \\
  	Y_2  & = {1 \over a_{12}} \left( {a_{22} - a_{11} \over 2} \right) X_2
		+{1 \over a_{12}} X_1  
 	\end{array}
 	\ .
 	\eqno(B7) 
$$
Since the exponential coefficient $z_i$ is a complex number, one an
split it into a real part $\rho_i$ and an imaginary part $\omega_i$,
$$
 	\begin{array}{ll}
	z_i      = & \rho_i + i \ \omega_i		\\
	\rho_i   = & Re [ S \pm (S^2 - D)^{1/2} ]	\\
	\omega_i = & Im [ S \pm (S^2 - D)^{1/2} ]  
 	\end{array}
 	\ .
	\eqno(B8)
$$
Physically, $\omega_i$ describes the frequency of the oscillations,
and $\rho_i$ denotes the growth (or negative damping) rate of the
perturbation. Using these variables, the time-dependent solution
is the classical solution of two coupled oscillators:  
$$
 	\begin{array}{ll}
  		{X(t)} &=  X_1 \exp^{\rho_i t} \cos (\omega_i t) + X_0 \\
  		{Y(t)} &=  Y_1 \exp^{\rho_i t} \cos (\omega_i t - \delta ) 
		\left[ {(\rho_i - a_{11})^2 + \omega_i^2 \over a_{12}^{1/2}}
		\right]^{1/2} + Y_0 \\
 	\end{array}
 	\ .
 	\eqno(B9) 
$$
The phase difference $\delta$ is
$$
	\tan{(\delta)} = {-\omega_i \over \rho_i - a_{11}} \ .
	\eqno(B10)
$$
The special case of an undamped oscillation ($\rho_i$=0) requires
$S=0$ and $D > 0$. The temporal functions $X(t)$ and $Y(t)$ of the
two coupled oscillators is then 
$$
 	\begin{array}{ll}
  		{X(t)} &=  X_1 \cos (\omega_i t) + X_0 \\
  		{Y(t)} &=  X_1 \cos (\omega_i t - \delta ) 
		\left( { - a_{21} \over a_{12}} \right)^{1/2} + Y_0 \\
		\omega_i &= \pm \sqrt{D} 
		= \pm (a_{11} a_{22} - a_{12} a_{21})^{1/2} \\	
		\tan{(\delta)} &= \omega_i / a_{11}
 	\end{array}
 	\ .
 	\eqno(B11) 
$$
which corresponds to a limit cycle with period $\omega_i$, 
fixpoint $(X_0, Y_0)$, and phase delay $\delta$.

\section*{ APPENDIX C: The Hopf Bifurcation }

Another nonlinear system that predicts limit-cycle behavior
is the so-called {\sl Hopf bifurcation} (Hopf 1942), which
is described in many textbooks (e.g., Schuster 1988).
A simple Hopf bifurcation generates a limit cycle starting
from a fixed point. A differential equation of the Hopf
bifurcation can be written in polar coordinates $[r, \theta]$,
$$
 	\begin{array}{ll}
  		{dr     / dt} &= - (\rho r + r^3) \\
  		{d\theta / dt} &= \omega 
 	\end{array}
 	\ .
	\eqno(C1)
$$
which has the following analytical solution,
$$
 	\begin{array}{ll}
	r^2(t) =& {\rho r_0^2 \exp{(-2 \rho t)}
	\over r_0^2 [ 1 - \exp{(-2 \rho t)}] + \rho }  \\
	\theta(t) =& \omega t
 	\end{array}
 	\ .
	\eqno(C2)
$$
for the initial conditions $r_0 = r (t=0)$ and $\theta(t=0)=0$.
For $\rho \ge 0$, the trajectory approaches the origin at
the fixed point $r_{\infty}=0$ and becomes stationary.

For negative values, $\rho < 0$, it converges to a limit cycle
$r_{\infty} = r(t=\infty) = \sqrt{|\rho|} > 0$.

The differential equation system in polar coordinates $[r, \theta]$
can be transformed into Cartesian coordinates $[X,Y]$ by using  
$$
 	\begin{array}{ll}
	X =& r \cos{(\omega t)} \\
	Y =& r \sin{(\omega t)} 
	\end{array}
 	\ ,
	\eqno(C3)
$$
which yields
$$
 	\begin{array}{ll}
	{dX / dt} =& -[\rho + (X^2 + Y^2)] X - Y \omega \\
	{dY / dt} =& -[\rho + (X^2 + Y^2)] Y + X \omega 
	\end{array}
 	\ .
	\eqno(C4)
$$
Linearizing with respect to the origin yields,
$$	
	{df \over dt} = A f  \ ,
	\eqno(C5)
$$
with $f=[\Delta X, \Delta Y]$ and the $A$ the matrix
$$
	A = \left(
	\begin{array}{ll}
	-\rho & -\omega \\
	\omega  & -\rho 
	\end{array}
	\right)
	\ ,
	\eqno(C6)
$$
which has the eignevalues,
$$
	\lambda = -\rho \pm i \ \omega 
	\ .
	\eqno(C7)
$$

\section*{ APPENDIX D: The Lotka-Volterra Equation System	}

The Lotka-Volterra equation system (Lotka 1925; Volterra 1931)
is a paradigm of a nonlinear dissipation process, with cyclic 
(oscillatory) behavior in some parameter space. 
In the simplest terms it can be written as
a coupled first-order, nonlinear, differential equation system
containing the time-dependent variables $X(t)$ and $Y(t)$, and
coefficients $k_1, k_2, k_3$,  
$$
 \begin{array}{ll}
  {dX / dt} &=   k_1 X   - k_2 X Y \\
  {dY / dt} &=   k_2 X Y - k_3 Y   
 \end{array}
 \eqno(D1) 
 \ ,
$$
with $k_1, k_2, k_3$ being positive coefficients.
The rate of change in the first variable $X(t)$ is
specified by a growth rate $k_1$, while the (negative) dissipation rate $k_2$ 
is coupled to the product of both variables, $X(t) Y(t)$.
The rate of change in the second variable $Y(t)$ is
specified by a decay rate $k_3$, while the (positive) dissipation rate $k_2$ 
has the opposite sign. In ecology (e.g., May 1974), the two variables
were designated to some predator and pray populations that compete
for life, such as foxes and rabbits. 

This equation system has a periodic solution, which is
called the limit cycle, also called a critical point or
attractor. Critical points occur when
$dX/dt=0$ and $dY/dt=0$, which yields a stationary point
in phase space at $(X_0, Y_0)$, representing a non-vanishing 
stationary solution,  
$$
 \begin{array}{ll}
  X_0 &= {(k_3 / k_2)} \\
  Y_0 &= {(k_1 / k_2)} 
 \end{array}
 \eqno(D2) 
$$

For small perturbations not too far off the limit cycle, we can
describe the cyclic dynamics with,
$$
 \begin{array}{ll}
  X(t) &= X_0 + x \exp{(\omega t )} \\
  Y(t) &= Y_0 + y \exp{(\omega t - \delta)} \\
 \end{array}
 \eqno(D3) 
$$
where the small amplitudes obey $|x/X_0| \ll 1$ and 
$|y \ Y_0| \ll 1$, and $\delta$ represents a phase delay. 
One can then derive the following dispersion relation,
$$
	\omega^2 + k_1 k_3 = 0 \ .
 	\eqno(D4) 
$$
The real part of the frequency is zero, $Re(\omega_n)=0$,
while the imaginary part characterizes an oscillation,
$Im(\omega)=\pm \sqrt{k_1 k_3}$. The dynamics is essentially
a circular motion in phase space $Y(X)$, which corresponds to
$X(t)=X_0 + x \sin{ \omega t}$ and 
$Y(t)=Y_0 + y \sin{ \omega (t-t_0)}$.  

Many nonlinear systems, however, are far off an equilibrium
state. In order to study the nonlinear behavior of the Lotka-Volterra
equation system, we can transform the variables in terms of the 
limit-cycle fixpoint $(X_0, Y_0)$,
$$
 \begin{array}{ll}
  X &= X_0 x = (k_3 / k_2) x \\
  Y &= Y_0 x = (k_1 / k_2) y 
 \end{array}
 \eqno(D5) 
$$
Inserting this parameterizaion (D5) into the original Lotka-Volterra
equation system (D1), multiplying them with $(y-1)$ and $(x-1)$,
and subtracting them from each other yields then
$$
	[(x-1) {dy \over dt} - (y-1) {dx \over dt} ]
	= k_3 (x-1)^2 y + k_1  (y-1)^2 x ]
	\eqno(D6)
$$
Substituting the variables $(x,y)$ with polar coordinates $(\rho, \omega)$
according to, 
$$
 \begin{array}{ll}
  (x-1) &= \rho \cos \omega \\
  (y-1) &= \rho \sin \omega \\
 \end{array}
 \eqno(D7) 
$$
yields then the function,
$$
	\Phi(\omega) := {d \omega \over dt} 
	= k_1 x \sin^2 \omega + k_3 y \cos^2 \omega
	\eqno(D8)
$$
which can be integrated to obtain the time dependence of the 
polar coordinate $\omega(t)$
$$
	\omega(t) = \int_0^{t'} \Phi (\omega) dt' \ ,
	\eqno(D9)
$$
The $(x,y)$ coordinates are found by the transcendental solution
of the Lotka-Volterra equation,
$$
	x^{-k_3} e^{k_3 x} = C \ 
	y^{k_1} e^{-k_1 y}
	\eqno(D10)
$$

\begin{figure}
\centerline{\includegraphics[width=0.9\textwidth,angle=0]{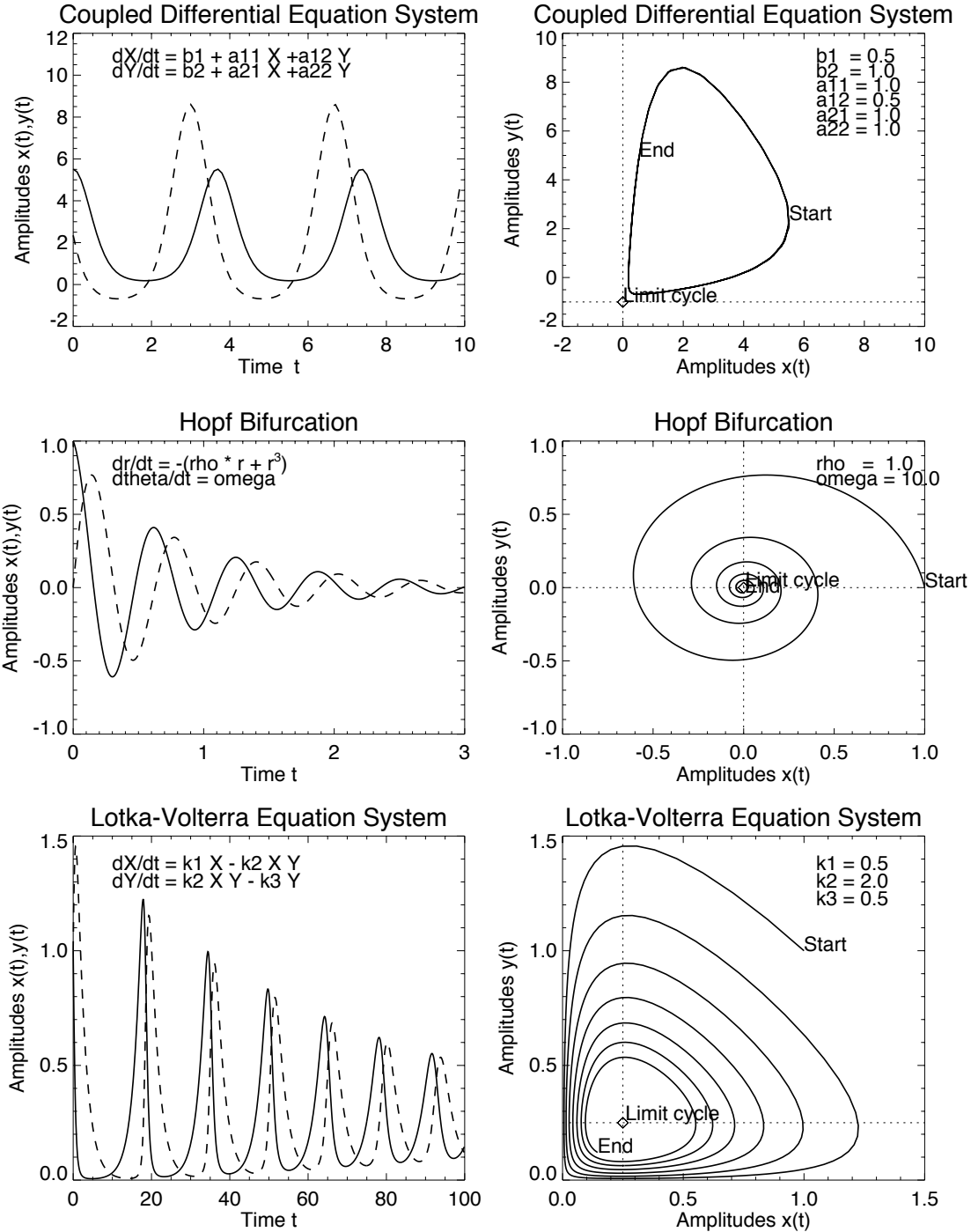}}
\caption{The dynamic behavior near a limit cycle is shown for
three different nonlinear systems: for coupled oscillators (top),
the Hopf bifurcation (middle), and the Lotka-Volterra equation system (bottom).
For each case the trajectories are shown in phase space $Y(X)$ (right panels), 
and as a function of time, $X(t)$ and $Y(t)$ (left panels), for the parameters
indicated in the right panels. The system starts to oscillate far away
from the limit cycle, but gradually approaches the attractor at the fixed point 
$(X_0, Y_0)$.}
\label{f_limitcycle}
\end{figure}

\begin{figure}
\centerline{\includegraphics[width=0.8\textwidth,angle=270]{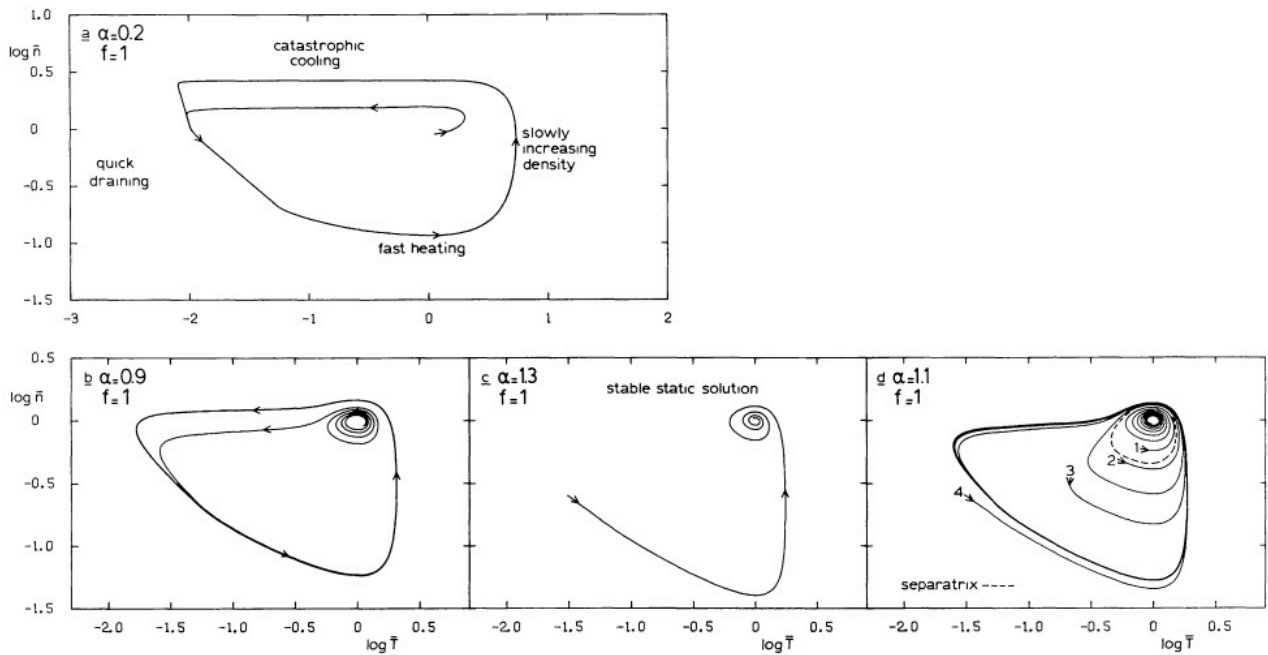}}
\caption{Solutions of the evaporation-condensation system (Eq.~24-25)
in the phase plane of dimensionless (logarithmic) temperature 
$\log(X)=\log(T/T_0)$ and electron density $log(n)=log(n/n_0)$
Cases (a) and (b) represent nonlinear oscillations near the
limit cycle, case (c) is a stable static solution. A separatrix
between stable and oscillatory solutions is indicated in (d)
[Kuin and Martens 1982].}
\label{f_kuin}
\end{figure}

\begin{figure}
\centerline{\includegraphics[width=0.8\textwidth,angle=0]{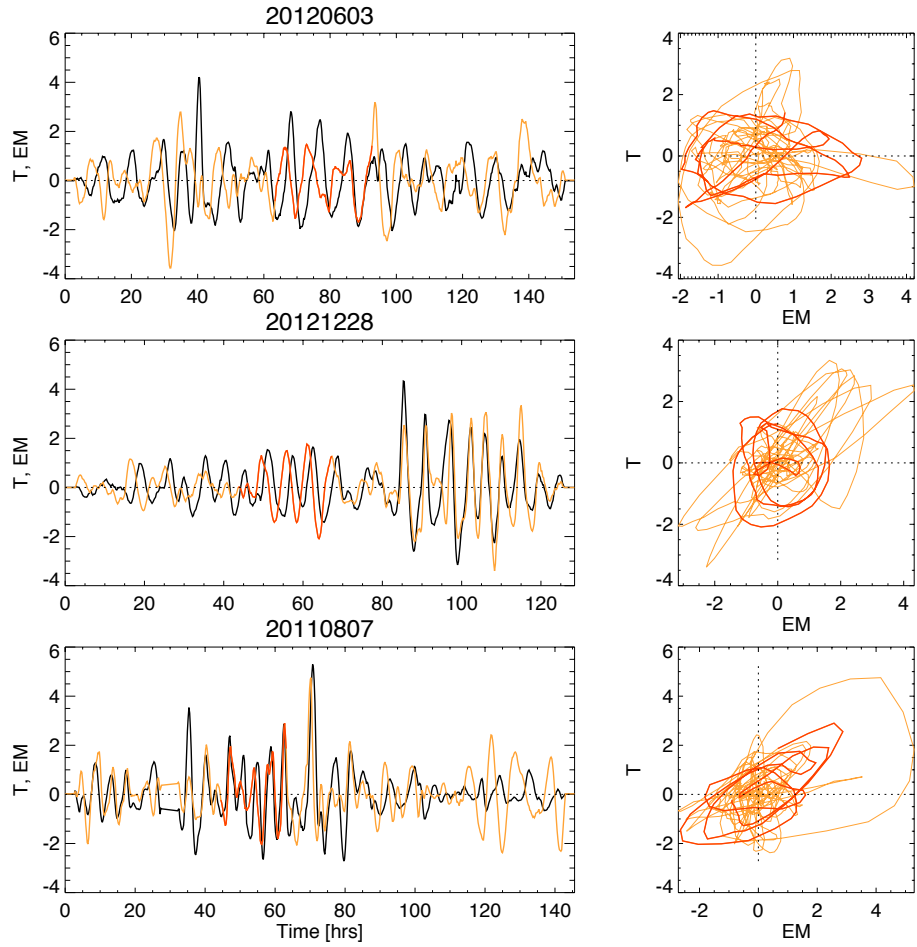}}
\caption{Smoothed time profiles of the emission measure $EM(t)$
(black in left panels) and the electron temperature $T_e(t)$ (red
in left panels, and phase diagram $T_e(EM)$ (right panels) of three
loop episodes observed in an active region with AIA/SDO. 
A moving average background has been
subtracted in all time profiles, and the amplitudes are normalized
by their standard deviation from the means. A quasi-stationary time 
interval with near-elliptical phase trajectories is colored with red.
The quasi-periodicity and the phase delay indicate a limit-cycle 
behavior of the evaporation-condensation cycle in solar flares 
[Froment et al.~2015].}
\label{f_froment}
\end{figure}

\begin{figure}
\centerline{\includegraphics[width=0.9\textwidth,angle=0]{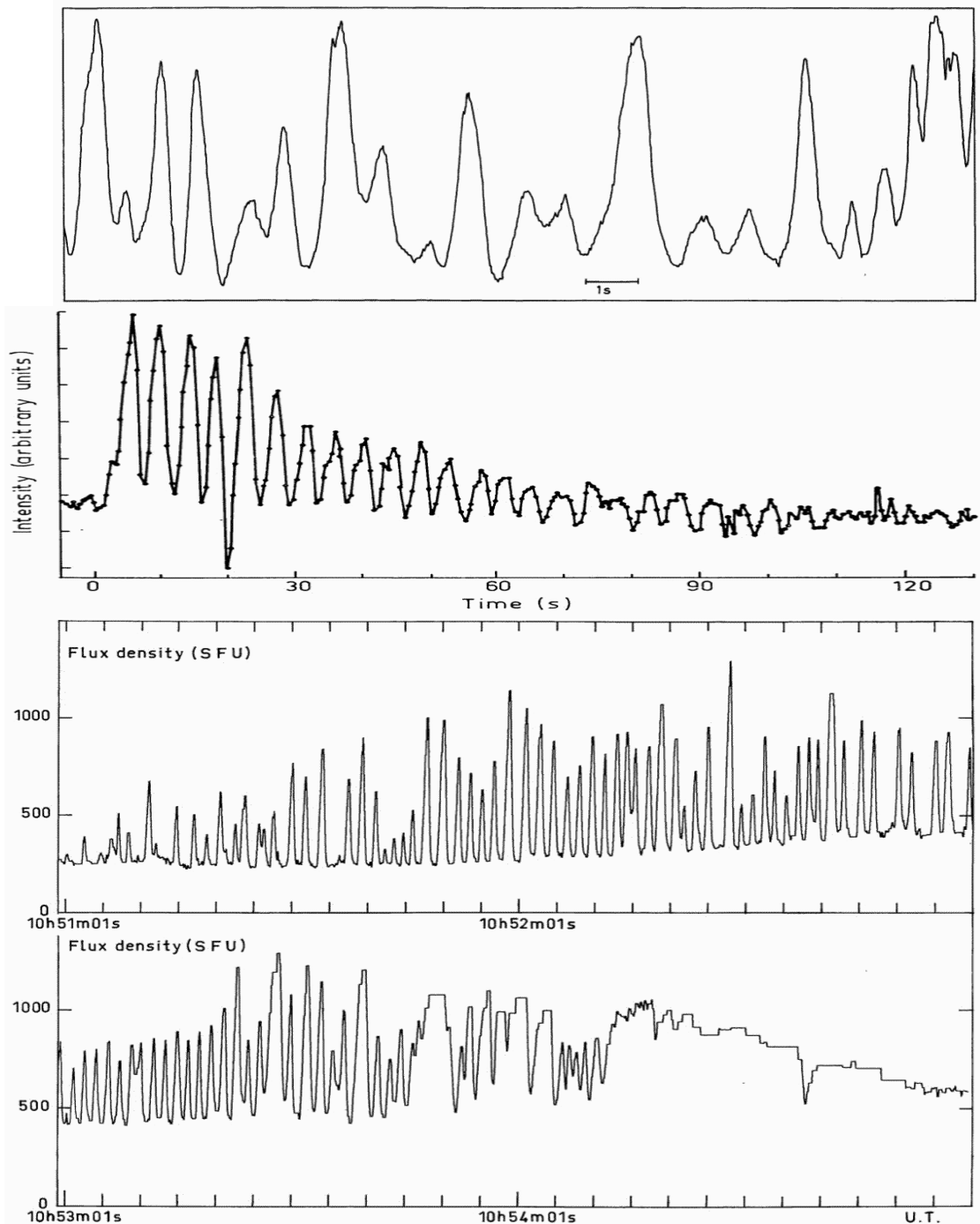}}
\caption{Three examples of solar radio burst oscillations:
{\sl Top:} Evidence for sub-harmonics (1:3) [Rosenberg 1970];
{\sl Second row:} Exponentially damped oscillation [McLean 
and Sheridan 1973]; {\sl Third and bottom row:} Metric radio
oscillations with a period of $P \approx 1.5$ s [Trottet et al.
1981].}
\label{f_radio_osc}
\end{figure}

\begin{figure}
\centerline{\includegraphics[width=0.8\textwidth,angle=270]{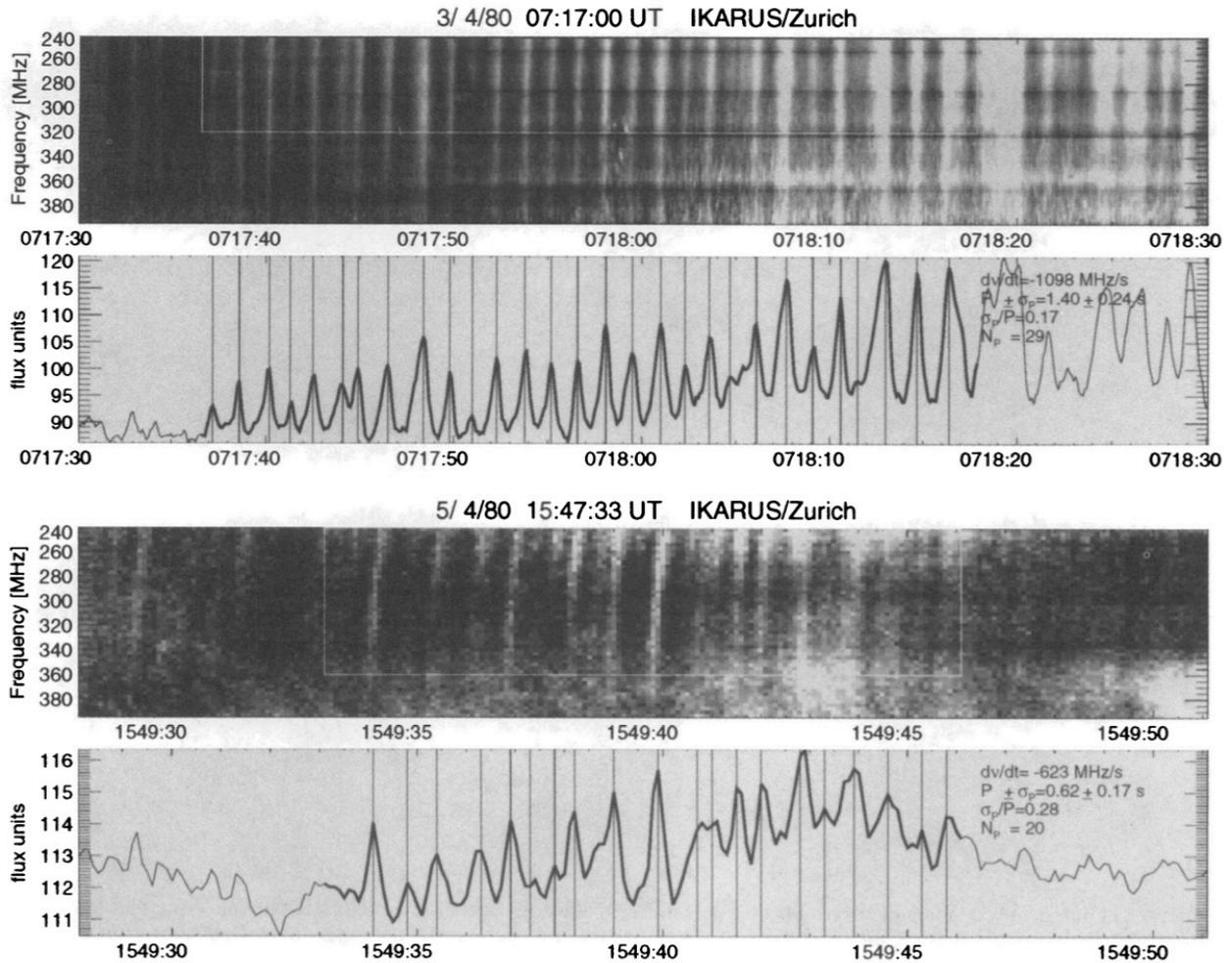}}
\caption{Dynamic spectra of solar radio pulsations (top two panels)
and a series of type III bursts (bottom two panels),
observed with the IKARUS/Zurich spectrograph in the frequency range
of 240-400 MHz. The quasi-periodic pulsation pattern is
characteristic for nonlinear systems with limit cycles, 
while the radio type III bursts appear to be produced
by a random process [Aschwanden et al.~1994].}
\label{f_zurich}
\end{figure}

\begin{figure}
\centerline{\includegraphics[width=0.9\textwidth,angle=0]{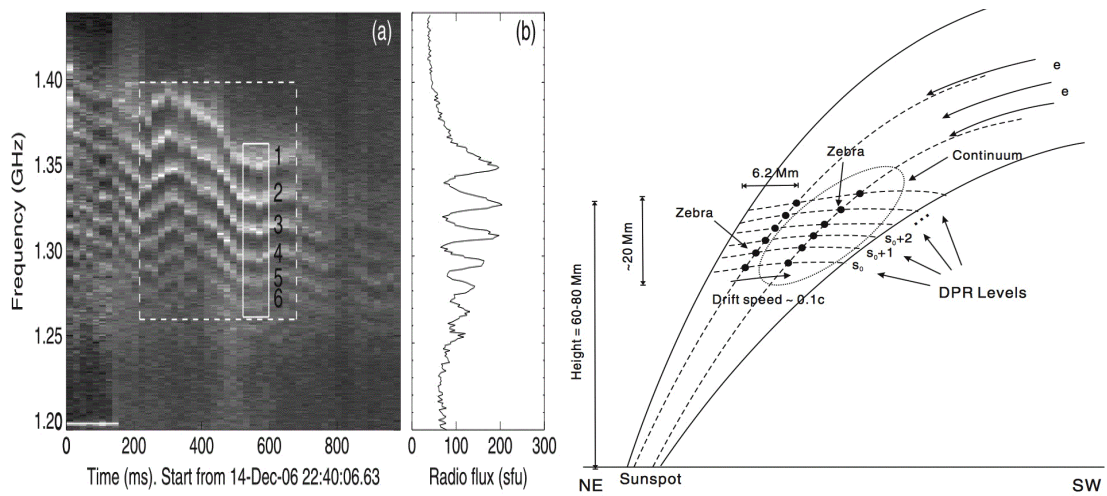}}
\caption{{\sl Left:} A zebra-pattern solar radio burst observed
on 2006 December 14, 22:40 UT. Six successive stripes with
decreasing frequency ar marked. {\sl Middle:} A time-averaged
flux profile as a function of the frequency, averaged over the
box shown in the left panel. {\sl Right:} Simplified spatial
model of the location of 6 different harmonics of the
gyrofrequency [Chen et al.~2011].}
\label{f_zebra}
\end{figure}

\begin{figure}
\centerline{\includegraphics[width=1.0\textwidth,angle=0]{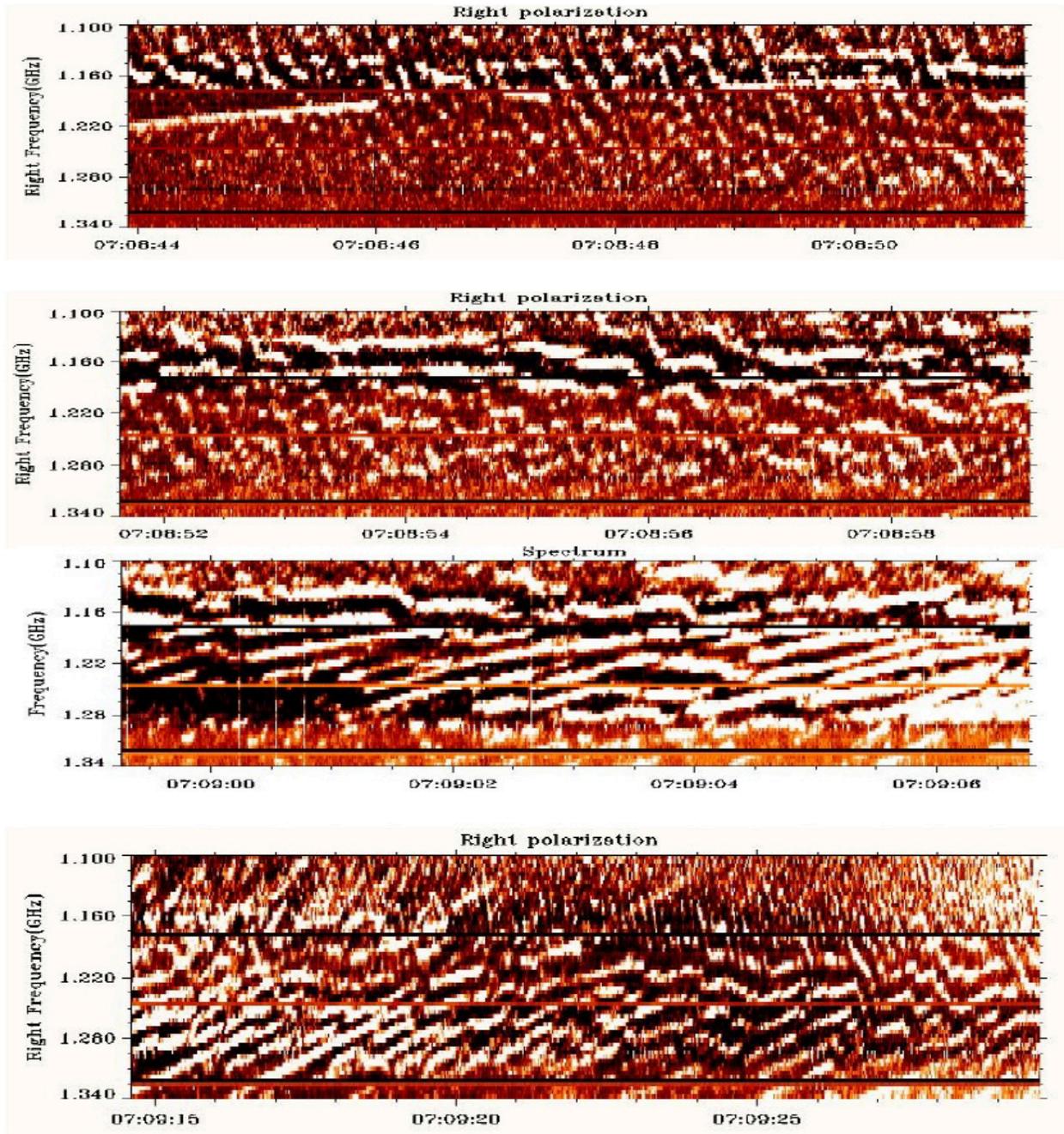}}
\caption{This dynamic spectrum (frequency versus time) shows the
evolution of zebra-type bursts during 46 s, observed on 
2004 December 1 with the Huairou radio station (Bejing).
Initial fiber-type bursts transform into zebra patterns, as well as
into decimetric millisecond spikes in the lower frequency range
of 1110-1160 MHz [Chernov et al.~2017].}
\label{f_chaos_zebra} 
\end{figure}

\begin{figure}
\centerline{\includegraphics[width=0.9\textwidth,angle=270]{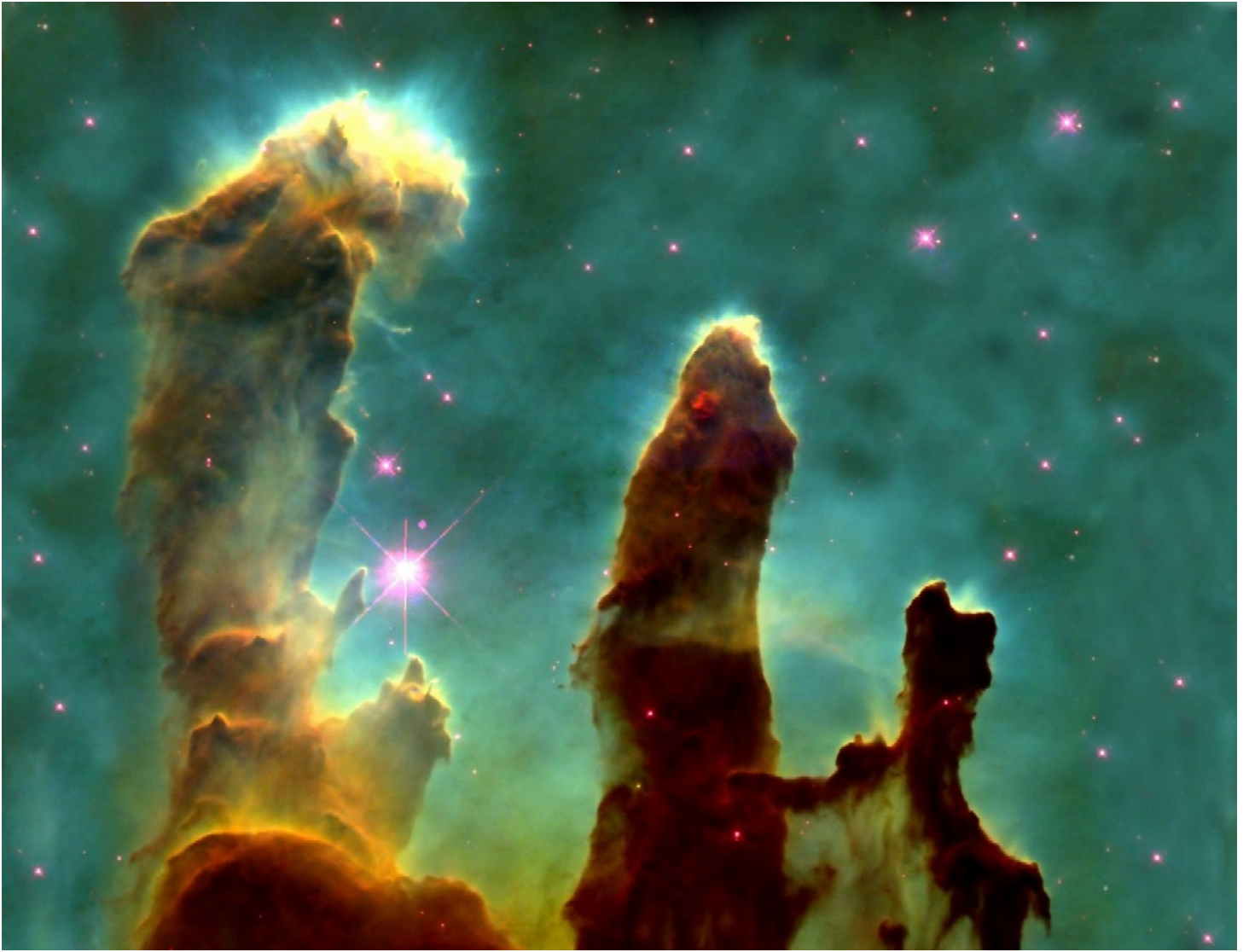}}
\caption{Star formation regions in the Eagle nebula form 
spatial structures in the shape of towering pillars 
[credit Hubble Space Telescope (HST), NASA].} 
\label{f_nebula_pillars}
\end{figure}

\begin{figure}
\centerline{\includegraphics[width=1.0\textwidth,angle=0]{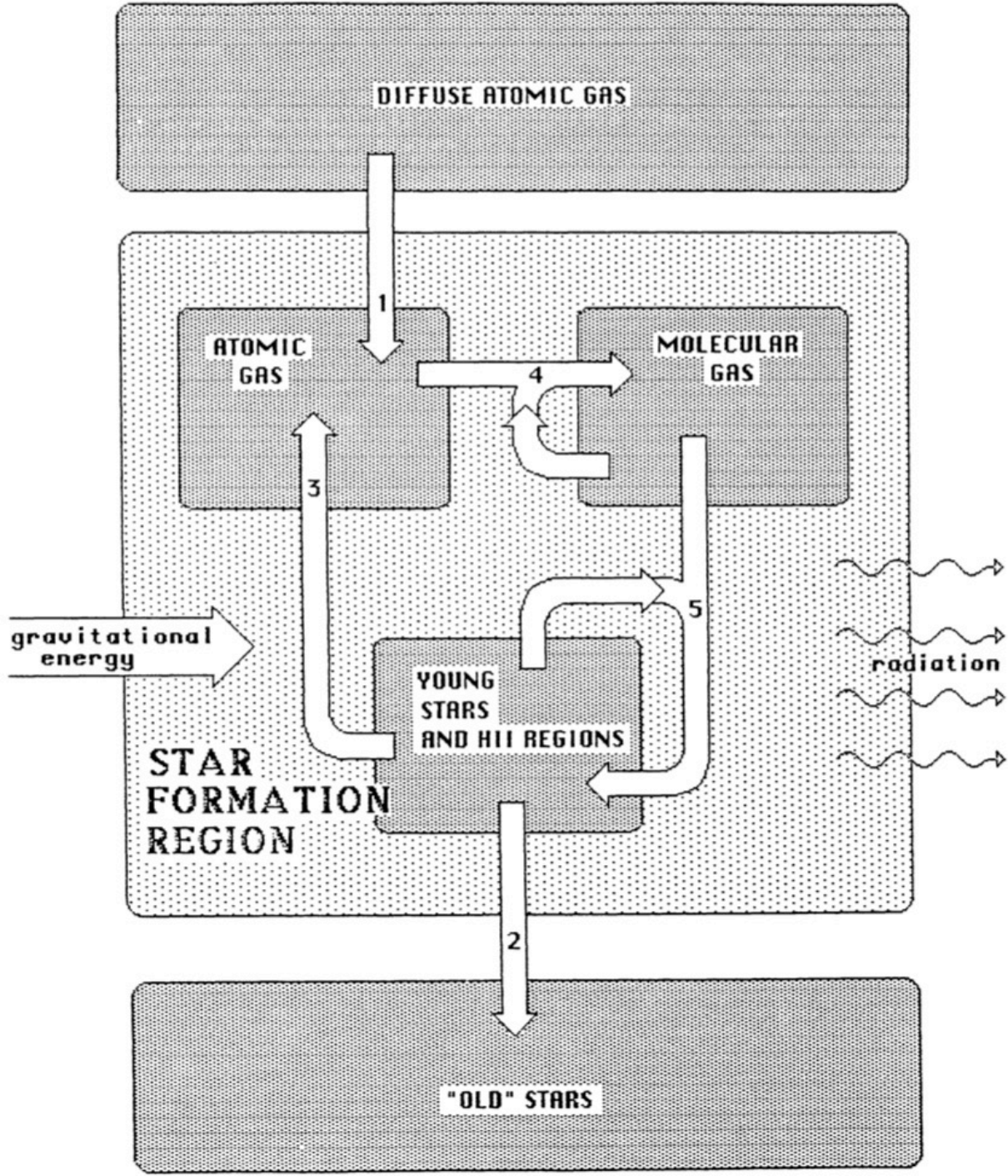}}
\caption{The self-organizing system of star formation amounts
globally to a transformation of diffuse atomic gas (supply 
reservoir) into ``old'' stars (waste reservoir). The internal
processes consist of: (1) gas inflow, (2) stellar evolution,
(3) stellar mass loss and recombination of ionized gas;
(4) production of molecular gas, and (5) triggered star
formation [Bodifee 1986].}
\label{f_bodifee}
\end{figure}

\begin{figure}
\centerline{\includegraphics[width=1.0\textwidth,angle=0]{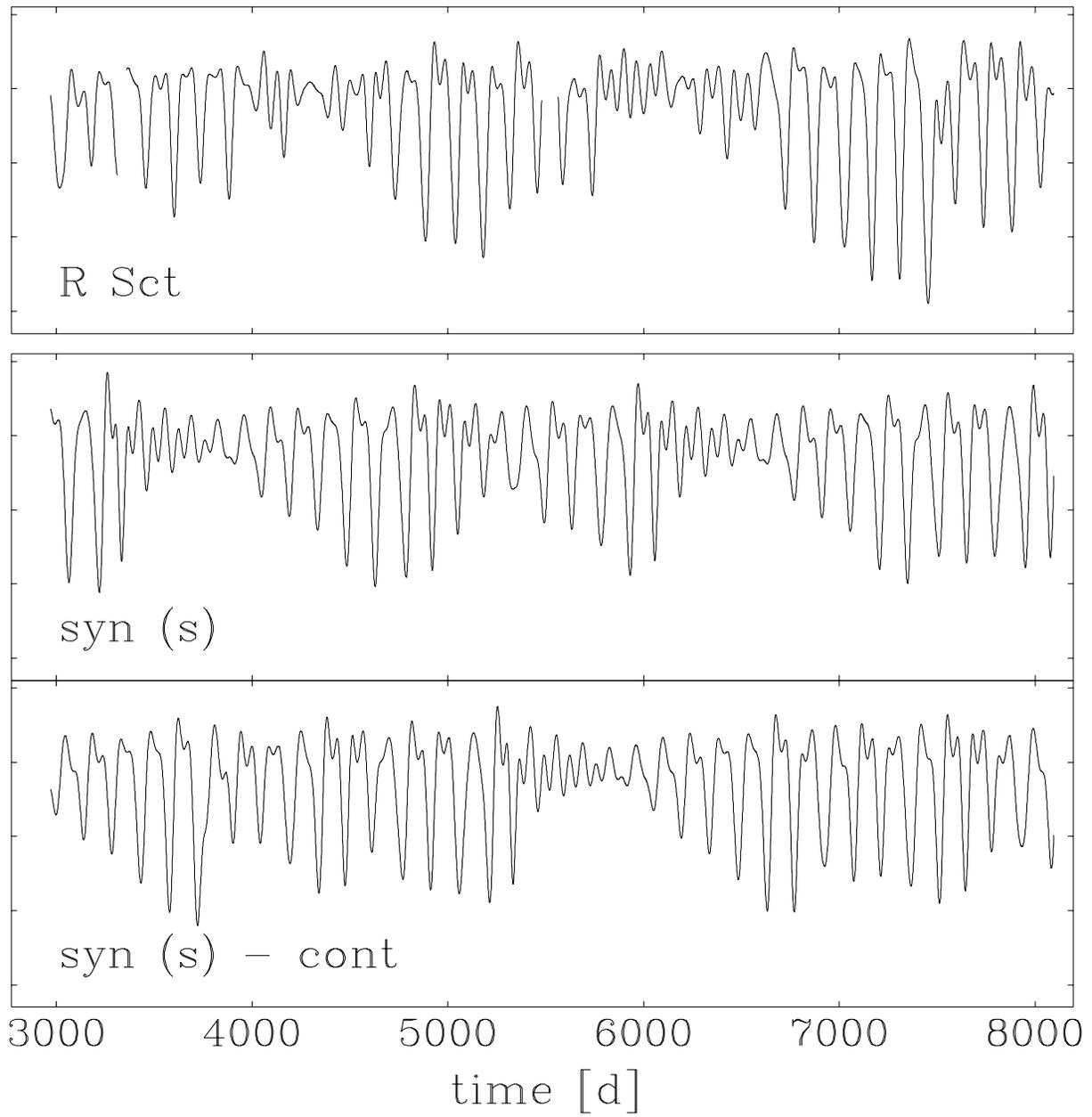}}
\caption{The smoothed light curve observed from the RV Tau-type star R Scuti
(top) and synthetically generated light curves with a model of a
low-dimensional strange attractor (middle and bottom)
[Buchler et al.~1996].}
\label{f_buchler}
\end{figure}

\begin{figure}
\centerline{\includegraphics[width=1.0\textwidth,angle=0]{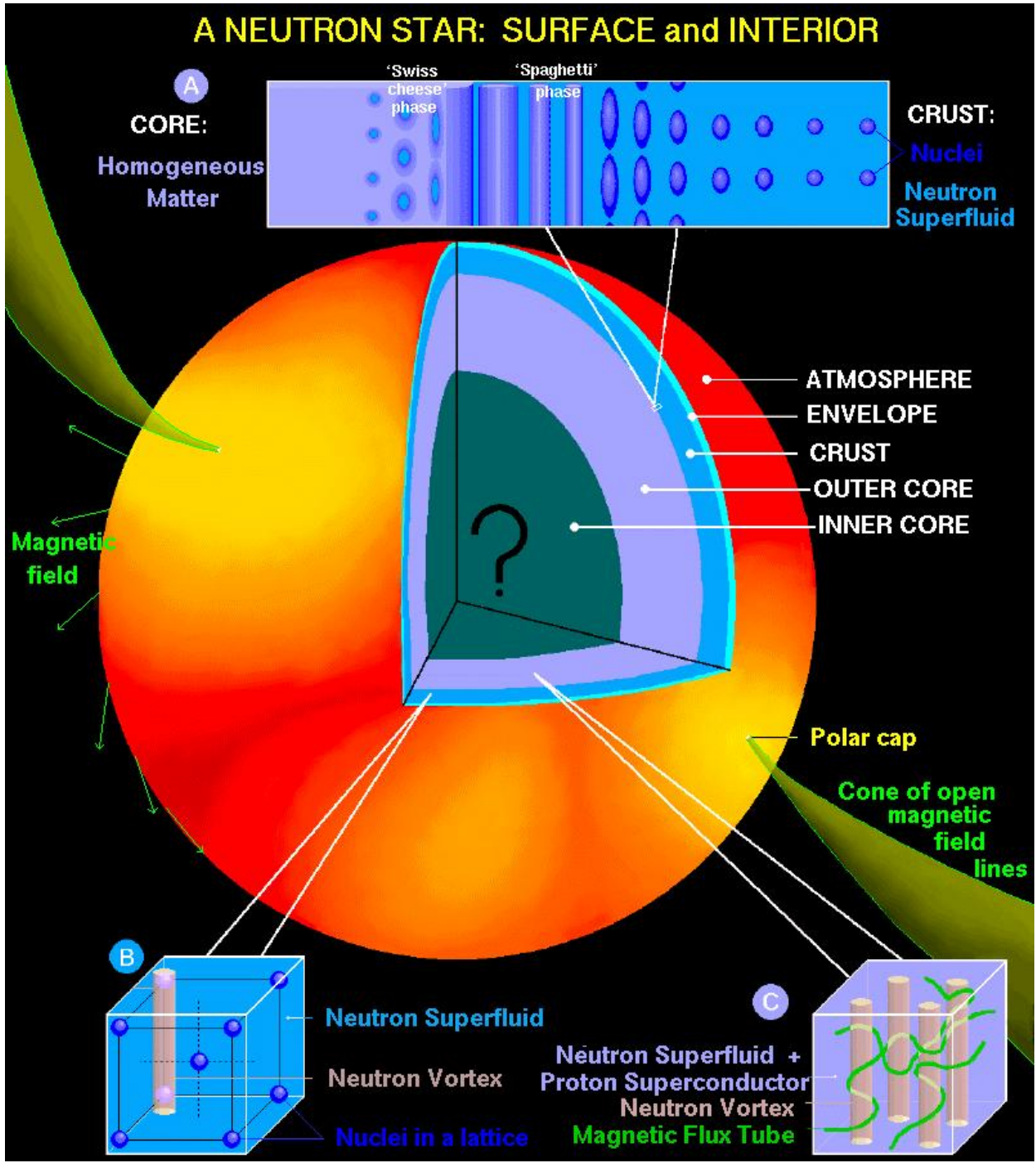}}
\caption{A cross-section of a neutron star shows the rich variety
of emergent quantum matter expected in its crust and core.
[credit: Matthew H. Schneps, Science Media Group, Harvard-Smithsonian 
Center for Astrophysics (CfA)].}
\label{f_neutron_star}
\end{figure}

\begin{figure}
\centerline{\includegraphics[width=0.9\textwidth,angle=270]{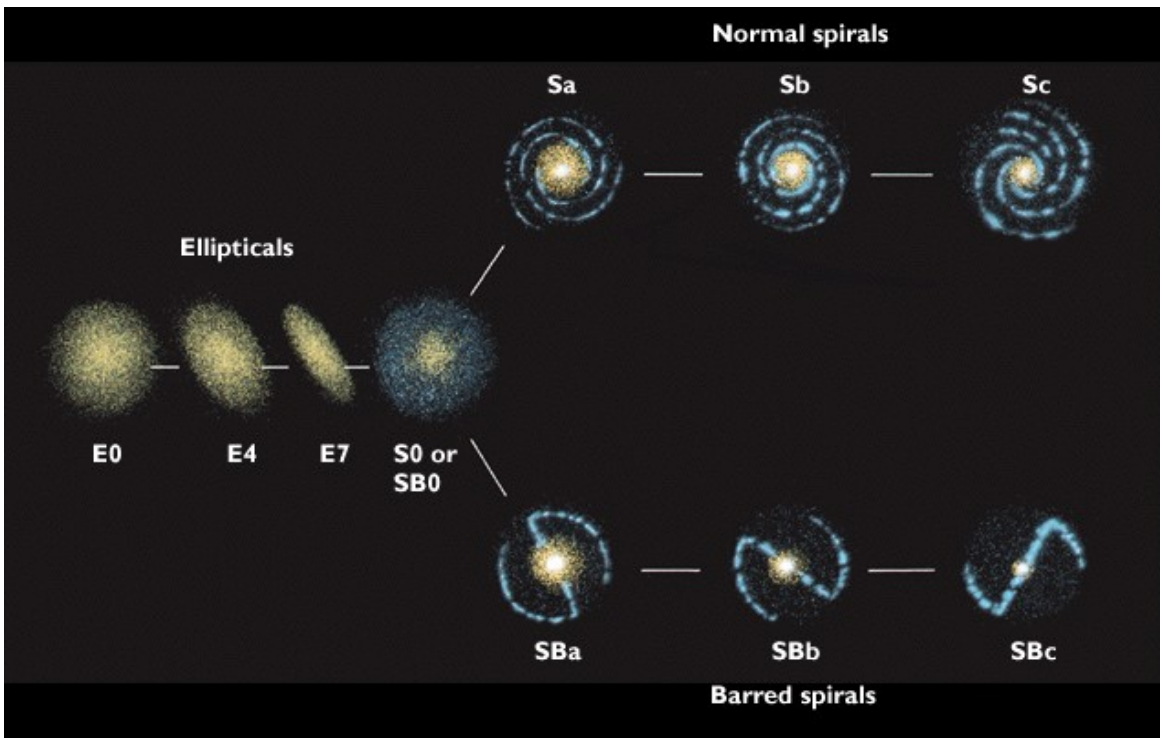}}
\caption{The Hubble galaxy classification reflects the morphology from
random-like clusters (E0) to spiral-structured ordering (Sc, SBc)
[credit www.physast.uga.edu].}
\label{f_galaxy_hubble}
\end{figure}

\begin{figure}
\centerline{\includegraphics[width=1.0\textwidth,angle=0]{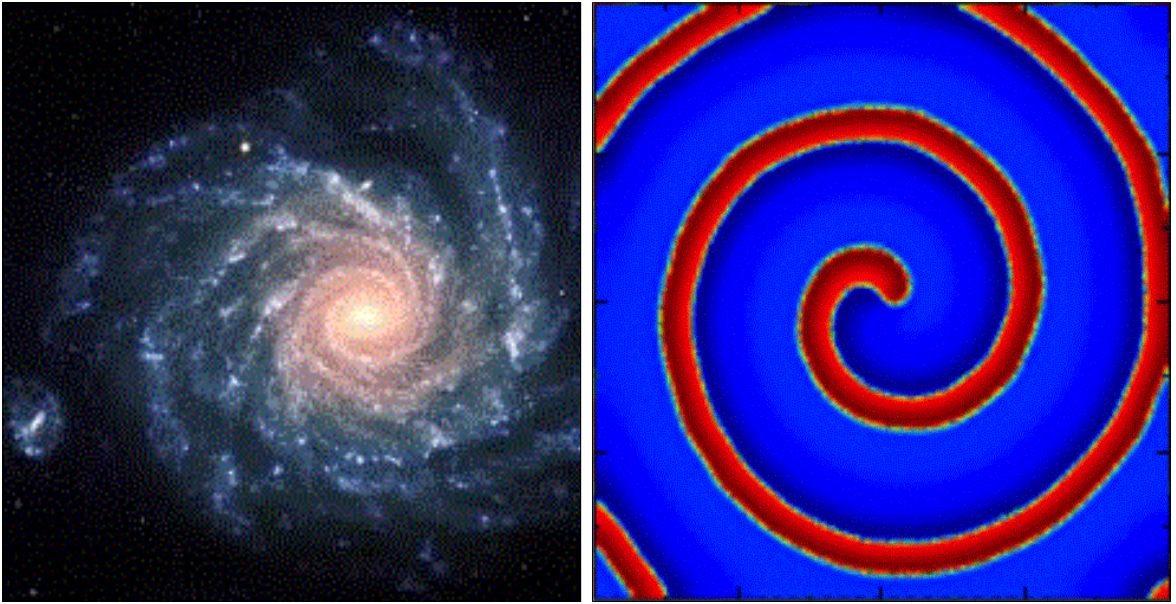}}
\caption{{\sl Left:} Spiral galaxy NGC 1232 [Credit: European Southern 
Observatory (ESO)];
{\sl Right:} Spiral pattern in two-component reaction-diffusion system of
Fitzhugh-Nagumo type [Credit: Wikipedia - Reaction-diffusion system].}
\label{f_galaxy_spiral}
\end{figure}

\begin{figure}
\centerline{\includegraphics[width=0.8\textwidth,angle=270]{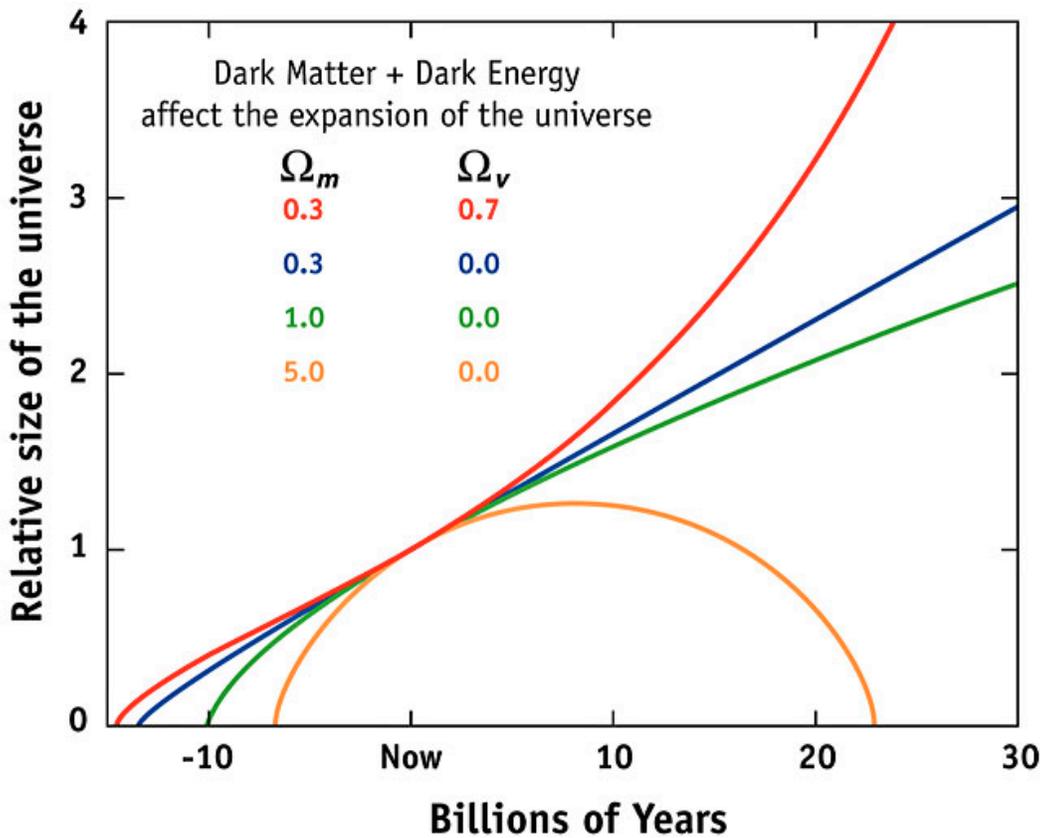}}
\caption{Four scenarios of the expansion or contraction of the universe:
the closed, high-density universe (orange), the critical-density universe,
the Einstein-de Sitter model (green), the open, low-density universe (blue),
and the universe in which a large fraction of the matter is in a form
of ``dark energy'' (the $\Lambda$CDM model; red), which is causing the 
expansion of the universe to accelerate [NASA/WMAP Science Team].}
\label{f_cosmology}
\end{figure}

\begin{figure}
\centerline{\includegraphics[width=0.9\textwidth,angle=270]{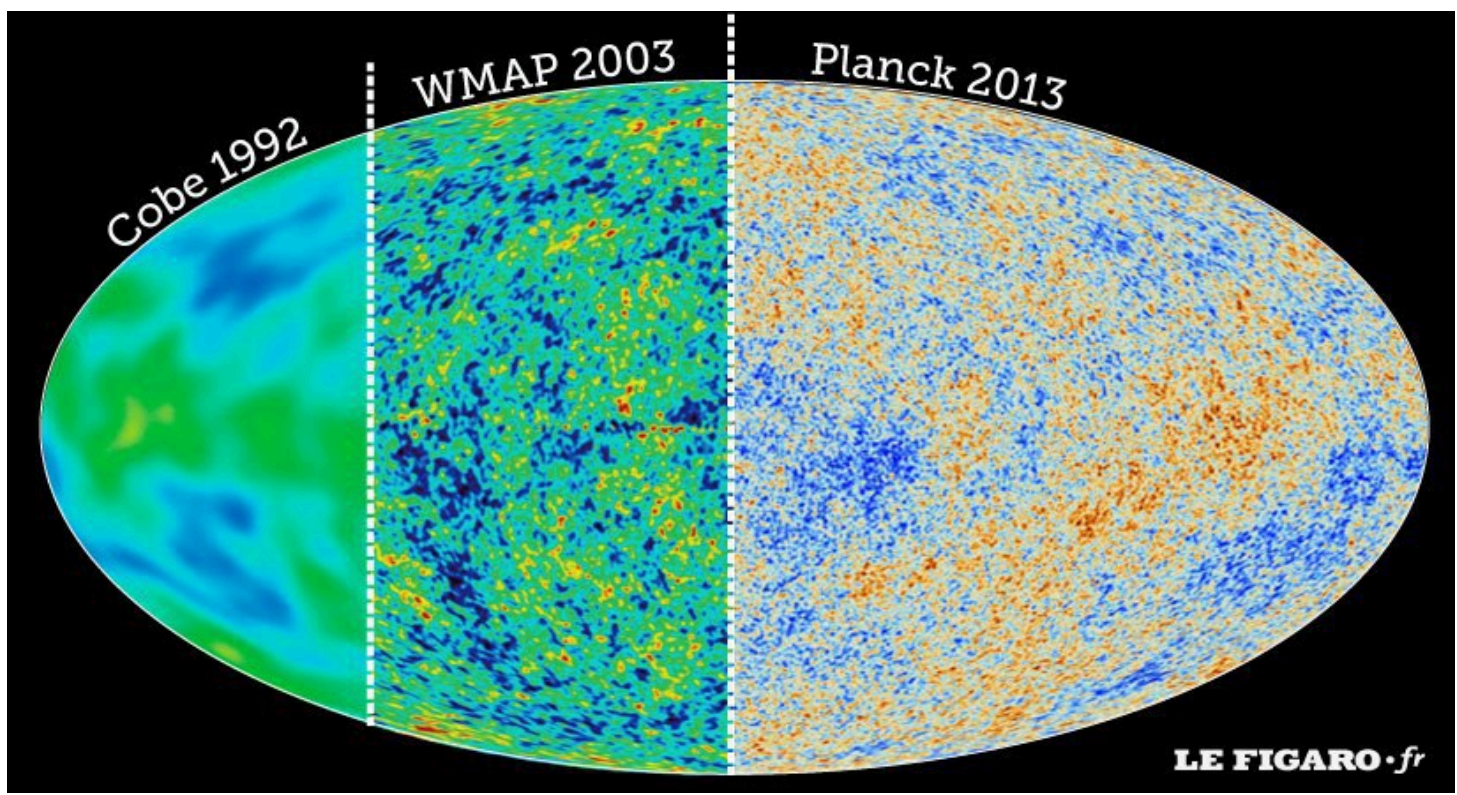}}
\caption{Measurements of the spatial distribution and anisotropy of
the cosmological microwave background radiation at 3 K with the
Cosmic Background Explorer (COBE) in 1992, the {\sl Wilkinson Microwave 
Anisotropy Probe (WMAP)} satellite in 2003, and the Planck satellite in 2013 
[Credit: COBE/NASA, Mather et al.~1991, Boggess et al.~1992;  
WMAP/NASA, Bennett et al.~2013, Hinshaw et al.~2013; and
Planck/ESA; Planck Collaboration 2016].}
\label{f_cobe}
\end{figure}

\begin{figure}
\centerline{\includegraphics[width=0.9\textwidth,angle=270]{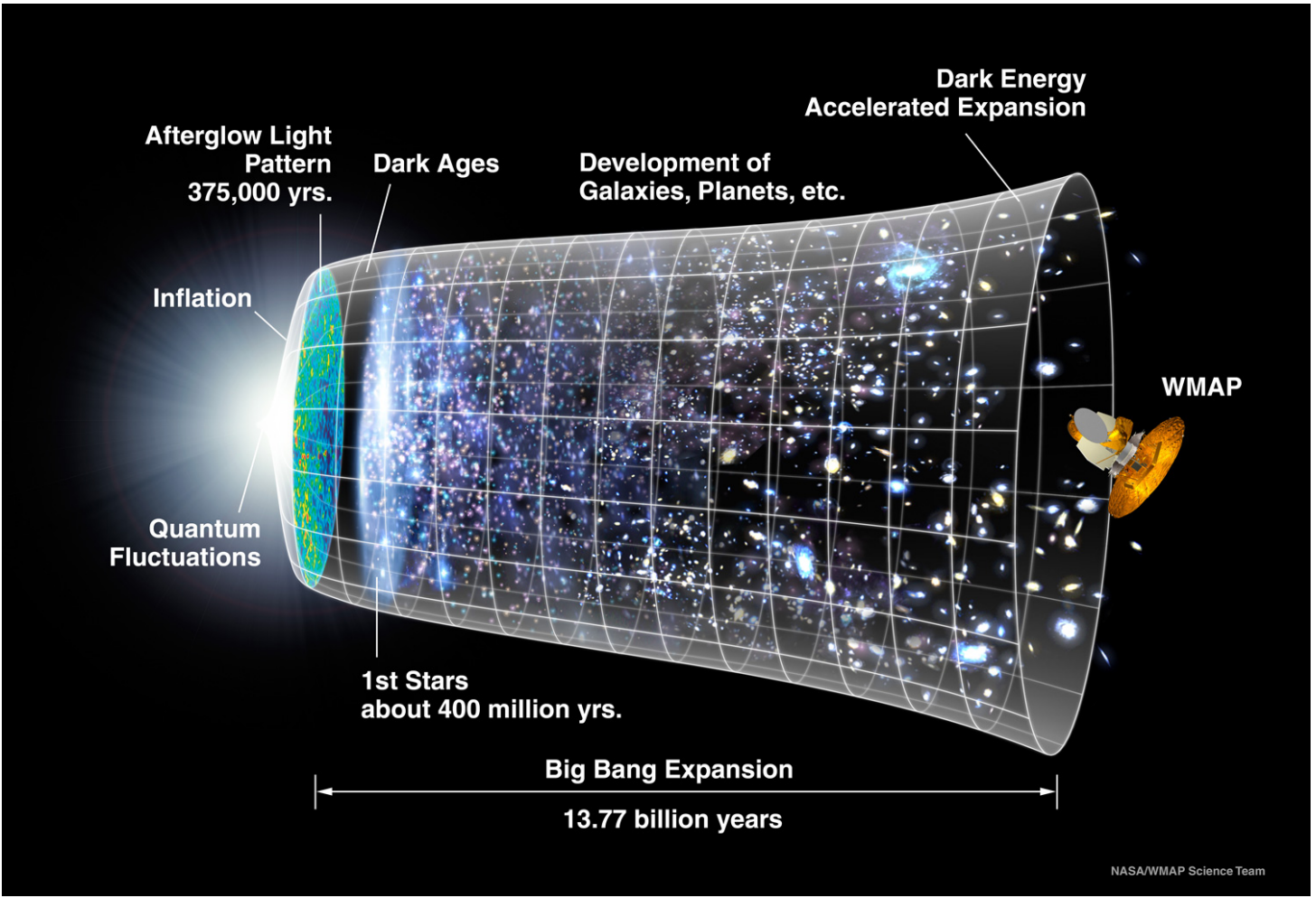}}
\caption{A representation of the evolution of the universe over 13.77
billion years. The far left depicts the earliest moment we can now probe, 
when a period of "inflation" produced a burst of exponential growth 
in the universe. (Size is depicted by the vertical extent of the grid). 
For the next several billion years, the expansion of the universe 
gradually slowed down as the matter in the universe pulled on itself 
via gravity. More recently, the expansion has begun to speed up again 
as the repulsive effects of dark energy have come to dominate the 
expansion of the universe. The afterglow light seen by WMAP was emitted 
about 375,000 years after inflation and has traversed the universe 
largely unimpeded since then. The conditions of earlier times are 
imprinted on this light; it also forms a backlight for later 
developments of the universe [Credit: NASA/WMAP Science Team].}
\label{f_inflation}
\end{figure}

\begin{figure}
\centerline{\includegraphics[width=1.0\textwidth,angle=0]{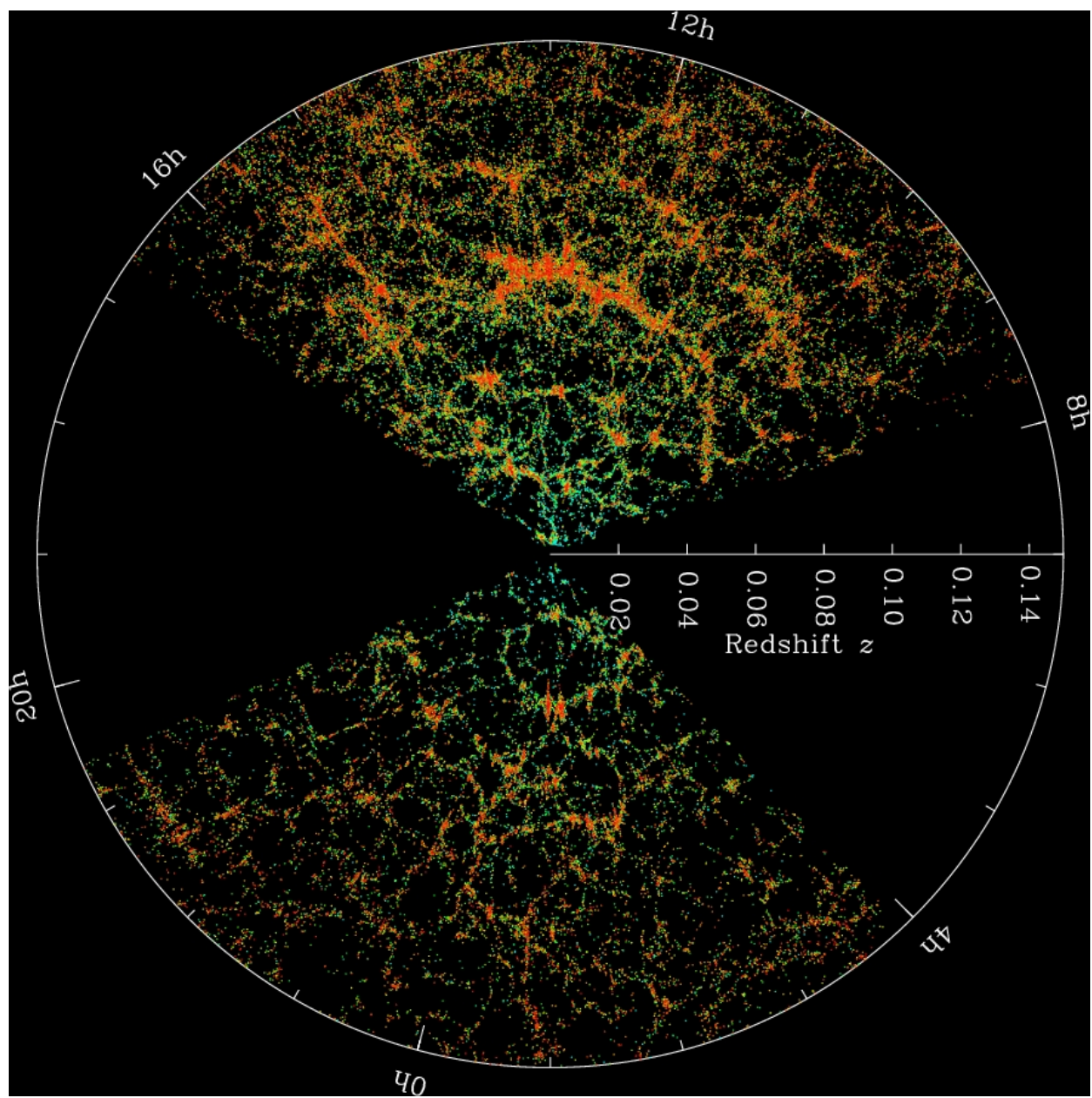}}
\caption{{\sl Sloan Digital Sky Survey (SDSS)} galaxy map: A slice 
of the universe showing the large-scale structure of galaxies.
Each dot is a galaxy: the color is the green-red color of that
galaxy [Credit: M. Blanton and Sloan Digital Sky Survey,
www.sdss.org].}
\label{f_blanton}
\end{figure}

\begin{figure}
\centerline{\includegraphics[width=0.9\textwidth,angle=0]{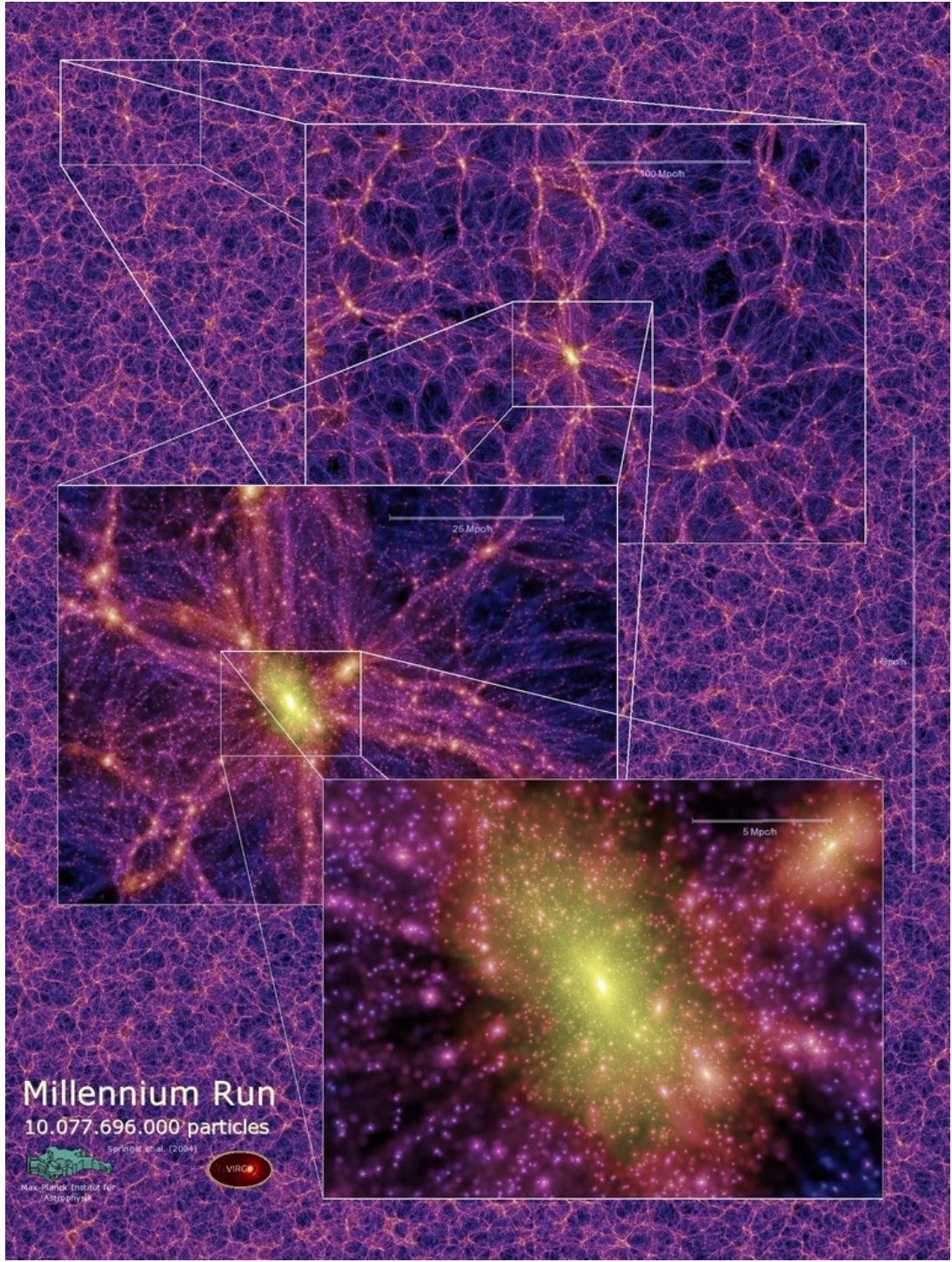}}
\caption{Numerical simulations of the dark matter density field on
various scales. Each individual image shows the projected dark
matter density field in a slab of thickness 15 h$^{-1}$ Mpc
(sliced from the periodic simulation  volume at an angle chosen
to avoid replicating structures in the lower two images),
color-coded by density and local dark matter velocity dispersion.
The zoom sequence displays consecutive enlargements by factors
of four, centered on one of the many galaxy cluster halos present
in the simulations [Credit: Millenium Simulation, Virgo Consortium,
Max-Planck-Institute for Astrophysics; Springel et al.~2005].}
\label{f_millenium_sim}
\end{figure}

\begin{figure}
\centerline{\includegraphics[width=1.0\textwidth,angle=0]{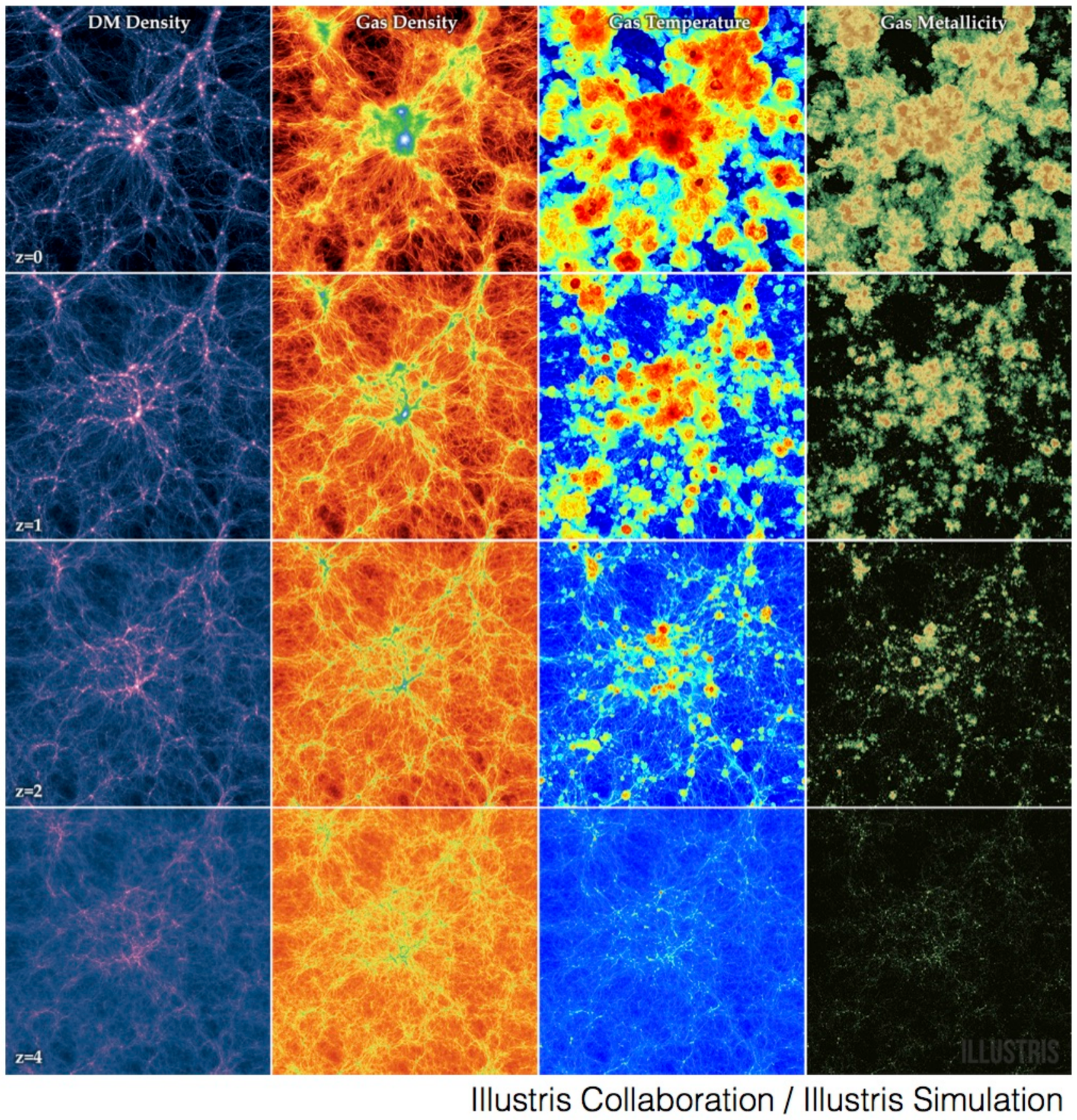}}
\caption{Illustris Simulation (IS) of the galaxy formation in our 
universe. In the $\Lambda$CDM model, galaxies build up their mass
hierarchically through the mergers of smaller galaxies to larger
ones in a cold dark matter-dominated universe. The time axis is
from bottom to top, and the 4 columns contain the dark matter
density (left), the gas density, the gas temperature, and the
gas metallicity (right column) [Credit: Illustris Collaboration,
Illustris Simulation; Vogelsberger et al.~2014].}
\label{f_unisim3}
\end{figure}

\begin{figure}
\centerline{\includegraphics[width=1.0\textwidth,angle=0]{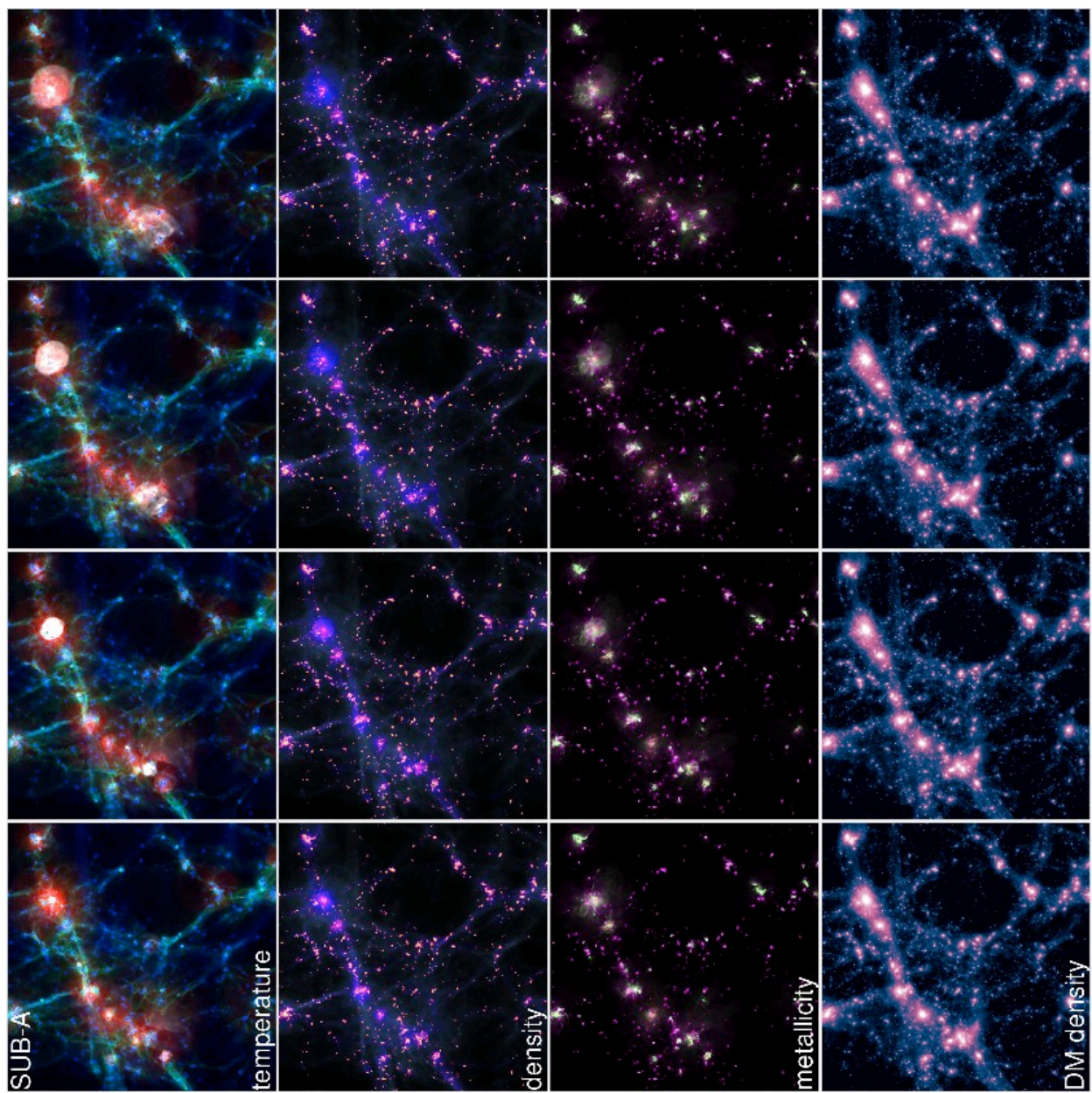}}
\caption{Short-time evolution of baryonic parameters 
(gas temperature, gas density, gas metallicity, Dark Matter density,
each one shown in a row, while the columns contain 4 different
time steps). The time resolution is less than 3 Myr. The more 
massive halo in the upper right shows strong AGN activity leading 
to heating and expansion of large amounts of gas
[Credit: Illustris Collaboration, Illustris Simulation; 
Vogelsberger et al.~2014].}
\label{f_unisim2}
\end{figure}

\begin{figure}
\centerline{\includegraphics[width=1.0\textwidth,angle=0]{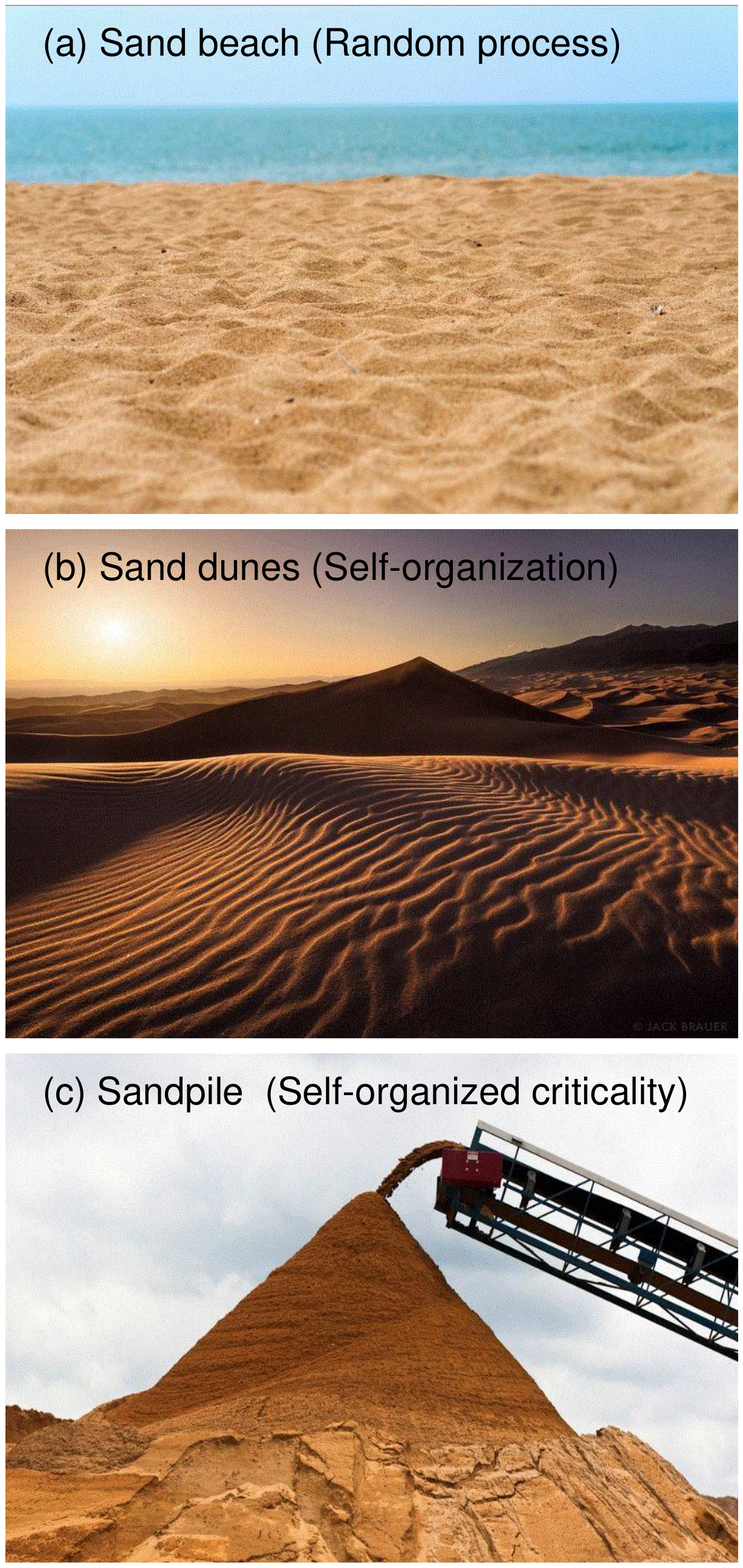}}
\caption{The same medium (for instance sand) can be subject of three
different dynamical processes, such as (a) random processes (sand beach),
(b) self-organization (sand dunes), or (c) self-organized criticality
(sand piles) [Credit: Google].}
\label{f_sand}
\end{figure}

We show a typical dynamic solution in Fig.~(\ref{f_limitcycle} bottom), computed by a numerical
code that obtains solutions for $X(t)$ and $Y(t)$ directly by
minimizing the coupled first-order, nonlinear, differential equation 
system given in Eq.~(D1), for the coefficient $k_1=0.5,
k_2=2.0, k_3=0.5$. The time-dependent solutions $X(t)$ and $Y(t)$
are shown in the bottom left panel of Fig.~(\ref{f_limitcycle}), and a phase plot $Y(X)$
is shown in the top right panel of Fig.~(\ref{f_limitcycle}). We see that the trajectory
in phase space is convergent towards the limit-cycle solution,
starting from highly nonlinear amplitude oscillations far off the
equilibrium, while the system asymptotically converges towards 
the attractor ($X_0, Y_0$), with gradually diminuishing amplitude.

\bigskip
\acknowledgements
{\sl Acknowledgements:}
The author acknowledges the hospitality and partial support for
two workshops on ``Self-Organized Criticality and Turbulence'' at the
{\sl International Space Science Institute (ISSI)} at Bern, Switzerland,
during October 15-19, 2012, and September 16-20, 2013, as well as
constructive and stimulating discussions (in alphabetical order)
with Robert Cameron, Sandra Chapman, Paul Charbonneau, Cavid Collins,
Daniel Fabrycky, Clara Froment,
Hermann Haken, Henrik Jeldtoft Jensen, Lucy McFadden, Maya Paczuski, 
Jens Juul Rasmussen, John Rundle, Loukas Vlahos, and Nick Watkins.
This work was partially supported by NASA contract NNX11A099G
``Self-organized criticality in solar physics''.

\clearpage


\clearpage

\begin{deluxetable}{llll}
\tablecaption{Self-organization processes in astrophysics: The symbols 
in the last column indicate the following system characteristics: 
LC= nonlinear systems with limit cycle(s),
I = instabilities, 
R = resonances, 
E = entropy, 
S = regular spatial pattern, 
T = regular temporal pattern, 
? = conjectural.
The stellar QPOs include also accretion disks and ``coronas'' of compact 
objects and supermassive black holes.}
\tablewidth{0pt}
\tablehead{
\colhead{Observed Phenomenon}&
\colhead{Driver Mechanism}&
\colhead{Feedback Mechanism}&
\colhead{Characteristics}\\
\colhead{}&
\colhead{}&
\colhead{}}
\startdata
Planetary spacing 	 	&gravity		& harmonic orbit resonances     & R, S, T  \\
Saturn rings and moons   	&gravity		& harmonic orbit resonances     & R, S, T  \\
Protoplanetary disks		&rotation		& Hall-shear instability        & I, S     \\
Jupiter's red spot		&temperature gradient 	& inverse MHD turbulent cascade & I, S     \\
Saturn's hexagon                &circumpolar jet-stream & diocotron instability         & I, S     \\
Planetary entropy		&solar radiation	& planetary infrared emission   & E        \\
Solar photospheric granulation  &temperature gradient 	& Rayleigh-B\'enard instability & I, S     \\
Solar magnetic fields  		&solar dynamo, rotation & bouyancy, kink instability	& I(?), S  \\
Solar magnetic Hale cycle 	&differential rotation 	& twisted magnetic field relaxation & LC, I(?), S, T \\
Solar flare loops               &chromospheric evaporation &coronal condensation 	& I, LC[?],T(?) \\
Solar radio pulsations 		&nonthermal particles 	& loss-cone instability 	& LC, I[?], T \\
Solar zebra radio bursts 	&nonthermal particles 	& double plasma resonance 	& R, S, IT[?] \\
			 	&nonthermal particles 	& Langmuir-whistler coalescence & R, S, IT[?] \\
Star formation			&gravity		& radiation and recombination 	& I, S, T  \\
Stellar quasi-periodic oscillations &rotation		& magneto-rotational instability& LC, I(?),T\\
Pulsar superfluid unpinning 	&rotation		& Magnus force  		& I(?), S[?]\\
Galaxy formation		&gravity, rotation 	& density waves, reaction-diffusion & S, I(?)  \\
Cosmology  			&Big Bang expansion 	& inflationary $\Lambda$CDM model & I(?),LC(?],S[?],T[?] \\
\enddata
\end{deluxetable}

\begin{deluxetable}{lll}
\tablecaption{Non-astrophysical self-organization processes.}
\tablewidth{0pt}
\tablehead{
\colhead{Field}&
\colhead{Phenomenon}&
\colhead{Reference}\\
\colhead{}&
\colhead{}&
\colhead{}}
\startdata
Ionosphere	& Stimulation of electromagnetic emission & Leyser (2001)         \\
		& Internal gravity waves	  	  & Aburjania et al.~(2013)\\
		& Acoustic gravity waves	 	  & Kaladze et al.~(2008) \\
Magnetosphere	& Substorm dynamics 			  & Sharma et al.~(2001)  \\
	  	& Substorm current sheet model 		  & Valdivia et al.~(2003)\\
		& Magnetospheric vortex formation         & Yoshida et al.~(2010) \\
		& 2D MHD transverse Kelvin-Helmholtz instability & Miura (1999)   \\
		& Turbulent relaxation of magnetic fields & Tetreault (1992a,b)   \\
Plasma physics  & Superconducting ring magnet vortex      & Yoshida et al.~2010   \\ 
		& Magnetic reconnection in laboratory 	  & Yamada et al.~(2010)  \\
		& Magnetic reconnection in laboratory 	  & Zweibel and Yamada (2009)\\
Physics		& Coupled pendulums			  & Tanaka et al.~(1997)  \\
		& Spontaneous magnetization		  & Boesiger et al.~(1978)\\
		& Laser					  & Zeiger and Kelley (1991)\\
		& Superconductivity			  & Vazifeh and Franz (2013)\\
                & Bose-Einstein condensation		  & Nagy et al.~(2008)    \\
Chemistry	& Molecular self-assembly 		  & Lehn (2002)           \\
	        & Supramolecular soft matter	 	  & M\"uller and Parisi (2015)\\
		& Reaction-diffusion systems		  & Kolmogorov et al.~(1937)\\
		& Oscillating reactions			  & Bray, W.C. (1921)	  \\
		& Oscillating catalytic reaction	  & Cox et al.~(1985)     \\
		& Liquid crystals			  & Rego e al.~(2010)     \\
		& Self-assembled monolayers		  & Love et al.~(2005)    \\
		& Langmuir-Blodgett films		  & Ritu (2016)           \\
		& Growth of SiGe nanostructures		  & Aqua et al.~(2013)    \\
Biology		& Biological systems 	 		  & Camazine et al.~(2001)\\
		& Pattern formation in slime molds and bacteria & Camazine et al.~(2001)\\
		& Feeding aggregations of bark beetles	  & Camazine et al.~(2001)\\
	        & Synchronized flashing among fireflies   & Camazine et al.~(2001)\\
		& Fish schooling                          & Camazine et al.~(2001)\\
		& Nectar source selection by honey bees   & Camazine et al.~(2001)\\
		& Trail formation in ants		  & Camazine et al.~(2001)\\
                & Swarm raids of army ants                & Camazine et al.~(2001)\\
		& Colony thermoregulation in honey bees   & Camazine et al.~(2001)\\
                & Comb patterns in honey bee colonies     & Camazine et al.~(2001)\\
		& Wall building by ants                   & Camazine et al.~(2001)\\
                & Termite mound building                  & Camazine et al.~(2001)\\
                & Construction algorithms in wasps        & Camazine et al.~(2001)\\
                & Dominance hierarchies in paper wasps    & Camazine et al.~(2001)\\ 
Social science 	& Social evolutionay systems 		  & Leydesdorff (1993)    \\
		& Learning algorithms 			  & Geach (2012)	  \\
	        & Coevolution in interdependent networks  & Wang et al.~(2014)    \\
Computer science& Cybernetics 				  & Ashby (1947)          \\
                & Cellular automata			  & Gacs (2000)           \\
		& Random graphs				  & Brooks (2009)         \\
		& Multi-agent systems			  & Kernbach (2008)       \\
		& Small-world networks			  & Watts and Strogatz (1998)\\
	        & Power grid network simulations          & Rohden et al.~(2012)  \\
	        & Self-organizing maps			  & Kohonen (1989)        \\	
		& Cloud computing systems		  & Zhang et al.~(2010)   \\
		& Moore's Law 		                  & Georgiev et al.~(2016)\\
\enddata
\end{deluxetable}

\begin{deluxetable}{llllllllll}
\tablecaption{Observed orbital periods and distances of the
planets from the Sun, and the nearest predicted harmonic orbit 
resonances $(H1:H2)$ or order of resonances $(l,k)$, orbital periods
$T[yr]$, observed and best-fit semi-major axes $a_{obs}$ and $a_{harm}$,
and ratios $a_{harm}/a_{obs}$.}
\tablewidth{0pt}
\tablehead{
\colhead{Planet}&
\colhead{number}&
\colhead{l}&
\colhead{k}&
\colhead{H1}&
\colhead{H2}&
\colhead{Period}&
\colhead{$a_{obs}$}&
\colhead{$a_{harm}$}&
\colhead{Ratio}\\
\colhead{}&
\colhead{}&
\colhead{}&
\colhead{}&
\colhead{}&
\colhead{}&
\colhead{[yr]}&
\colhead{[AU]}&
\colhead{[AU]}&
\colhead{$a_{harm}/a_{obs}$}}
\startdata
Mercury & 1 & 3 & 2 & 5 & 2 &  0.241 & 0.39 &  0.391 & 1.002 \\
Venus   & 2 & 2 & 3 & 5 & 3 &  0.615 & 0.72 &  0.711 & 0.988 \\
Earth   & 3 & 1 & 1 & 2 & 1 &  1.000 & 1.00 &  0.958 & 0.957 \\
Mars    & 4 & 3 & 2 & 5 & 2 &  1.881 & 1.52 &  1.504 & 0.989 \\
Ceres   & 5 & 3 & 2 & 5 & 2 &  4.601 & 2.77 &  2.823 & 1.019 \\
Jupiter & 6 & 3 & 2 & 5 & 2 & 11.862 & 5.20 &  5.179 & 0.996 \\
Saturn  & 7 & 2 & 1 & 3 & 1 & 29.457 & 9.54 &  9.225 & 0.967 \\
Uranus  & 8 & 1 & 1 & 2 & 1 & 84.018 & 19.19 & 18.943 & 0.987 \\
Neptune & 9 & 1 & 2 & 3 & 2 & 164.78 & 30.07 & 30.129 & 1.002 \\
Pluto   &10 &   &   &   &   & 284.40 & 39.48 & ...... & ..... \\
Mean	&   &   &   &   &   &        &       &        & $0.99\pm0.02$ \\
\enddata
\end{deluxetable}

\begin{deluxetable}{llll}
\tablecaption{Characteristics of randomness, self-organization, and
self-organized criticality systems or processes.}
\tablewidth{0pt}
\tablehead{
\colhead{Parameter}&
\colhead{Randomness}&
\colhead{Self-organizing}&
\colhead{Self-organized criticality}\\
\colhead{}&
\colhead{process}&
\colhead{process}&
\colhead{process}}
\startdata
Dynamics :		& events	& limit cycle 	  & avalanches	\\
Temporal structure :    & intermittent  & quasi-periodic  & intermittent, scale-free \\
Temporal size distribution : & exponential & quantized       & power law \\
Spatial structure :	& random	& ordered	  & fractal, scale-free	\\
Spatial size distribution : & Gaussian	& quantized       & power law \\
Entropy evolution : 	& increasing	& decreasing	  & invariant	\\
Physical condition :    & independency  & positive feedback & critical threshold \\
\enddata
\end{deluxetable}
\clearpage


\clearpage

\clearpage

\clearpage

\clearpage

\clearpage

\clearpage

\clearpage

\clearpage

\clearpage

\clearpage

\clearpage

\clearpage

\clearpage

\clearpage

\clearpage

\clearpage

\clearpage

\clearpage

\clearpage

\clearpage

\clearpage

\clearpage

\clearpage

\clearpage

\clearpage

\clearpage

\clearpage

\clearpage

\clearpage

\clearpage

\clearpage

\clearpage

\clearpage

\clearpage

\clearpage

\clearpage

\clearpage

\clearpage

\end{document}